\documentclass[superscriptaddress,twocolumn,10pt]{revtex4}

\usepackage{xr}
\usepackage{filecontents}

\usepackage[utf8]{inputenc}
\usepackage{microtype}
\usepackage{amsmath}												
\usepackage{amsfonts}
\usepackage{amssymb}												
\usepackage[hidelinks]{hyperref}
\usepackage{epstopdf}
\usepackage{xcolor}
\usepackage{graphicx}
\graphicspath{{img/}}

\newcommand{\B}{\boldsymbol}

\newcommand{\etal}{\textit{et al.}~}
\newcommand{\beq}{\begin{eqnarray}}
\newcommand{\eeq}{\end{eqnarray}}
\newcommand{\unsim}{{\sim}}

\usepackage{pdfpages}

\begin{document}
\title{Early life imprints the hierarchy of T cell clone sizes}
\author{Mario U. Gaimann}
\affiliation{Lewis-Sigler Institute for Integrative Genomics, Princeton University}
\affiliation{Arnold Sommerfeld Center for Theoretical Physics and Center for NanoScience, Department of Physics, Ludwig-Maximilians-Universität München}
\author{Maximilian Nguyen}
\affiliation{Lewis-Sigler Institute for Integrative Genomics, Princeton University}
\author{Jonathan Desponds}
\affiliation{NSF-Simons Center for Quantitative Biology, Northwestern University}
\author{Andreas Mayer}
\affiliation{Lewis-Sigler Institute for Integrative Genomics, Princeton University}

\begin{abstract}
    The adaptive immune system responds to pathogens by selecting clones of cells with specific receptors.
    While clonal selection in response to particular antigens has been studied in detail, it is unknown how a lifetime of exposures to many antigens collectively shape the immune repertoire.
    Here, through mathematical modeling and statistical analyses of T cell receptor sequencing data we demonstrate that clonal expansions during a perinatal time window leave a long-lasting imprint on the human T cell repertoire.
We demonstrate how the empirical scaling law relating the rank of the largest clones to their size can emerge from clonal growth during repertoire formation. We statistically identify early founded clones and find that they are indeed highly enriched among the largest clones. This enrichment persists even after decades of human aging, in a way that is quantitatively predicted by a model of fluctuating clonal selection.
Our work presents a quantitative theory of human T cell dynamics compatible with the statistical laws of repertoire organization and provides a mechanism for how early clonal dynamics imprint the hierarchy of T cell clone sizes with implications for pathogen defense and autoimmunity.
 \end{abstract}

\maketitle

\section{Introduction}

The hallmark of adaptive immunity is the generation of diversity through genetic recombination and clonal selection. Their interplay balances the breadth and specificity of the $\unsim 10^{12}$ T cells in the human body (Fig.~\ref{fig_statistics}A) \cite{Arstila1999,Farber2014}:
The genetic recombination of the T cell  receptor (TCR) locus, termed VDJ recombination, generates an enormous potential diversity of receptors ranging from early estimates of $\unsim 10^{15}$ \cite{Davis1988} to more recent estimates of $\unsim 10^{61}$ \cite{Mora2016} different possible receptor TCR$\alpha\beta$ heterodimers. Clonal selection expands the number of specific cells during an infection for effector functions, a fraction of which are retained over prolonged periods of time as immune memory \cite{Ahmed1996,Farber2014}.

Much progress has been made deciphering the mechanisms of regulation and control of T cell dynamics over the last decades \cite{Antia2005,Sallusto2010,Farber2014}.
However, much of that progress has focused on the dynamics of subsets of T cells specific to a particular antigen and has come from experiments in mice. An important open question is how exposures to many antigens over a human lifetime collectively shape our T cell repertoire \cite{Farber2014,Davis2018}.

High-throughput repertoire sequencing enables direct surveys of the diversity and clonal composition of T cells from human blood or tissue samples and thus promises to provide quantitative answers to this question 
\cite{Robins2009,Thomas2014,Britanova2016,Emerson2017,Oakes2017,Thome2016,Robins2010,Qi2014,Lindau2019,Joshi2019}.
However, while the TCR locus provides a natural barcode for clonal lineages due to its large diversity, this same diversity also makes inferring past clonal dynamics a challenging inverse problem, in particular given practical limitations on sequencing depth and temporal resolution in longitudinal studies.
Mathematical modeling can help address this challenge by solving the forward problem of linking clonal dynamics to emergent statistical patterns \cite{Desponds2016,Lythe2016,Dessalles2019,Altan-Bonnet2019,Greef2020}. Comparing patterns to data can provide insights about dynamics from static snapshots of repertoire organization in different individuals.
A particularly striking such pattern has been the observation of power-law scaling of clone sizes spanning several orders of magnitude \cite{Robins2009,Thomas2014,Desponds2016,Britanova2016,Emerson2017,Oakes2017}. In a typical sample of T cells from peripheral blood a large fraction (more than half in some individuals) of clones are only seen once within $10^5 - 10^7$ sampled sequences. At the same time the most abundant clones typically account for more than 1\% of all sequencing reads, equivalent to a clone size of $\unsim 10^{10}$ cells when extrapolating to the full repertoire. It is unknown when these large clonal expansions happen, and more broadly what determines the hierarchy of clone sizes.

Here, we use cohort and longitudinal human TCR repertoire sequencing data \cite{Britanova2016,Emerson2017,Chu2019,Lindau2019} to develop a statistical theory of T cell dynamics. We find that clonal expansions during repertoire formation establish clone size scaling, and we show that clonal selection pressures during adult life only slowly reshape the initial hierarchy.

\section{Results}

\subsection{A scaling law of human T cell repertoire organization}
\label{secresultsa}

\begin{figure*}
    \centering
     \includegraphics{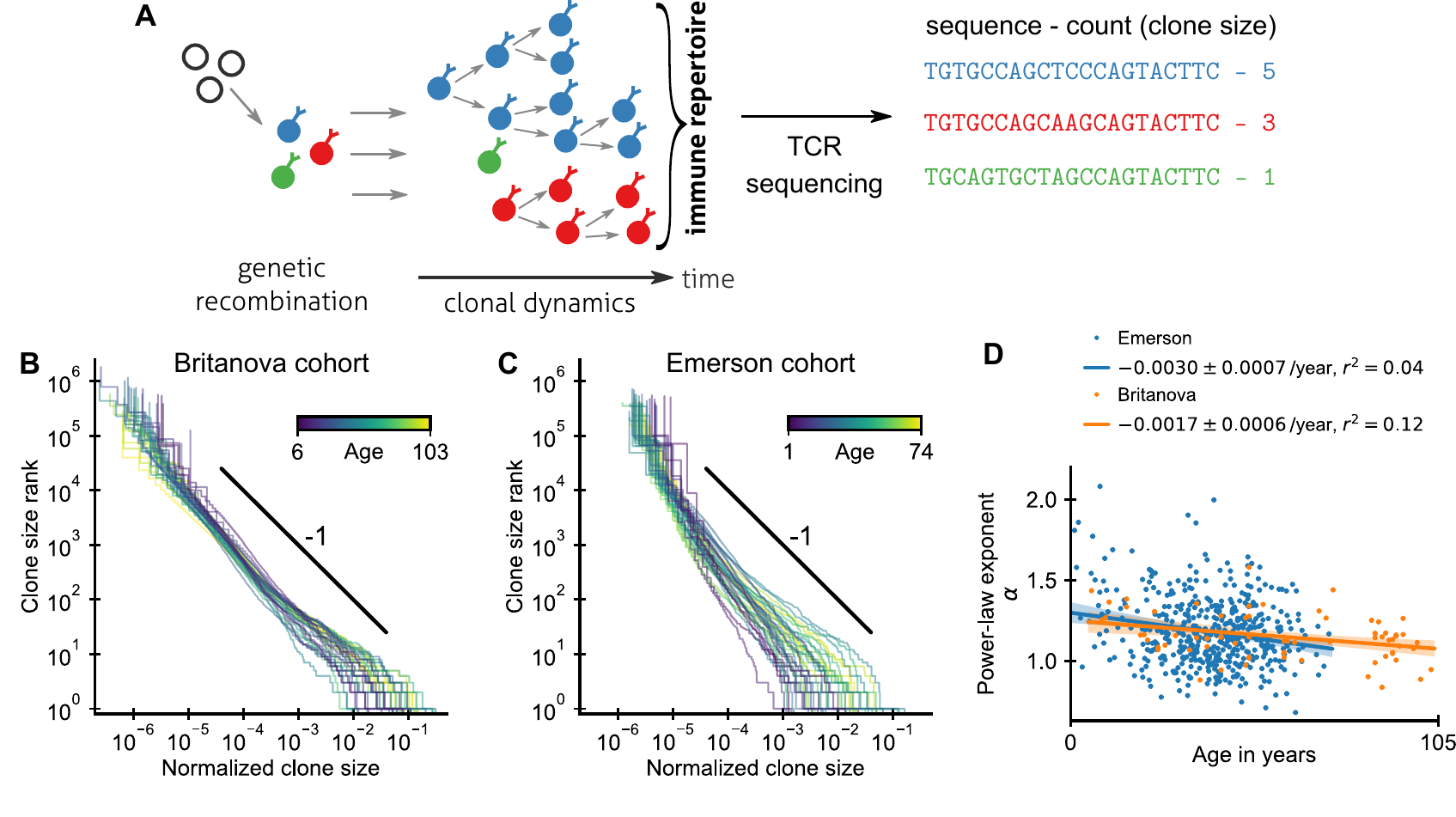}
    \caption{{\bf Statistics of human T cell repertoire organization.}
    (A) T cells with highly diverse receptors are created from progenitor cells through genetic recombination (left), which then undergo clonal selection (middle) together shaping the immune repertoire. The T cell receptor (TCR) locus acts as a natural barcode for clonal lineages, which can be read out by sequencing (right). 
    (B, C) Clone size distributions in two large cohort studies of human blood samples using disparate sequencing protocols display a power-law relationship between the rank and size of the largest clones.
Each line shows the clone size distribution in an individual. Ages are color coded as indicated in the legend. The black line shows a power law with a slope of -1 for visual comparison.
    Clone sizes were normalized by the total number of reads and by the memory cell fraction to account for variations in sampling depth and in the subset composition of peripheral blood, respectively (Fig.~\ref{fig_clonesizes_stepbystep}).
    Only a single individual is displayed per two-year age bracket to improve visibility.
    (D) Power-law exponents as a function of the age (legend: linear regression slope and coefficient of determination).
    Data sources: B, D \cite{Britanova2016}, C, D \cite{Emerson2017}.
  } \label{fig_statistics}
\end{figure*}

An important statistic to summarize repertoire organization is the clone size distribution, which tabulates the number of clones found at different multiplicities within a repertoire or sample. Multiple previous studies have shown that these distributions are heavy-tailed \cite{Robins2009,Thomas2014,Britanova2016,Emerson2017,Oakes2017}, but potential confounding by noise introduced during the sequencing process has remain debated \cite{Altan-Bonnet2019} and systematic analyses of how variable these distributions are across healthy individuals have been lacking.
To fill these gaps we reanalyzed data from two large-scale cohort repertoire sequencing studies, which used fundamentally different sequencing pipelines and thus have different sources of noise (Material and Methods).
Both studies sequenced the locus coding for the hypervariable TCR CDR3$-\beta$ chain from peripheral blood samples of healthy human volunteers spanning a large range of ages (Fig.~\ref{fig_cohortages}).

After normalizing clone sizes to account for variations in sampling depth and subset composition (Fig.~\ref{fig_clonesizes_stepbystep}), we found that the tails of the clone size distributions collapsed to the same statistical law across individuals and cohorts (Fig.~\ref{fig_statistics}B,C): Ranking clones by decreasing size, the rank of the largest clones approximately scales with their size $C$ as a power law,
\begin{equation} \label{eqscaling}
    \mathrm{rank} \sim C^{-\alpha},
\end{equation}
where $\alpha$ is a scaling exponent.
To quantify the apparent similarity of the scaling relationship we determined $\alpha$ for each sample by maximum likelihood estimation. Only a small fraction of all T cells are sampled, which poses a challenge because subsampling a power law leads to deviations from scaling at small clone sizes \cite{Stumpf2005}. To overcome this challenge we used a trimming procedure and excluded clones smaller than a minimal size from the fitting, which decreases bias arising from subsampling (SI Text~\ref{subsampling}).
Determined in this subsampling-robust manner the fitted power-law exponents agree remarkably well within the range of ages covered by both cohorts (Fig.~\ref{fig_statistics}D); with $\alpha = 1.17 \pm 0.03$ (mean $\pm$ standard error (SE)) and $\alpha = 1.18 \pm 0.01$ in the Britanova and Emerson cohort, respectively.
Moreover, the fitted exponents varied little between individuals in both cohorts; with a sample standard deviation of fitted exponents of $0.14$ and $0.21$, respectively.
The agreement of the mean exponents is noteworthy given the different sequencing pipelines and provides strong evidence that the scaling relationship (Eq.~\ref{eqscaling}) is a true feature of the clone size distribution and not of the measurement process. 

\begin{figure*}
 \begin{center}
     \includegraphics{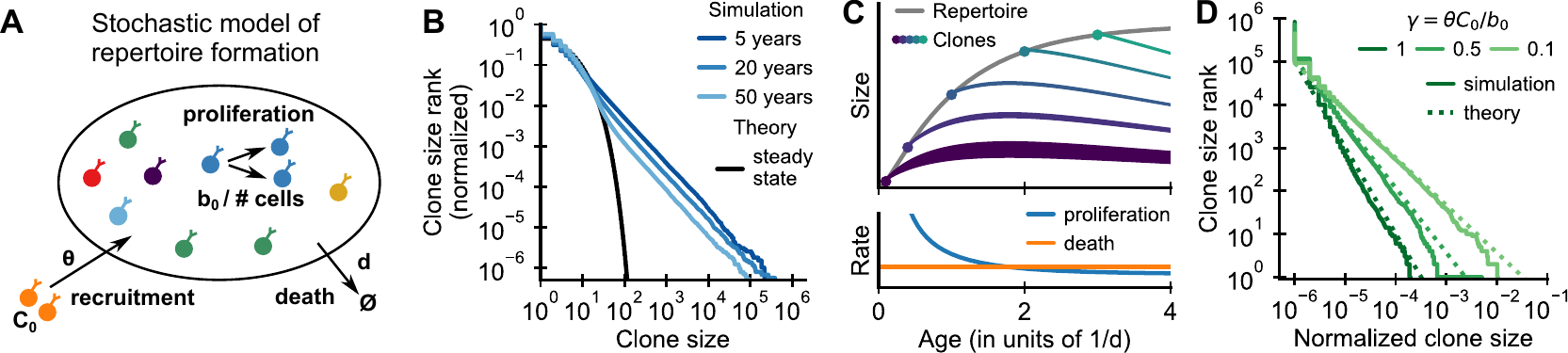}
 \end{center}
    \caption{{\bf Emergence of power-law scaling of clone sizes in a minimal model of repertoire formation.} (A) Sketch of the stochastic dynamics of recruitment, proliferation, and death of T cells. Proliferation is inversely proportional to total repertoire size modeling increasing competition during repertoire growth.
    (B) Clone size distributions in simulated repertoires display power-law scaling (blue lines), in contrast to steady-state predictions (black line, SI Text~Eq.~\ref{eqneutral}). 
    (C) Illustration of the mechanism: Early in life rates of proliferation exceed clonal turnover (lower panel). As the total repertoire size increases (grey line, upper panel) the proliferation rate decreases due to increased competition. The dynamics of selected clones after their recruitment marked by a dot is indicated by colored lines (upper panel). The line position shows the cumulative size of all prior clones, while the line width indicates the size of the clone (not to scale). The earlier a clones is recruited the larger it expands during the period of overall repertoire growth. 
    (D) Dependence of the clone size distribution on parameters. Simulated repertoires at 5 years of age were subsampled to $10^6$ cells to mimick the experimental sampling depth (solid lines). The simulated data closely follow predictions from a continuum theory of repertoire formation (dashed lines).
    Parameters: (B,D) $d=0.2$/year, $C_0=1$, $\theta = 10^6$/year; (B) $\gamma=0.1$ (implying $b_0 = 10^7$/year). 
  } \label{fig_model}
\end{figure*}

What drives the emergence of a power-law distributed hierarchy of clone sizes? Given the reproducibility of the scaling law across individuals we might hope for a statistical explanation independent of the precise antigenic history that drives the expansion of specific cells in an individual. To test hypotheses about mechanisms underlying scaling we describe repertoire dynamics using a general mathematical framework based on effective stochastic rate equations for the recruitment of new clones, and the proliferation and death of already existing clones within a T cell compartment (Material and Methods). In macroecology, where such reductionist approaches have a long history, simple neutral models within this framework have had surprising success in describing species abundance distributions only accounting for demographic stochasticity \cite{Volkov2003}, but this source of variability is insufficient to account for the observed breadth of T cell clone sizes \cite{Desponds2016,Greef2020} (for a detailed discussion see SI Text~\ref{si_neutral}). The failure of this null model has prompted a search for other mechanisms that explain scaling.

To constrain this search we analyzed how fitted exponents varied with age. In particular, we expected a substantially steeper tail in young individuals based on a finite time solution we derived for a previously proposed model of how power-law scaling can emerge from the cumulative effect of temporal fluctuations in clonal growth rates \cite{Desponds2016} (SI Text~\ref{si_ffconvergence}).
While exponents overall decreased slightly with age, the dependence on age accounted for surprisingly little variation in both cohorts (Fig.~\ref{fig_statistics}D and Fig.~\ref{fig_exponent_cmv}).  
Notably, scaling is established within the first decade of life, with significant clone size variability existing as early as at birth (Fig.~\ref{fig_clonesizes_britanova_cordblood}), defying previous model predictions.

\subsection{A mechanism for the emergence of scaling during repertoire formation}
\label{secresultsb}

We hypothesized that scaling might result from clonal expansions during repertoire formation, which would naturally explain the early onset of scaling.
Our hypothesis is based on experimental evidence in mice \cite{Campion2002,Min2003} and human \cite{Rufer1999,Schonland2003} that repertoire formation is driven not only by increased thymic output, but also by large proliferative expansion of some T cell clones.
Additionally, multiple studies \cite{Hammarlund2003,Pogorelyy2017,Tanno2020} have shown that some T cell clones can persist over multiple decades, which suggested to us that clonal turnover might be sufficiently slow (see also SI Text~\ref{si_neutral_timescale}) for transient expansionary dynamics early in life to shape repertoire organization over prolonged periods of time.

\begin{figure*}
    \centering
     \includegraphics{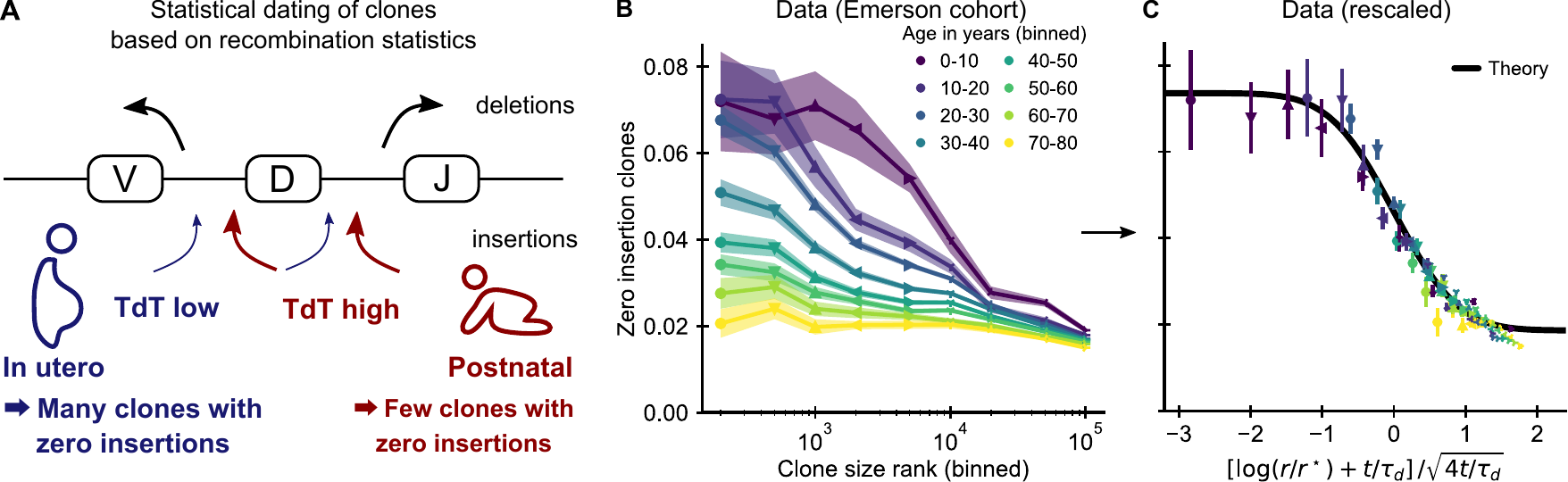}
    \caption{{\bf Statistical dating of clones reveals that early expansions have a long-lasting effect.}
    (A) Genetic recombination of a TCR involves the choice of a V, D, and J region among multiple genomically-encoded templates as well as the deletion and insertion of nucleotides at both the VD and DJ junction. The enzyme TdT, which is responsible for nucleotide insertions, is not expressed during early fetal development. This allows a statistical dating of clonal ages, as clones with zero insertions at both junctions constitute a much larger fraction of all clones during a fetal and perinatal time window.
    (B) Fraction ($\pm$ SE) of clones with zero insertions as a function of age and clone size. Clones are binned by their size into non-overlapping bins (rank 1 to 500, 501 to 1000, and so on; upper values are indicated on the x-axis).
    (C) Same data as in B displayed with a rescaled x-axis using fitted parameters $\tau_d = 9.1\pm0.5\, \mathrm{years}, r^\star = 1.2\pm0.2 \cdot 10^4$. The data collapses onto a sigmoidal function predicted by theory (SI Text~Eq.~\ref{eqerfc}) with fitted $p_{0, -} = 0.074\pm0.004$, $p_{0,+} = 0.0187\pm0.0005$ (black line).
    Data source: \cite{Emerson2017}.
    }
\label{figzeroinsertion}
\end{figure*}

To test our hypothesis we constructed a minimal model of repertoire formation based on known T cell biology (Fig.~\ref{fig_model}A). Following previous work \cite{Bains2009,Lythe2016} we assume that the proliferation rate $b$ is inversely proportional to the total number of cells already in the repertoire to model increased proliferation early in life. This dependence of proliferation rate on repertoire size arises in a simple mechanistic model of T cell competition (SI Text~\ref{si_1oNmechanism}). For simplicity we further assume that the rates of cellular death $d$ and recruitment of new clones $\theta$ are constant. Importantly, recruitment of new clones and total expansion of already existing clones maintain a constant ratio throughout development under these assumptions in line with findings that the fraction of cells with T cell receptor excision circles, which are diluted during peripheral division, is constant during fetal development \cite{Schonland2003} and infancy \cite{Bains2009}.

We simulated the model starting from an empty repertoire and found that large clones displayed power-law scaling (Fig.~\ref{fig_model}B blue lines). The simulation results contrast with steady state predictions (Fig.~\ref{fig_model}B black line), where the model effectively reduces to the neutral null model introduced earlier (SI Text~\ref{si_steadystate_repertoireformation}). Thus we find that repertoire formation can produce transient but long-lasting power-law scaling of clone sizes.

To obtain intuitive insight into how scaling is established, we developed a continuum theory of clonal dynamics during repertoire growth (SI Text~\ref{si_continuum_theory}).
We find that the clone size $C_i$ of the $i$-th clone recruited at time $t_i$ follows a subexponential growth law $C_i(t) = C_0 \, (t/t_i)^{1/(1+\gamma)}$, where $\gamma$ is the ratio of the contribution of recruitment and proliferation to overall compartment growth. Clones recruited early grow large deterministically until competition lowers proliferation rates below the death rate (Fig.~\ref{fig_model}C, lower panel). Different clones are recruited at different times and thus have more or less time to grow (Fig.~\ref{fig_model}C, upper panel), which leads to a clone size distribution that follows power-law scaling with an exponent
$
    \alpha = 1+\gamma.
$ 
We note that this origin of the power-law scaling is closely related to a well-known generative mechanism for power-laws first studied by Yule \cite{Yule1924} (for a detailed discussion see SI Text~\ref{si_preferentialattachment}).

The predicted exponent closely matches simulation results for different values of $\gamma$ (Fig.~\ref{fig_model}D dashed lines).
Intuitively, when recruitment rates are higher clones founded early have less time to outgrow later competitors, and thus the power law is steeper ($\alpha$ is larger).
Importantly, in the biological parameter regime in which proliferation dominates, $\gamma < 1$, the exponent is compatible with experiments (Fig.~\ref{fig_statistics}B-E). We thus find, that the model -- without fine tuning of parameters -- reproduces the observed scaling exponent.

To expose a basic mechanism capable of producing broad clone size distributions we have kept the model deliberately simple.
More detailed models demonstrate the conditions and limits on the generalizability of this mechanism  (SI Text~\ref{secmodelrelaxations}). Variable recruitment sizes only affect the distribution of small clones (SI Text~\ref{fig_introsizevar}); while a saturation of proliferation rates, or competition between subsets of T cells for specific resources maintain distributions at small and intermediate sizes while leading to cutoffs for the largest clones (Fig.~\ref{fig_relaxations} and Fig.~\ref{figspecific}).

\subsection{Long-lived incumbency advantage shows early expansions imprint clone size hierarchy}
\label{secresultsc}

Our proposed theory for the rapid emergence of scaling predicts that large clones have expanded massively during repertoire formation. 
To test this prediction we need to trace the dynamics of early founded clones. To this end, we exploit a change in the recombination statistics taking place during fetal development \cite{Feeney1991,Rechavi2015,Park2020} (Fig.~\ref{figzeroinsertion}A). While T cells are produced by the thymus from the late first trimester the enzyme terminal deoxynucleotidyl transferase (TdT), which inserts non-templated nucleotides during VDJ recombination, is not expressed until the mid second trimester \cite{Park2020}. Therefore many more T cells in fetal and neonatal blood have zero insertions than expected by the adult recombination statistics \cite{Rechavi2015}. This enables a statistical dating of individual clones in a repertoire based on their sequence \cite{Sethna2017,Pogorelyy2017}.

\begin{figure*}
    \centering
    \includegraphics{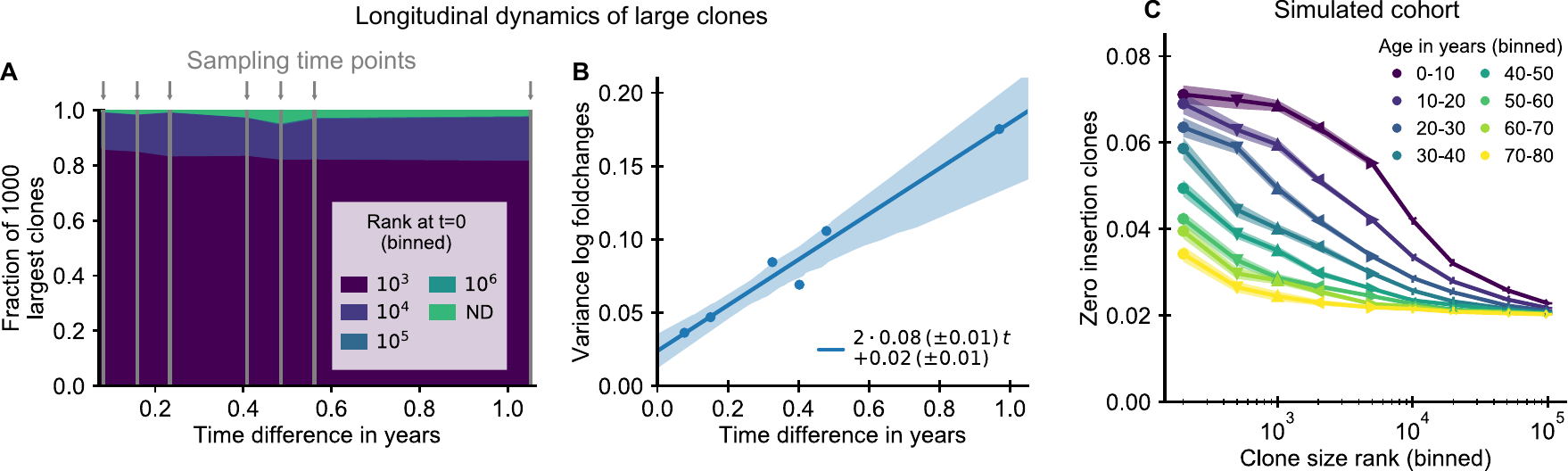}
    \caption{{\bf The small magnitude of longitudinal clone size fluctuations implies a slow reordering of the clone size hierarchy.}
    (A,B) Longitudinal clonal dynamics in a healthy adult over a one year time span.
    (A) Fraction of the 1000 largest clones that fall within a specific clone size rank bin at the earliest time point. A small number of clones was not detected at all at the first time point (ND) likely representing recently expanded clones. All other clones were already among the largest clones initially.
    (B) Variance of log-foldchanges in clone size as a function of time difference for the 250 largest clones. 
    (C) Fraction of clones with zero insertions as a function of age and clone size in a simulated cohort using a magnitude of clonal growth rate fluctuations inferred from the longitudinal data.
    Data source: \cite{Chu2019}.
    }
\label{figlongitudinalmain}
\end{figure*}

If our model is correct we expect abundant clones to be more likely to have zero insertions than smaller clones. 
Analyzing data from the Emerson cohort we find that zero insertion clones are indeed highly enriched within the most abundant clones (Fig.~\ref{figzeroinsertion}B). This generalizes a previous report of such an enrichment within the naive compartment \cite{Pogorelyy2017}. The large cohort size allows us to perform a fine-grained analysis of how the fraction of zero-insertion clones depends on clonal abundance and age. We find that enrichment is particularly pronounced in the young and decreases with age at different speeds depending on clone size. Among the largest clones many more still have zero insertions than expected under the adult recombination statistics even multiple decades after repertoire formation. This suggests that the incumbent large clones created during repertoire formation are only slowly replaced by clones expanding later in life.
Additional analyses rule out other potential explanations for the relation between insertion statistics and clonal abundance.
Firstly, sequences with zero insertions are similarly enriched among the largest clones in productive and unproductive sequences (Fig.~\ref{fig_zeroinsertion_out}) demonstrating that convergent selection pressures during adult life are not a primary source of the higher abundance of these clones.
Secondly, while abundant clones are also enriched for sequences with known antigen specificity (Fig.~\ref{fig_invdjdb}) and sequences likely to be convergently recombined (Fig.~\ref{fig_pgen}), these enrichment do not show the same striking dependence on age.
Furthermore, we find that zero insertion clones were consistently less enriched in individuals infected by cytomegalovirus (Fig.~\ref{fig_zeroinsertion_cmv}), in contrast to the hypothesis that this infection might drive their expansion \cite{Pogorelyy2017}. 
Taken together, these analyses support the conclusion that dynamics during the perinatal time window of repertoire formation leave a long-lasting imprint on the T cell clonal hierarchy well into adulthood.

\subsection{Longitudinal clone size fluctuations predict the dynamics of the clone size hierarchy with aging}

Building on this successful validation of a core prediction of our theory we asked whether we could leverage the detailed pattern of enrichments at different ranks and ages to quantify how much being part of the wave of early expansions determines the fate of a clone relative to other sources of clone size variability.
To this end we extended our model beyond repertoire formation and allowed clonal proliferation rates to fluctuate over time to model the net effect of clonal selection by changing antigenic stimuli during adult life \cite{Desponds2016} (Eq.~\ref{eqbirthfluctuating}).

To determine a biologically plausible fluctuation strength we analyzed the variability of clone sizes over time in a longitudinal study of T cell dynamics \cite{Chu2019}. We first analyzed to what extent recently expanded clones contribute to the tail of the clone sizes, and found that only a small fraction of the largest clones in any sample were not already large at the earliest time point (Fig.~\ref{figlongitudinalmain}A and Fig.~\ref{fig_longitudinal_provenance}). To minimize confounding by transient dynamics affecting these clones, we excluded these clones from further analysis. We found that large clones had remarkably stable abundances over time, which we quantified by calculating the variance of log-foldchanges in clone size between the second and every subsequent time point (Fig.~\ref{figlongitudinalmain}B and Fig.~\ref{fig_longitudinal_msd}). The variability of clone sizes increased linearly over time as expected theoretically, from which we determined a magnitude of net growth rate fluctuations compatible with the slope of increase (SI Text~\ref{si_longitudinal_modeling}).

Using the fitted fluctuation strength we constructed an {\it in silico} cohort of individuals of different ages according to the extended model (SI Text~\ref{secsimulations}). In short, we computationally assigned each newly recruited clone to have zero insertions in a way that mimicks the change in fetal recombination statistics, and we simulated memory repertoire dynamics based on the combined effect of early expansion and fluctuating clonal selection. The enrichment of zero insertion clones in the simulated cohort (Fig.~\ref{figlongitudinalmain}C) closely recapitulated the empirical findings using plausible parameter values (SI Text~\ref{si_parameters}). Notably, the more long-lasting enrichment of zero insertion clones among the very largest clones is also found in the simulated cohort, and the timescales over which the enrichment decays agree remarkably well. 

For a direct comparison between theory and experiment we mathematically analyzed how fluctuating selection reorders the initially established clone size hierarchy. The analytical results suggest a two-parameter rescaling of the enrichment of zero insertion clones as a general test of our theory  (SI Text~\ref{seclongtermdynamics}). The two parameters of the theory, $\tau_d$ and $r^\star$, can be fitted from the enrichment data (SI Text~\ref{sec_dataanalysis}). Rescaling the data with the fitted parameters leads to a collapse of all data points onto a single curve predicted by theory for both the simulated (Fig.~\ref{fig_mastercurve_model_collapse}) and experimental cohort (Fig.~\ref{figzeroinsertion}C). The fitted parameters quantify key features of long-term repertoire dynamics, with $\tau_d$ characterizing the timescale over which fluctuations change the clone size hierarchy, and $r^\star$ being related to the number of clones recruited during early repertoire growth. In line with the long-lived enrichment of zero insertion clones, the fitting reveals a remarkably slow timescale of about a decade over which the clone size hierarchy is reordered during healthy aging. The fitted $r^\star$ indicates that early repertoire formation involves the expansion of a large number of different clones.
Overall, the agreement between theory and data demonstrates that our model quantitatively captures how early expansions and ongoing fluctuating selection together shape the clone size hierarchy.

\section{Discussion}

The evolution of the adaptive immune system has endowed vertebrates with the ability to adapt to pathogens that evolve on a timescale faster than host reproduction \cite{Mayer2016}. However, this ability comes with a cost: every generation needs to rebuild immune memory anew. As the organism first comes into contact with the outside world it quickly needs to train its adaptive immune system to tolerate innocuous antigens and build up immune memory against pathogens. Here, we have shown that this process of rapid adaptation leaves a long-lasting imprint on the organization of the human T cell repertoire. More broadly, we propose a theory of repertoire dynamics that quantitatively describes how early expansions during repertoire formation combine with a lifetime of exposures to cumulatively shape the T cell hierarchy. Notably, we find that the T cell repertoire is remarkably stable over time in adult individuals outside of the punctuated expansions and contractions of specific clones in acute responses. Our study demonstrates that despite its vast complexity repertoire dynamics is partially predictable by quantitative models. The model predictions can help guide future longitudinal studies, which in turn will allow refinements of modeling assumptions. The current work thus provides a stepping stone towards a detailed quantitative understanding of T cell dynamics that we hope will ultimately power the rational development of immunodiagnostics and therapeutics.

The general mechanism we describe for imprinting in the adaptive immune system provides a unified lens through which to view a number of converging lines of evidence about how a developmental time window shapes adaptive immunity \cite{Guerau-de-arellano2009,Farber2014,Gostic2016,Constantinides2019,Li2019a,Davenport2020,Hong2020}.
In our model, overall repertoire growth early in life amplifies the effect of any early exposures, as they lead to much larger clonal expansions than similar exposures happening after the homeostatic repertoire size is reached.
We thus expect early pathogen exposures to be particularly potent, as has been observed in influenza, where disease severity across age cohorts for different strains depends on the first exposure \cite{Gostic2016}. Conversely, we expect the presence of tolerizing factors early in life to be particularly crucial during repertoire formation to avoid autoimmunity, as has been observed for the autoimmune regulator gene AIRE, for which expression is only essential during a perinatal time window \cite{Guerau-de-arellano2009}.

A limitation of datasets used in this study is that they do not provide direct information about the phenotypic characteristics of cells belonging to different clones. Repertoire sequencing of phenotypically sorted blood samples shows that the largest clones predominantly consist of cells with memory phenotype (SI Text~\ref{phenotypes}). This indirectly suggests that the clonal expansions during repertoire formation produce memory cells as we have assumed in our simulated cohort (Fig.~\ref{figlongitudinalmain}C). Supporting this interpretation, a substantial number of memory cells circulate in the blood quickly following birth \cite{Shearer2003} and recent evidence suggests that memory-like T cells are already generated in the human intestine even before birth \cite{Li2019a}. However alternatively, early expansions could also set up a broad distribution of naive T cell clone sizes \cite{Greef2020}, whose hierarchy would then need to be roughly maintained during the transition into memory to be compatible with the observed impact of early expansions on the hierarchy of the most abundant clones. Advances in single-cell technologies linking TCR sequencing and cellular phenotyping could help differentiate between these scenarios in the future.

An important questions raised by our work is which antigens drive the expansion of early T cell clones. To address this question it will be necessary to determine the exposures that imprint the abundance of these clones, as has been done recently for mucosal-associated invariant T cells \cite{Constantinides2019}, a subset of non-conventional T cells. Going forward, the highly abundant clones with sequences close to the genetically inherited gene templates resulting from the absence of TdT expression during early fetal development are a particularly interesting target of study. They might constitute an evolutionarily controlled set of innate-like defenses within the adaptive immune system. Determining what imprints their abundances will help resolve the question of whether their large abundances are simply a byproduct of rapid repertoire formation or whether these clones serve particular functions. 

\section{Material and Methods}

\subsection{Repertoire sequencing data}

We analyzed T cell repertoire sequencing data from the two largest published cohort studies of healthy human volunteers by Britanova \etal \cite{Britanova2016} and Emerson \etal \cite{Emerson2017} and from a longitudinal study by Chu \etal \cite{Chu2019}, detailed descriptions of which are provided in the SI Text Extended Methods.

In short, Britanova \etal \cite{Britanova2016} sequenced reverse transcribed mRNA with added unique molecular identifiers (UMIs), while Emerson \etal \cite{Emerson2017} and Chu \etal \cite{Chu2019} sequenced genomic DNA coding for this region without the addition of UMIs. These approaches have complementary strengths: The addition of UMIs allows to correct for stochasticity during polymerase chain reaction (PCR) amplification and sequencing artifacts, while DNA sequencing removes the influence of cell-to-cell gene expression heterogeneity. 

\subsection{Mathematical framework}
We describe T cell dynamics using the following general set of stochastic rate equations. The class of models we consider are known in the mathematical literature as birth-death-immigration models. The number of cells $C_i$, $i = 1, \dots, M$ of each of the $M$ clones in the repertoire changes according to
\begin{align}
    \text{proliferation:} \quad C_i &\xrightarrow{b_i(\B C, \B X, t) C_i} C_i + 1, \label{eqbirth}\\
    \text{death:} \quad C_i &\xrightarrow{d_i(\B C, \B X, t) C_i} C_i -1, \label{eqdeath}
\end{align}
where the rate of proliferation $b_i(\B C, \B X, t)$ or cell death $d_i(\B C, \B X, t)$ generally can depend on the repertoire composition $\B C$, on the time $t$, and on the state of the environment $\B X(t)$ representing e.g. the levels of different antigens and cytokines in the organism at a given time.
We furthermore consider that new clones are added at rate $\theta(\B X, t)$ at a size $C_0$, 
\begin{align}
    \text{recruitment:} \quad &\xrightarrow{\theta(\B X, t)} C_{M+1} = C_0. \label{eqimmigration}
\end{align}
This recruitment represents thymic output and antigen-driven differentiation of naive cells for the naive and memory compartment, respectively. 

In Sec.~\ref{secresultsb} we study the influence of repertoire formation on clone sizes under the following assumptions:
\begin{equation}
    b_i(\B C, \B X, t) = b_0/N, \; d_i(\B C, \B X, t) = d, \; \theta(\B X, t) = \theta
\end{equation}
where $N(t)=\sum_{j=0}^{M(t)} C_j(t)$ is the total repertoire size.
In Sec.~\ref{secresultsc} we modify this model by adding a noise term that describes the effective influence of environmental variations on clonal proliferation,
\begin{equation} \label{eqbirthfluctuating}
    b_i(\B C, \B X, t) = b_0/N + \sqrt{2} \sigma \eta_i(t),
\end{equation}
where $\langle \eta_i(t) \eta_j(t') \rangle = \delta_{ij} \delta(t-t')$. 

\vspace{2mm}
{\bf Acknowledgements.}
We thank William Bialek, Curtis Callan, Ivana Cvijovic, Yuval Elhanati, Simone Mayer, Mikhail Pogorelyy, and Ned Wingreen for discussions and comments on the manuscript. This work was supported by a DAAD RISE Worldwide fellowship (MUG), the NSF-Simons Center for Quantitative Biology under grants Simons Foundation SFARI/597491-RWC and National Science Foundation 17764421 (JD), and a Lewis–Sigler fellowship (AM).

\vspace{2mm}
{\bf Author contributions.} AM conceptualized the problem and supervised research, MUG and AM wrote the draft manuscript, all authors performed analytic calculations, simulations, and reviewed the manuscript.

\bibliography{library}

\begin{thebibliography}{47}
\expandafter\ifx\csname natexlab\endcsname\relax\def\natexlab#1{#1}\fi
\expandafter\ifx\csname bibnamefont\endcsname\relax
  \def\bibnamefont#1{#1}\fi
\expandafter\ifx\csname bibfnamefont\endcsname\relax
  \def\bibfnamefont#1{#1}\fi
\expandafter\ifx\csname citenamefont\endcsname\relax
  \def\citenamefont#1{#1}\fi
\expandafter\ifx\csname url\endcsname\relax
  \def\url#1{\texttt{#1}}\fi
\expandafter\ifx\csname urlprefix\endcsname\relax\def\urlprefix{URL }\fi
\providecommand{\bibinfo}[2]{#2}
\providecommand{\eprint}[2][]{\url{#2}}

\bibitem[{\citenamefont{Arstila et~al.}(1999)\citenamefont{Arstila, Casrouge,
  Even, Kanellopoulos, and Kourilsky}}]{Arstila1999}
\bibinfo{author}{\bibfnamefont{T.~P.} \bibnamefont{Arstila}},
  \bibinfo{author}{\bibfnamefont{A.}~\bibnamefont{Casrouge}},
  \bibinfo{author}{\bibfnamefont{J.}~\bibnamefont{Even}},
  \bibinfo{author}{\bibfnamefont{J.}~\bibnamefont{Kanellopoulos}},
  \bibnamefont{and}
  \bibinfo{author}{\bibfnamefont{P.}~\bibnamefont{Kourilsky}},
  \bibinfo{journal}{Science} \textbf{\bibinfo{volume}{286}},
  \bibinfo{pages}{958} (\bibinfo{year}{1999}).

\bibitem[{\citenamefont{Farber et~al.}(2014)\citenamefont{Farber, Yudanin, and
  Restifo}}]{Farber2014}
\bibinfo{author}{\bibfnamefont{D.~L.} \bibnamefont{Farber}},
  \bibinfo{author}{\bibfnamefont{N.~A.} \bibnamefont{Yudanin}},
  \bibnamefont{and} \bibinfo{author}{\bibfnamefont{N.~P.}
  \bibnamefont{Restifo}}, \bibinfo{journal}{Nature Reviews Immunology}
  \textbf{\bibinfo{volume}{14}}, \bibinfo{pages}{24} (\bibinfo{year}{2014}).

\bibitem[{\citenamefont{Davis and Bjorkman}(1988)}]{Davis1988}
\bibinfo{author}{\bibfnamefont{M.~M.} \bibnamefont{Davis}} \bibnamefont{and}
  \bibinfo{author}{\bibfnamefont{P.~J.} \bibnamefont{Bjorkman}},
  \bibinfo{journal}{Nature} \textbf{\bibinfo{volume}{334}},
  \bibinfo{pages}{395} (\bibinfo{year}{1988}).

\bibitem[{\citenamefont{Mora and Walczak}(2016)}]{Mora2016}
\bibinfo{author}{\bibfnamefont{T.}~\bibnamefont{Mora}} \bibnamefont{and}
  \bibinfo{author}{\bibfnamefont{A.}~\bibnamefont{Walczak}},
  \bibinfo{journal}{arXiv preprint arXiv:1604.00487} pp. \bibinfo{pages}{1--10}
  (\bibinfo{year}{2016}).

\bibitem[{\citenamefont{Ahmed and Gray}(1996)}]{Ahmed1996}
\bibinfo{author}{\bibfnamefont{R.}~\bibnamefont{Ahmed}} \bibnamefont{and}
  \bibinfo{author}{\bibfnamefont{D.}~\bibnamefont{Gray}},
  \bibinfo{journal}{Science} \textbf{\bibinfo{volume}{272}},
  \bibinfo{pages}{54} (\bibinfo{year}{1996}).

\bibitem[{\citenamefont{Antia et~al.}(2005)\citenamefont{Antia, Ganusov, and
  Ahmed}}]{Antia2005}
\bibinfo{author}{\bibfnamefont{R.}~\bibnamefont{Antia}},
  \bibinfo{author}{\bibfnamefont{V.~V.} \bibnamefont{Ganusov}},
  \bibnamefont{and} \bibinfo{author}{\bibfnamefont{R.}~\bibnamefont{Ahmed}},
  \bibinfo{journal}{Nature Reviews Immunology} \textbf{\bibinfo{volume}{5}},
  \bibinfo{pages}{101} (\bibinfo{year}{2005}).

\bibitem[{\citenamefont{Sallusto et~al.}(2010)\citenamefont{Sallusto,
  Lanzavecchia, Araki, and Ahmed}}]{Sallusto2010}
\bibinfo{author}{\bibfnamefont{F.}~\bibnamefont{Sallusto}},
  \bibinfo{author}{\bibfnamefont{A.}~\bibnamefont{Lanzavecchia}},
  \bibinfo{author}{\bibfnamefont{K.}~\bibnamefont{Araki}}, \bibnamefont{and}
  \bibinfo{author}{\bibfnamefont{R.}~\bibnamefont{Ahmed}},
  \bibinfo{journal}{Immunity} \textbf{\bibinfo{volume}{33}},
  \bibinfo{pages}{451} (\bibinfo{year}{2010}).

\bibitem[{\citenamefont{Davis and Brodin}(2018)}]{Davis2018}
\bibinfo{author}{\bibfnamefont{M.~M.} \bibnamefont{Davis}} \bibnamefont{and}
  \bibinfo{author}{\bibfnamefont{P.}~\bibnamefont{Brodin}},
  \bibinfo{journal}{Annual review of immunology} \textbf{\bibinfo{volume}{36}}
  (\bibinfo{year}{2018}).

\bibitem[{\citenamefont{Robins et~al.}(2009)\citenamefont{Robins, Campregher,
  Srivastava, Wacher, Turtle, Kahsai, Riddell, Warren, and
  Carlson}}]{Robins2009}
\bibinfo{author}{\bibfnamefont{H.~S.} \bibnamefont{Robins}},
  \bibinfo{author}{\bibfnamefont{P.~V.} \bibnamefont{Campregher}},
  \bibinfo{author}{\bibfnamefont{S.~K.} \bibnamefont{Srivastava}},
  \bibinfo{author}{\bibfnamefont{A.}~\bibnamefont{Wacher}},
  \bibinfo{author}{\bibfnamefont{C.~J.} \bibnamefont{Turtle}},
  \bibinfo{author}{\bibfnamefont{O.}~\bibnamefont{Kahsai}},
  \bibinfo{author}{\bibfnamefont{S.~R.} \bibnamefont{Riddell}},
  \bibinfo{author}{\bibfnamefont{E.~H.} \bibnamefont{Warren}},
  \bibnamefont{and} \bibinfo{author}{\bibfnamefont{C.~S.}
  \bibnamefont{Carlson}}, \bibinfo{journal}{Blood}
  \textbf{\bibinfo{volume}{114}}, \bibinfo{pages}{4099} (\bibinfo{year}{2009}).

\bibitem[{\citenamefont{Thomas et~al.}(2014)\citenamefont{Thomas, Best,
  Cinelli, Reich-Zeliger, Gal, Shifrut, Madi, Friedman, Shawe-Taylor, and
  Chain}}]{Thomas2014}
\bibinfo{author}{\bibfnamefont{N.}~\bibnamefont{Thomas}},
  \bibinfo{author}{\bibfnamefont{K.}~\bibnamefont{Best}},
  \bibinfo{author}{\bibfnamefont{M.}~\bibnamefont{Cinelli}},
  \bibinfo{author}{\bibfnamefont{S.}~\bibnamefont{Reich-Zeliger}},
  \bibinfo{author}{\bibfnamefont{H.}~\bibnamefont{Gal}},
  \bibinfo{author}{\bibfnamefont{E.}~\bibnamefont{Shifrut}},
  \bibinfo{author}{\bibfnamefont{A.}~\bibnamefont{Madi}},
  \bibinfo{author}{\bibfnamefont{N.}~\bibnamefont{Friedman}},
  \bibinfo{author}{\bibfnamefont{J.}~\bibnamefont{Shawe-Taylor}},
  \bibnamefont{and} \bibinfo{author}{\bibfnamefont{B.}~\bibnamefont{Chain}},
  \bibinfo{journal}{Bioinformatics} \textbf{\bibinfo{volume}{30}}
  (\bibinfo{year}{2014}).

\bibitem[{\citenamefont{Britanova et~al.}(2016)\citenamefont{Britanova, Shugay,
  Merzlyak, Staroverov, Putintseva, Turchaninova, Mamedov, Pogorelyy, Bolotin,
  Izraelson et~al.}}]{Britanova2016}
\bibinfo{author}{\bibfnamefont{O.~V.} \bibnamefont{Britanova}},
  \bibinfo{author}{\bibfnamefont{M.}~\bibnamefont{Shugay}},
  \bibinfo{author}{\bibfnamefont{E.~M.} \bibnamefont{Merzlyak}},
  \bibinfo{author}{\bibfnamefont{D.~B.} \bibnamefont{Staroverov}},
  \bibinfo{author}{\bibfnamefont{E.~V.} \bibnamefont{Putintseva}},
  \bibinfo{author}{\bibfnamefont{M.~A.} \bibnamefont{Turchaninova}},
  \bibinfo{author}{\bibfnamefont{I.~Z.} \bibnamefont{Mamedov}},
  \bibinfo{author}{\bibfnamefont{M.~V.} \bibnamefont{Pogorelyy}},
  \bibinfo{author}{\bibfnamefont{D.~A.} \bibnamefont{Bolotin}},
  \bibinfo{author}{\bibfnamefont{M.}~\bibnamefont{Izraelson}},
  \bibnamefont{et~al.}, \bibinfo{journal}{The Journal of Immunology}
  \textbf{\bibinfo{volume}{196}}, \bibinfo{pages}{5005} (\bibinfo{year}{2016}).

\bibitem[{\citenamefont{Emerson et~al.}(2017)\citenamefont{Emerson, DeWitt,
  Vignali, Gravley, Hu, Osborne, Desmarais, Klinger, Carlson, Hansen
  et~al.}}]{Emerson2017}
\bibinfo{author}{\bibfnamefont{R.~O.} \bibnamefont{Emerson}},
  \bibinfo{author}{\bibfnamefont{W.~S.} \bibnamefont{DeWitt}},
  \bibinfo{author}{\bibfnamefont{M.}~\bibnamefont{Vignali}},
  \bibinfo{author}{\bibfnamefont{J.}~\bibnamefont{Gravley}},
  \bibinfo{author}{\bibfnamefont{J.~K.} \bibnamefont{Hu}},
  \bibinfo{author}{\bibfnamefont{E.~J.} \bibnamefont{Osborne}},
  \bibinfo{author}{\bibfnamefont{C.}~\bibnamefont{Desmarais}},
  \bibinfo{author}{\bibfnamefont{M.}~\bibnamefont{Klinger}},
  \bibinfo{author}{\bibfnamefont{C.~S.} \bibnamefont{Carlson}},
  \bibinfo{author}{\bibfnamefont{J.~A.} \bibnamefont{Hansen}},
  \bibnamefont{et~al.}, \bibinfo{journal}{Nature Genetics} pp.
  \bibinfo{pages}{1--10} (\bibinfo{year}{2017}).

\bibitem[{\citenamefont{Oakes et~al.}(2017)\citenamefont{Oakes, Heather, Best,
  Byng-Maddick, Husovsky, Ismail, Joshi, Maxwell, Noursadeghi, Riddell
  et~al.}}]{Oakes2017}
\bibinfo{author}{\bibfnamefont{T.}~\bibnamefont{Oakes}},
  \bibinfo{author}{\bibfnamefont{J.~M.} \bibnamefont{Heather}},
  \bibinfo{author}{\bibfnamefont{K.}~\bibnamefont{Best}},
  \bibinfo{author}{\bibfnamefont{R.}~\bibnamefont{Byng-Maddick}},
  \bibinfo{author}{\bibfnamefont{C.}~\bibnamefont{Husovsky}},
  \bibinfo{author}{\bibfnamefont{M.}~\bibnamefont{Ismail}},
  \bibinfo{author}{\bibfnamefont{K.}~\bibnamefont{Joshi}},
  \bibinfo{author}{\bibfnamefont{G.}~\bibnamefont{Maxwell}},
  \bibinfo{author}{\bibfnamefont{M.}~\bibnamefont{Noursadeghi}},
  \bibinfo{author}{\bibfnamefont{N.}~\bibnamefont{Riddell}},
  \bibnamefont{et~al.}, \bibinfo{journal}{Frontiers in Immunology}
  \textbf{\bibinfo{volume}{8}}, \bibinfo{pages}{1} (\bibinfo{year}{2017}).

\bibitem[{\citenamefont{Thome et~al.}(2016)\citenamefont{Thome, Grinshpun,
  Kumar, Kubota, Ohmura, Lerner, Sempowski, Shen, and Farber}}]{Thome2016}
\bibinfo{author}{\bibfnamefont{J.~J.~C.} \bibnamefont{Thome}},
  \bibinfo{author}{\bibfnamefont{B.}~\bibnamefont{Grinshpun}},
  \bibinfo{author}{\bibfnamefont{B.~V.} \bibnamefont{Kumar}},
  \bibinfo{author}{\bibfnamefont{M.}~\bibnamefont{Kubota}},
  \bibinfo{author}{\bibfnamefont{Y.}~\bibnamefont{Ohmura}},
  \bibinfo{author}{\bibfnamefont{H.}~\bibnamefont{Lerner}},
  \bibinfo{author}{\bibfnamefont{G.~D.} \bibnamefont{Sempowski}},
  \bibinfo{author}{\bibfnamefont{Y.}~\bibnamefont{Shen}}, \bibnamefont{and}
  \bibinfo{author}{\bibfnamefont{D.~L.} \bibnamefont{Farber}},
  \bibinfo{journal}{Science Immunology} \textbf{\bibinfo{volume}{1}},
  \bibinfo{pages}{eaah6506} (\bibinfo{year}{2016}).

\bibitem[{\citenamefont{Robins et~al.}(2010)\citenamefont{Robins, Srivastava,
  Campregher, Turtle, Andriesen, Riddell, Carlson, and Warren}}]{Robins2010}
\bibinfo{author}{\bibfnamefont{H.~S.} \bibnamefont{Robins}},
  \bibinfo{author}{\bibfnamefont{S.~K.} \bibnamefont{Srivastava}},
  \bibinfo{author}{\bibfnamefont{P.~V.} \bibnamefont{Campregher}},
  \bibinfo{author}{\bibfnamefont{C.~J.} \bibnamefont{Turtle}},
  \bibinfo{author}{\bibfnamefont{J.}~\bibnamefont{Andriesen}},
  \bibinfo{author}{\bibfnamefont{S.~R.} \bibnamefont{Riddell}},
  \bibinfo{author}{\bibfnamefont{C.~S.} \bibnamefont{Carlson}},
  \bibnamefont{and} \bibinfo{author}{\bibfnamefont{E.~H.}
  \bibnamefont{Warren}}, \bibinfo{journal}{Science Translational Medicine}
  \textbf{\bibinfo{volume}{2}}, \bibinfo{pages}{47ra64} (\bibinfo{year}{2010}).

\bibitem[{\citenamefont{Qi et~al.}(2014)\citenamefont{Qi, Liu, Cheng,
  Glanville, Zhang, Lee, Olshen, Weyand, Boyd, and Goronzy}}]{Qi2014}
\bibinfo{author}{\bibfnamefont{Q.}~\bibnamefont{Qi}},
  \bibinfo{author}{\bibfnamefont{Y.}~\bibnamefont{Liu}},
  \bibinfo{author}{\bibfnamefont{Y.}~\bibnamefont{Cheng}},
  \bibinfo{author}{\bibfnamefont{J.}~\bibnamefont{Glanville}},
  \bibinfo{author}{\bibfnamefont{D.}~\bibnamefont{Zhang}},
  \bibinfo{author}{\bibfnamefont{J.-Y.} \bibnamefont{Lee}},
  \bibinfo{author}{\bibfnamefont{R.~a.} \bibnamefont{Olshen}},
  \bibinfo{author}{\bibfnamefont{C.~M.} \bibnamefont{Weyand}},
  \bibinfo{author}{\bibfnamefont{S.~D.} \bibnamefont{Boyd}}, \bibnamefont{and}
  \bibinfo{author}{\bibfnamefont{J.~J.} \bibnamefont{Goronzy}},
  \bibinfo{journal}{Proceedings of the National Academy of Sciences of the
  United States of America} \textbf{\bibinfo{volume}{111}},
  \bibinfo{pages}{13139} (\bibinfo{year}{2014}).

\bibitem[{\citenamefont{Lindau et~al.}(2019)\citenamefont{Lindau, Mukherjee,
  Gutschow, Vignali, Warren, Riddell, Makar, Turtle, and Robins}}]{Lindau2019}
\bibinfo{author}{\bibfnamefont{P.}~\bibnamefont{Lindau}},
  \bibinfo{author}{\bibfnamefont{R.}~\bibnamefont{Mukherjee}},
  \bibinfo{author}{\bibfnamefont{M.~V.} \bibnamefont{Gutschow}},
  \bibinfo{author}{\bibfnamefont{M.}~\bibnamefont{Vignali}},
  \bibinfo{author}{\bibfnamefont{E.~H.} \bibnamefont{Warren}},
  \bibinfo{author}{\bibfnamefont{S.~R.} \bibnamefont{Riddell}},
  \bibinfo{author}{\bibfnamefont{K.~W.} \bibnamefont{Makar}},
  \bibinfo{author}{\bibfnamefont{C.~J.} \bibnamefont{Turtle}},
  \bibnamefont{and} \bibinfo{author}{\bibfnamefont{H.~S.}
  \bibnamefont{Robins}}, \bibinfo{journal}{The Journal of Immunology}
  \textbf{\bibinfo{volume}{202}}, \bibinfo{pages}{476} (\bibinfo{year}{2019}).

\bibitem[{\citenamefont{Joshi et~al.}(2019)\citenamefont{Joshi, Massy, Ismail,
  Reading, Uddin, Woolston, Hatipoglu, Oakes, Rosenthal, Peacock
  et~al.}}]{Joshi2019}
\bibinfo{author}{\bibfnamefont{K.}~\bibnamefont{Joshi}},
  \bibinfo{author}{\bibfnamefont{M.~R.~D.} \bibnamefont{Massy}},
  \bibinfo{author}{\bibfnamefont{M.}~\bibnamefont{Ismail}},
  \bibinfo{author}{\bibfnamefont{J.~L.} \bibnamefont{Reading}},
  \bibinfo{author}{\bibfnamefont{I.}~\bibnamefont{Uddin}},
  \bibinfo{author}{\bibfnamefont{A.}~\bibnamefont{Woolston}},
  \bibinfo{author}{\bibfnamefont{E.}~\bibnamefont{Hatipoglu}},
  \bibinfo{author}{\bibfnamefont{T.}~\bibnamefont{Oakes}},
  \bibinfo{author}{\bibfnamefont{R.}~\bibnamefont{Rosenthal}},
  \bibinfo{author}{\bibfnamefont{T.}~\bibnamefont{Peacock}},
  \bibnamefont{et~al.}, \bibinfo{journal}{Nature Medicine}
  (\bibinfo{year}{2019}).

\bibitem[{\citenamefont{Desponds et~al.}(2016)\citenamefont{Desponds, Mora, and
  Walczak}}]{Desponds2016}
\bibinfo{author}{\bibfnamefont{J.}~\bibnamefont{Desponds}},
  \bibinfo{author}{\bibfnamefont{T.}~\bibnamefont{Mora}}, \bibnamefont{and}
  \bibinfo{author}{\bibfnamefont{A.~M.} \bibnamefont{Walczak}},
  \bibinfo{journal}{Proceedings of the National Academy of Sciences}
  \textbf{\bibinfo{volume}{113}}, \bibinfo{pages}{274} (\bibinfo{year}{2016}).

\bibitem[{\citenamefont{Lythe et~al.}(2016)\citenamefont{Lythe, Callard, Hoare,
  and Molina-Par{\'{i}}s}}]{Lythe2016}
\bibinfo{author}{\bibfnamefont{G.}~\bibnamefont{Lythe}},
  \bibinfo{author}{\bibfnamefont{R.~E.} \bibnamefont{Callard}},
  \bibinfo{author}{\bibfnamefont{R.~L.} \bibnamefont{Hoare}}, \bibnamefont{and}
  \bibinfo{author}{\bibfnamefont{C.}~\bibnamefont{Molina-Par{\'{i}}s}},
  \bibinfo{journal}{Journal of Theoretical Biology}
  \textbf{\bibinfo{volume}{389}}, \bibinfo{pages}{214} (\bibinfo{year}{2016}).

\bibitem[{\citenamefont{Dessalles et~al.}(2019)\citenamefont{Dessalles,
  D'Orsogna, and Chou}}]{Dessalles2019}
\bibinfo{author}{\bibfnamefont{R.}~\bibnamefont{Dessalles}},
  \bibinfo{author}{\bibfnamefont{M.}~\bibnamefont{D'Orsogna}},
  \bibnamefont{and} \bibinfo{author}{\bibfnamefont{T.}~\bibnamefont{Chou}},
  \bibinfo{journal}{arXiv preprint arXiv:1906.07463}  (\bibinfo{year}{2019}).

\bibitem[{\citenamefont{Altan-Bonnet et~al.}(2019)\citenamefont{Altan-Bonnet,
  Mora, and Walczak}}]{Altan-Bonnet2019}
\bibinfo{author}{\bibfnamefont{G.}~\bibnamefont{Altan-Bonnet}},
  \bibinfo{author}{\bibfnamefont{T.}~\bibnamefont{Mora}}, \bibnamefont{and}
  \bibinfo{author}{\bibfnamefont{A.~M.} \bibnamefont{Walczak}},
  \bibinfo{journal}{Physics Reports} \textbf{\bibinfo{volume}{61}},
  \bibinfo{pages}{1} (\bibinfo{year}{2019}).

\bibitem[{\citenamefont{Greef et~al.}(2020)\citenamefont{Greef, Oakes,
  Gerritsen, Ismail, Heather, Hermsen, Chain, Boer, James, Hermsen
  et~al.}}]{Greef2020}
\bibinfo{author}{\bibfnamefont{P.~C.~D.} \bibnamefont{Greef}},
  \bibinfo{author}{\bibfnamefont{T.}~\bibnamefont{Oakes}},
  \bibinfo{author}{\bibfnamefont{B.}~\bibnamefont{Gerritsen}},
  \bibinfo{author}{\bibfnamefont{M.}~\bibnamefont{Ismail}},
  \bibinfo{author}{\bibfnamefont{J.~M.} \bibnamefont{Heather}},
  \bibinfo{author}{\bibfnamefont{R.}~\bibnamefont{Hermsen}},
  \bibinfo{author}{\bibfnamefont{B.}~\bibnamefont{Chain}},
  \bibinfo{author}{\bibfnamefont{R.~J.~D.} \bibnamefont{Boer}},
  \bibinfo{author}{\bibfnamefont{M.}~\bibnamefont{James}},
  \bibinfo{author}{\bibfnamefont{R.}~\bibnamefont{Hermsen}},
  \bibnamefont{et~al.}, \bibinfo{journal}{eLife} \textbf{\bibinfo{volume}{9}},
  \bibinfo{pages}{e49900} (\bibinfo{year}{2020}).

\bibitem[{\citenamefont{Chu et~al.}(2019)\citenamefont{Chu, Bi, Emerson,
  Sherwood, Birnbaum, Robins, and Alm}}]{Chu2019}
\bibinfo{author}{\bibfnamefont{N.~D.} \bibnamefont{Chu}},
  \bibinfo{author}{\bibfnamefont{H.~S.} \bibnamefont{Bi}},
  \bibinfo{author}{\bibfnamefont{R.~O.} \bibnamefont{Emerson}},
  \bibinfo{author}{\bibfnamefont{A.~M.} \bibnamefont{Sherwood}},
  \bibinfo{author}{\bibfnamefont{M.~E.} \bibnamefont{Birnbaum}},
  \bibinfo{author}{\bibfnamefont{H.~S.} \bibnamefont{Robins}},
  \bibnamefont{and} \bibinfo{author}{\bibfnamefont{E.~J.} \bibnamefont{Alm}},
  \bibinfo{journal}{BMC Immunology} \textbf{\bibinfo{volume}{20}},
  \bibinfo{pages}{1} (\bibinfo{year}{2019}).

\bibitem[{\citenamefont{Stumpf et~al.}(2005)\citenamefont{Stumpf, Wiuf, and
  May}}]{Stumpf2005}
\bibinfo{author}{\bibfnamefont{M.~P.} \bibnamefont{Stumpf}},
  \bibinfo{author}{\bibfnamefont{C.}~\bibnamefont{Wiuf}}, \bibnamefont{and}
  \bibinfo{author}{\bibfnamefont{R.~M.} \bibnamefont{May}},
  \bibinfo{journal}{Proceedings of the National Academy of Sciences}
  \textbf{\bibinfo{volume}{102}}, \bibinfo{pages}{4221 }
  (\bibinfo{year}{2005}).

\bibitem[{\citenamefont{Volkov et~al.}(2003)\citenamefont{Volkov, Banavar,
  Hubbell, and Maritan}}]{Volkov2003}
\bibinfo{author}{\bibfnamefont{I.}~\bibnamefont{Volkov}},
  \bibinfo{author}{\bibfnamefont{J.~R.} \bibnamefont{Banavar}},
  \bibinfo{author}{\bibfnamefont{S.~P.} \bibnamefont{Hubbell}},
  \bibnamefont{and} \bibinfo{author}{\bibfnamefont{A.}~\bibnamefont{Maritan}},
  \bibinfo{journal}{Nature} pp. \bibinfo{pages}{1035--1037}
  (\bibinfo{year}{2003}).

\bibitem[{\citenamefont{{Le Campion} et~al.}(2002)\citenamefont{{Le Campion},
  Bourgeois, Lambolez, Martin, L{\'{e}}aument, Dautigny, Tanchot, P{\'{e}}nit,
  and Lucas}}]{Campion2002}
\bibinfo{author}{\bibfnamefont{A.}~\bibnamefont{{Le Campion}}},
  \bibinfo{author}{\bibfnamefont{C.}~\bibnamefont{Bourgeois}},
  \bibinfo{author}{\bibfnamefont{F.}~\bibnamefont{Lambolez}},
  \bibinfo{author}{\bibfnamefont{B.}~\bibnamefont{Martin}},
  \bibinfo{author}{\bibfnamefont{S.}~\bibnamefont{L{\'{e}}aument}},
  \bibinfo{author}{\bibfnamefont{N.}~\bibnamefont{Dautigny}},
  \bibinfo{author}{\bibfnamefont{C.}~\bibnamefont{Tanchot}},
  \bibinfo{author}{\bibfnamefont{C.}~\bibnamefont{P{\'{e}}nit}},
  \bibnamefont{and} \bibinfo{author}{\bibfnamefont{B.}~\bibnamefont{Lucas}},
  \bibinfo{journal}{Proceedings of the National Academy of Sciences}
  \textbf{\bibinfo{volume}{99}}, \bibinfo{pages}{4538} (\bibinfo{year}{2002}).

\bibitem[{\citenamefont{Min et~al.}(2003)\citenamefont{Min, McHugh, Sempowski,
  Mackall, Foucras, and Paul}}]{Min2003}
\bibinfo{author}{\bibfnamefont{B.}~\bibnamefont{Min}},
  \bibinfo{author}{\bibfnamefont{R.}~\bibnamefont{McHugh}},
  \bibinfo{author}{\bibfnamefont{G.~D.} \bibnamefont{Sempowski}},
  \bibinfo{author}{\bibfnamefont{C.}~\bibnamefont{Mackall}},
  \bibinfo{author}{\bibfnamefont{G.}~\bibnamefont{Foucras}}, \bibnamefont{and}
  \bibinfo{author}{\bibfnamefont{W.~E.} \bibnamefont{Paul}},
  \bibinfo{journal}{Immunity} \textbf{\bibinfo{volume}{18}},
  \bibinfo{pages}{131} (\bibinfo{year}{2003}).

\bibitem[{\citenamefont{Rufer et~al.}(1999)\citenamefont{Rufer,
  Br{\"{u}}mmendorf, Kolvraa, Bischoff, Christensen, Wadsworth, Schulzer, and
  Lansdorp}}]{Rufer1999}
\bibinfo{author}{\bibfnamefont{B.~N.} \bibnamefont{Rufer}},
  \bibinfo{author}{\bibfnamefont{T.~H.} \bibnamefont{Br{\"{u}}mmendorf}},
  \bibinfo{author}{\bibfnamefont{S.}~\bibnamefont{Kolvraa}},
  \bibinfo{author}{\bibfnamefont{C.}~\bibnamefont{Bischoff}},
  \bibinfo{author}{\bibfnamefont{K.}~\bibnamefont{Christensen}},
  \bibinfo{author}{\bibfnamefont{L.}~\bibnamefont{Wadsworth}},
  \bibinfo{author}{\bibfnamefont{M.}~\bibnamefont{Schulzer}}, \bibnamefont{and}
  \bibinfo{author}{\bibfnamefont{P.~M.} \bibnamefont{Lansdorp}},
  \bibinfo{journal}{Journal of Experimental Medicine}
  \textbf{\bibinfo{volume}{190}}, \bibinfo{pages}{157} (\bibinfo{year}{1999}).

\bibitem[{\citenamefont{Schonland et~al.}(2003)\citenamefont{Schonland, Zimmer,
  Lopez-Benitez, Widmann, Ramin, Goronzy, and Weyand}}]{Schonland2003}
\bibinfo{author}{\bibfnamefont{S.~O.} \bibnamefont{Schonland}},
  \bibinfo{author}{\bibfnamefont{J.~K.} \bibnamefont{Zimmer}},
  \bibinfo{author}{\bibfnamefont{C.~M.} \bibnamefont{Lopez-Benitez}},
  \bibinfo{author}{\bibfnamefont{T.}~\bibnamefont{Widmann}},
  \bibinfo{author}{\bibfnamefont{K.~D.} \bibnamefont{Ramin}},
  \bibinfo{author}{\bibfnamefont{J.}~\bibnamefont{Goronzy}}, \bibnamefont{and}
  \bibinfo{author}{\bibfnamefont{C.~M.} \bibnamefont{Weyand}},
  \bibinfo{journal}{Blood} \textbf{\bibinfo{volume}{102}},
  \bibinfo{pages}{1428} (\bibinfo{year}{2003}).

\bibitem[{\citenamefont{Hammarlund et~al.}(2003)\citenamefont{Hammarlund,
  Lewis, Hansen, Strelow, Nelson, Sexton, Hanifin, and
  Slifka}}]{Hammarlund2003}
\bibinfo{author}{\bibfnamefont{E.}~\bibnamefont{Hammarlund}},
  \bibinfo{author}{\bibfnamefont{M.~W.} \bibnamefont{Lewis}},
  \bibinfo{author}{\bibfnamefont{S.~G.} \bibnamefont{Hansen}},
  \bibinfo{author}{\bibfnamefont{L.~I.} \bibnamefont{Strelow}},
  \bibinfo{author}{\bibfnamefont{J.~A.} \bibnamefont{Nelson}},
  \bibinfo{author}{\bibfnamefont{G.~J.} \bibnamefont{Sexton}},
  \bibinfo{author}{\bibfnamefont{J.~M.} \bibnamefont{Hanifin}},
  \bibnamefont{and} \bibinfo{author}{\bibfnamefont{M.~K.}
  \bibnamefont{Slifka}}, \bibinfo{journal}{Nature Medicine}
  \textbf{\bibinfo{volume}{9}}, \bibinfo{pages}{1131} (\bibinfo{year}{2003}).

\bibitem[{\citenamefont{Pogorelyy et~al.}(2017)\citenamefont{Pogorelyy,
  Elhanati, Marcou, Sycheva, Komech, Nazarov, Britanova, Chudakov, Mamedov,
  Lebedev et~al.}}]{Pogorelyy2017}
\bibinfo{author}{\bibfnamefont{M.~V.} \bibnamefont{Pogorelyy}},
  \bibinfo{author}{\bibfnamefont{Y.}~\bibnamefont{Elhanati}},
  \bibinfo{author}{\bibfnamefont{Q.}~\bibnamefont{Marcou}},
  \bibinfo{author}{\bibfnamefont{A.~L.} \bibnamefont{Sycheva}},
  \bibinfo{author}{\bibfnamefont{E.~A.} \bibnamefont{Komech}},
  \bibinfo{author}{\bibfnamefont{V.~I.} \bibnamefont{Nazarov}},
  \bibinfo{author}{\bibfnamefont{O.~V.} \bibnamefont{Britanova}},
  \bibinfo{author}{\bibfnamefont{D.~M.} \bibnamefont{Chudakov}},
  \bibinfo{author}{\bibfnamefont{I.~Z.} \bibnamefont{Mamedov}},
  \bibinfo{author}{\bibfnamefont{Y.~B.} \bibnamefont{Lebedev}},
  \bibnamefont{et~al.}, \bibinfo{journal}{PLoS Computational Biology}
  \textbf{\bibinfo{volume}{13}}, \bibinfo{pages}{e1005572}
  (\bibinfo{year}{2017}).

\bibitem[{\citenamefont{Tanno et~al.}(2020)\citenamefont{Tanno, Gould,
  Mcdaniel, Cao, Tanno, and Durrett}}]{Tanno2020}
\bibinfo{author}{\bibfnamefont{H.}~\bibnamefont{Tanno}},
  \bibinfo{author}{\bibfnamefont{T.~M.} \bibnamefont{Gould}},
  \bibinfo{author}{\bibfnamefont{J.~R.} \bibnamefont{Mcdaniel}},
  \bibinfo{author}{\bibfnamefont{W.}~\bibnamefont{Cao}},
  \bibinfo{author}{\bibfnamefont{Y.}~\bibnamefont{Tanno}}, \bibnamefont{and}
  \bibinfo{author}{\bibfnamefont{R.~E.} \bibnamefont{Durrett}},
  \bibinfo{journal}{Proceedings of the National Academy of Sciences}
  \textbf{\bibinfo{volume}{117}} (\bibinfo{year}{2020}).

\bibitem[{\citenamefont{Bains et~al.}(2009)\citenamefont{Bains, Antia, Callard,
  and Yates}}]{Bains2009}
\bibinfo{author}{\bibfnamefont{I.}~\bibnamefont{Bains}},
  \bibinfo{author}{\bibfnamefont{R.}~\bibnamefont{Antia}},
  \bibinfo{author}{\bibfnamefont{R.}~\bibnamefont{Callard}}, \bibnamefont{and}
  \bibinfo{author}{\bibfnamefont{A.~J.} \bibnamefont{Yates}},
  \bibinfo{journal}{Blood} \textbf{\bibinfo{volume}{113}},
  \bibinfo{pages}{5480} (\bibinfo{year}{2009}).

\bibitem[{\citenamefont{Yule}(1924)}]{Yule1924}
\bibinfo{author}{\bibfnamefont{G.~U.} \bibnamefont{Yule}},
  \bibinfo{journal}{Phil. Trans. B} \textbf{\bibinfo{volume}{213}},
  \bibinfo{pages}{21} (\bibinfo{year}{1924}).

\bibitem[{\citenamefont{Feeney}(1991)}]{Feeney1991}
\bibinfo{author}{\bibfnamefont{B.~A.~J.} \bibnamefont{Feeney}},
  \bibinfo{journal}{Journal of Experimental Medicine}
  \textbf{\bibinfo{volume}{174}} (\bibinfo{year}{1991}).

\bibitem[{\citenamefont{Rechavi et~al.}(2015)\citenamefont{Rechavi, Lev, Lee,
  Simon, Yinon, Lipitz, Amariglio, Weisz, Notarangelo, and
  Somech}}]{Rechavi2015}
\bibinfo{author}{\bibfnamefont{E.}~\bibnamefont{Rechavi}},
  \bibinfo{author}{\bibfnamefont{A.}~\bibnamefont{Lev}},
  \bibinfo{author}{\bibfnamefont{Y.~N.} \bibnamefont{Lee}},
  \bibinfo{author}{\bibfnamefont{A.~J.} \bibnamefont{Simon}},
  \bibinfo{author}{\bibfnamefont{Y.}~\bibnamefont{Yinon}},
  \bibinfo{author}{\bibfnamefont{S.}~\bibnamefont{Lipitz}},
  \bibinfo{author}{\bibfnamefont{N.}~\bibnamefont{Amariglio}},
  \bibinfo{author}{\bibfnamefont{B.}~\bibnamefont{Weisz}},
  \bibinfo{author}{\bibfnamefont{L.~D.} \bibnamefont{Notarangelo}},
  \bibnamefont{and} \bibinfo{author}{\bibfnamefont{R.}~\bibnamefont{Somech}},
  \bibinfo{journal}{Science Translational Medicine}
  \textbf{\bibinfo{volume}{7}}, \bibinfo{pages}{1} (\bibinfo{year}{2015}).

\bibitem[{\citenamefont{Park et~al.}(2020)\citenamefont{Park, Jardine,
  Gottgens, Teichmann, and Haniffa}}]{Park2020}
\bibinfo{author}{\bibfnamefont{J.-E.} \bibnamefont{Park}},
  \bibinfo{author}{\bibfnamefont{L.}~\bibnamefont{Jardine}},
  \bibinfo{author}{\bibfnamefont{B.}~\bibnamefont{Gottgens}},
  \bibinfo{author}{\bibfnamefont{S.~A.} \bibnamefont{Teichmann}},
  \bibnamefont{and} \bibinfo{author}{\bibfnamefont{M.}~\bibnamefont{Haniffa}},
  \bibinfo{journal}{Science} \textbf{\bibinfo{volume}{603}},
  \bibinfo{pages}{600} (\bibinfo{year}{2020}).

\bibitem[{\citenamefont{Sethna et~al.}(2017)\citenamefont{Sethna, Elhanati,
  Dudgeon, Callan, Levine, Mora, and Walczak}}]{Sethna2017}
\bibinfo{author}{\bibfnamefont{Z.}~\bibnamefont{Sethna}},
  \bibinfo{author}{\bibfnamefont{Y.}~\bibnamefont{Elhanati}},
  \bibinfo{author}{\bibfnamefont{C.~S.} \bibnamefont{Dudgeon}},
  \bibinfo{author}{\bibfnamefont{C.~G.} \bibnamefont{Callan}},
  \bibinfo{author}{\bibfnamefont{A.~J.} \bibnamefont{Levine}},
  \bibinfo{author}{\bibfnamefont{T.}~\bibnamefont{Mora}}, \bibnamefont{and}
  \bibinfo{author}{\bibfnamefont{A.~M.} \bibnamefont{Walczak}},
  \bibinfo{journal}{Proceedings of the National Academy of Sciences}
  \textbf{\bibinfo{volume}{114}}, \bibinfo{pages}{201700241}
  (\bibinfo{year}{2017}).

\bibitem[{\citenamefont{Mayer et~al.}(2016)\citenamefont{Mayer, Mora, Rivoire,
  and Walczak}}]{Mayer2016}
\bibinfo{author}{\bibfnamefont{A.}~\bibnamefont{Mayer}},
  \bibinfo{author}{\bibfnamefont{T.}~\bibnamefont{Mora}},
  \bibinfo{author}{\bibfnamefont{O.}~\bibnamefont{Rivoire}}, \bibnamefont{and}
  \bibinfo{author}{\bibfnamefont{A.~M.} \bibnamefont{Walczak}},
  \bibinfo{journal}{Proceedings of the National Academy of Sciences}
  \textbf{\bibinfo{volume}{113}}, \bibinfo{pages}{8630} (\bibinfo{year}{2016}).

\bibitem[{\citenamefont{Guerau-de Arellano et~al.}(2009)\citenamefont{Guerau-de
  Arellano, Martinic, Benoist, and Mathis}}]{Guerau-de-arellano2009}
\bibinfo{author}{\bibfnamefont{M.}~\bibnamefont{Guerau-de Arellano}},
  \bibinfo{author}{\bibfnamefont{M.}~\bibnamefont{Martinic}},
  \bibinfo{author}{\bibfnamefont{C.}~\bibnamefont{Benoist}}, \bibnamefont{and}
  \bibinfo{author}{\bibfnamefont{D.}~\bibnamefont{Mathis}},
  \bibinfo{journal}{Journal of Experimental Medicine}
  \textbf{\bibinfo{volume}{206}}, \bibinfo{pages}{1245} (\bibinfo{year}{2009}).

\bibitem[{\citenamefont{Gostic et~al.}(2016)\citenamefont{Gostic, Ambrose,
  Worobey, and Lloyd-Smith}}]{Gostic2016}
\bibinfo{author}{\bibfnamefont{K.~M.} \bibnamefont{Gostic}},
  \bibinfo{author}{\bibfnamefont{M.}~\bibnamefont{Ambrose}},
  \bibinfo{author}{\bibfnamefont{M.}~\bibnamefont{Worobey}}, \bibnamefont{and}
  \bibinfo{author}{\bibfnamefont{J.~O.} \bibnamefont{Lloyd-Smith}},
  \bibinfo{journal}{Science} \textbf{\bibinfo{volume}{354}},
  \bibinfo{pages}{722} (\bibinfo{year}{2016}).

\bibitem[{\citenamefont{Constantinides
  et~al.}(2019)\citenamefont{Constantinides, Link, Tamoutounour, Wong,
  Perez-Chaparro, Han, Chen, Li, Farhat, Weckel et~al.}}]{Constantinides2019}
\bibinfo{author}{\bibfnamefont{M.~G.} \bibnamefont{Constantinides}},
  \bibinfo{author}{\bibfnamefont{V.~M.} \bibnamefont{Link}},
  \bibinfo{author}{\bibfnamefont{S.}~\bibnamefont{Tamoutounour}},
  \bibinfo{author}{\bibfnamefont{A.~C.} \bibnamefont{Wong}},
  \bibinfo{author}{\bibfnamefont{P.~J.} \bibnamefont{Perez-Chaparro}},
  \bibinfo{author}{\bibfnamefont{S.-J.} \bibnamefont{Han}},
  \bibinfo{author}{\bibfnamefont{Y.~E.} \bibnamefont{Chen}},
  \bibinfo{author}{\bibfnamefont{K.}~\bibnamefont{Li}},
  \bibinfo{author}{\bibfnamefont{S.}~\bibnamefont{Farhat}},
  \bibinfo{author}{\bibfnamefont{A.}~\bibnamefont{Weckel}},
  \bibnamefont{et~al.}, \bibinfo{journal}{Science}
  \textbf{\bibinfo{volume}{6624}} (\bibinfo{year}{2019}).

\bibitem[{\citenamefont{Li et~al.}(2019)\citenamefont{Li, van Unen, Abdelaal,
  Guo, Kasatskaya, Ladell, McLaren, Egorov, Izraelson, {Chuva de Sousa Lopes}
  et~al.}}]{Li2019a}
\bibinfo{author}{\bibfnamefont{N.}~\bibnamefont{Li}},
  \bibinfo{author}{\bibfnamefont{V.}~\bibnamefont{van Unen}},
  \bibinfo{author}{\bibfnamefont{T.}~\bibnamefont{Abdelaal}},
  \bibinfo{author}{\bibfnamefont{N.}~\bibnamefont{Guo}},
  \bibinfo{author}{\bibfnamefont{S.~A.} \bibnamefont{Kasatskaya}},
  \bibinfo{author}{\bibfnamefont{K.}~\bibnamefont{Ladell}},
  \bibinfo{author}{\bibfnamefont{J.~E.} \bibnamefont{McLaren}},
  \bibinfo{author}{\bibfnamefont{E.~S.} \bibnamefont{Egorov}},
  \bibinfo{author}{\bibfnamefont{M.}~\bibnamefont{Izraelson}},
  \bibinfo{author}{\bibfnamefont{S.~M.} \bibnamefont{{Chuva de Sousa Lopes}}},
  \bibnamefont{et~al.}, \bibinfo{journal}{Nature Immunology}
  \textbf{\bibinfo{volume}{20}}, \bibinfo{pages}{301} (\bibinfo{year}{2019}).

\bibitem[{\citenamefont{Davenport et~al.}(2020)\citenamefont{Davenport, Smith,
  and Rudd}}]{Davenport2020}
\bibinfo{author}{\bibfnamefont{M.~P.} \bibnamefont{Davenport}},
  \bibinfo{author}{\bibfnamefont{N.~L.} \bibnamefont{Smith}}, \bibnamefont{and}
  \bibinfo{author}{\bibfnamefont{B.~D.} \bibnamefont{Rudd}},
  \bibinfo{journal}{Nature Reviews Immunology}  (\bibinfo{year}{2020}).

\bibitem[{\citenamefont{Hong et~al.}(2020)\citenamefont{Hong, Lim, Carvalho,
  Annicelli, Ip, and Medzhitov}}]{Hong2020}
\bibinfo{author}{\bibfnamefont{J.~Y.} \bibnamefont{Hong}},
  \bibinfo{author}{\bibfnamefont{J.}~\bibnamefont{Lim}},
  \bibinfo{author}{\bibfnamefont{F.}~\bibnamefont{Carvalho}},
  \bibinfo{author}{\bibfnamefont{C.}~\bibnamefont{Annicelli}},
  \bibinfo{author}{\bibfnamefont{W.~K.~E.} \bibnamefont{Ip}}, \bibnamefont{and}
  \bibinfo{author}{\bibfnamefont{R.}~\bibnamefont{Medzhitov}},
  \bibinfo{journal}{Cell} \textbf{\bibinfo{volume}{180}}, \bibinfo{pages}{847}
  (\bibinfo{year}{2020}).

\bibitem[{\citenamefont{Shearer et~al.}(2003)\citenamefont{Shearer, Rosenblatt,
  Gelman, Oymopito, Plaeger, Stiehm, Wara, Douglas, Luzuriaga, McFarland
  et~al.}}]{Shearer2003}
\bibinfo{author}{\bibfnamefont{W.~T.} \bibnamefont{Shearer}},
  \bibinfo{author}{\bibfnamefont{H.~M.} \bibnamefont{Rosenblatt}},
  \bibinfo{author}{\bibfnamefont{R.~S.} \bibnamefont{Gelman}},
  \bibinfo{author}{\bibfnamefont{R.}~\bibnamefont{Oymopito}},
  \bibinfo{author}{\bibfnamefont{S.}~\bibnamefont{Plaeger}},
  \bibinfo{author}{\bibfnamefont{E.~R.} \bibnamefont{Stiehm}},
  \bibinfo{author}{\bibfnamefont{D.~W.} \bibnamefont{Wara}},
  \bibinfo{author}{\bibfnamefont{S.~D.} \bibnamefont{Douglas}},
  \bibinfo{author}{\bibfnamefont{K.}~\bibnamefont{Luzuriaga}},
  \bibinfo{author}{\bibfnamefont{E.~J.} \bibnamefont{McFarland}},
  \bibnamefont{et~al.}, \bibinfo{journal}{Journal of Allergy and Clinical
  Immunology} \textbf{\bibinfo{volume}{112}}, \bibinfo{pages}{973}
  (\bibinfo{year}{2003}).

\end{thebibliography}


\begin{thebibliography}{46}
\expandafter\ifx\csname natexlab\endcsname\relax\def\natexlab#1{#1}\fi
\expandafter\ifx\csname bibnamefont\endcsname\relax
  \def\bibnamefont#1{#1}\fi
\expandafter\ifx\csname bibfnamefont\endcsname\relax
  \def\bibfnamefont#1{#1}\fi
\expandafter\ifx\csname citenamefont\endcsname\relax
  \def\citenamefont#1{#1}\fi
\expandafter\ifx\csname url\endcsname\relax
  \def\url#1{\texttt{#1}}\fi
\expandafter\ifx\csname urlprefix\endcsname\relax\def\urlprefix{URL }\fi
\providecommand{\bibinfo}[2]{#2}
\providecommand{\eprint}[2][]{\url{#2}}

\bibitem[{\citenamefont{Britanova et~al.}(2016)\citenamefont{Britanova, Shugay,
  Merzlyak, Staroverov, Putintseva, Turchaninova, Mamedov, Pogorelyy, Bolotin,
  Izraelson et~al.}}]{Britanova2016}
\bibinfo{author}{\bibfnamefont{O.~V.} \bibnamefont{Britanova}},
  \bibinfo{author}{\bibfnamefont{M.}~\bibnamefont{Shugay}},
  \bibinfo{author}{\bibfnamefont{E.~M.} \bibnamefont{Merzlyak}},
  \bibinfo{author}{\bibfnamefont{D.~B.} \bibnamefont{Staroverov}},
  \bibinfo{author}{\bibfnamefont{E.~V.} \bibnamefont{Putintseva}},
  \bibinfo{author}{\bibfnamefont{M.~A.} \bibnamefont{Turchaninova}},
  \bibinfo{author}{\bibfnamefont{I.~Z.} \bibnamefont{Mamedov}},
  \bibinfo{author}{\bibfnamefont{M.~V.} \bibnamefont{Pogorelyy}},
  \bibinfo{author}{\bibfnamefont{D.~A.} \bibnamefont{Bolotin}},
  \bibinfo{author}{\bibfnamefont{M.}~\bibnamefont{Izraelson}},
  \bibnamefont{et~al.}, \bibinfo{journal}{The Journal of Immunology}
  \textbf{\bibinfo{volume}{196}}, \bibinfo{pages}{5005} (\bibinfo{year}{2016}).

\bibitem[{\citenamefont{Emerson et~al.}(2017)\citenamefont{Emerson, DeWitt,
  Vignali, Gravley, Hu, Osborne, Desmarais, Klinger, Carlson, Hansen
  et~al.}}]{Emerson2017}
\bibinfo{author}{\bibfnamefont{R.~O.} \bibnamefont{Emerson}},
  \bibinfo{author}{\bibfnamefont{W.~S.} \bibnamefont{DeWitt}},
  \bibinfo{author}{\bibfnamefont{M.}~\bibnamefont{Vignali}},
  \bibinfo{author}{\bibfnamefont{J.}~\bibnamefont{Gravley}},
  \bibinfo{author}{\bibfnamefont{J.~K.} \bibnamefont{Hu}},
  \bibinfo{author}{\bibfnamefont{E.~J.} \bibnamefont{Osborne}},
  \bibinfo{author}{\bibfnamefont{C.}~\bibnamefont{Desmarais}},
  \bibinfo{author}{\bibfnamefont{M.}~\bibnamefont{Klinger}},
  \bibinfo{author}{\bibfnamefont{C.~S.} \bibnamefont{Carlson}},
  \bibinfo{author}{\bibfnamefont{J.~A.} \bibnamefont{Hansen}},
  \bibnamefont{et~al.}, \bibinfo{journal}{Nature Genetics} pp.
  \bibinfo{pages}{1--10} (\bibinfo{year}{2017}).

\bibitem[{\citenamefont{Sylwester et~al.}(2005)\citenamefont{Sylwester,
  Mitchell, Edgar, Taormina, Pelte, Ruchti, Sleath, Grabstein, Hosken, Kern
  et~al.}}]{Sylwester2005}
\bibinfo{author}{\bibfnamefont{A.~W.} \bibnamefont{Sylwester}},
  \bibinfo{author}{\bibfnamefont{B.~L.} \bibnamefont{Mitchell}},
  \bibinfo{author}{\bibfnamefont{J.~B.} \bibnamefont{Edgar}},
  \bibinfo{author}{\bibfnamefont{C.}~\bibnamefont{Taormina}},
  \bibinfo{author}{\bibfnamefont{C.}~\bibnamefont{Pelte}},
  \bibinfo{author}{\bibfnamefont{F.}~\bibnamefont{Ruchti}},
  \bibinfo{author}{\bibfnamefont{P.~R.} \bibnamefont{Sleath}},
  \bibinfo{author}{\bibfnamefont{K.~H.} \bibnamefont{Grabstein}},
  \bibinfo{author}{\bibfnamefont{N.~A.} \bibnamefont{Hosken}},
  \bibinfo{author}{\bibfnamefont{F.}~\bibnamefont{Kern}}, \bibnamefont{et~al.},
  \bibinfo{journal}{Journal of Experimental Medicine}
  \textbf{\bibinfo{volume}{202}}, \bibinfo{pages}{673} (\bibinfo{year}{2005}).

\bibitem[{\citenamefont{Lindau et~al.}(2019)\citenamefont{Lindau, Mukherjee,
  Gutschow, Vignali, Warren, Riddell, Makar, Turtle, and Robins}}]{Lindau2019}
\bibinfo{author}{\bibfnamefont{P.}~\bibnamefont{Lindau}},
  \bibinfo{author}{\bibfnamefont{R.}~\bibnamefont{Mukherjee}},
  \bibinfo{author}{\bibfnamefont{M.~V.} \bibnamefont{Gutschow}},
  \bibinfo{author}{\bibfnamefont{M.}~\bibnamefont{Vignali}},
  \bibinfo{author}{\bibfnamefont{E.~H.} \bibnamefont{Warren}},
  \bibinfo{author}{\bibfnamefont{S.~R.} \bibnamefont{Riddell}},
  \bibinfo{author}{\bibfnamefont{K.~W.} \bibnamefont{Makar}},
  \bibinfo{author}{\bibfnamefont{C.~J.} \bibnamefont{Turtle}},
  \bibnamefont{and} \bibinfo{author}{\bibfnamefont{H.~S.}
  \bibnamefont{Robins}}, \bibinfo{journal}{The Journal of Immunology}
  \textbf{\bibinfo{volume}{202}}, \bibinfo{pages}{476} (\bibinfo{year}{2019}).

\bibitem[{\citenamefont{Klein and Flanagan}(2016)}]{Klein2016}
\bibinfo{author}{\bibfnamefont{S.~L.} \bibnamefont{Klein}} \bibnamefont{and}
  \bibinfo{author}{\bibfnamefont{K.~L.} \bibnamefont{Flanagan}},
  \bibinfo{journal}{Nature Reviews Immunology} \textbf{\bibinfo{volume}{16}}
  (\bibinfo{year}{2016}).

\bibitem[{\citenamefont{Shugay et~al.}(2017)\citenamefont{Shugay, Bagaev,
  Zvyagin, Vroomans, Crawford, Dolton, Komech, Sycheva, Koneva, Egorov
  et~al.}}]{Shugay2017}
\bibinfo{author}{\bibfnamefont{M.}~\bibnamefont{Shugay}},
  \bibinfo{author}{\bibfnamefont{D.~V.} \bibnamefont{Bagaev}},
  \bibinfo{author}{\bibfnamefont{I.~V.} \bibnamefont{Zvyagin}},
  \bibinfo{author}{\bibfnamefont{R.~M.} \bibnamefont{Vroomans}},
  \bibinfo{author}{\bibfnamefont{J.~C.} \bibnamefont{Crawford}},
  \bibinfo{author}{\bibfnamefont{G.}~\bibnamefont{Dolton}},
  \bibinfo{author}{\bibfnamefont{E.~A.} \bibnamefont{Komech}},
  \bibinfo{author}{\bibfnamefont{A.~L.} \bibnamefont{Sycheva}},
  \bibinfo{author}{\bibfnamefont{A.~E.} \bibnamefont{Koneva}},
  \bibinfo{author}{\bibfnamefont{E.~S.} \bibnamefont{Egorov}},
  \bibnamefont{et~al.}, \bibinfo{journal}{Nucleic Acids Research} pp.
  \bibinfo{pages}{1--9} (\bibinfo{year}{2017}).

\bibitem[{\citenamefont{Sethna et~al.}(2019)\citenamefont{Sethna, Elhanati,
  Callan, Walczak, and Mora}}]{Sethna2019a}
\bibinfo{author}{\bibfnamefont{Z.}~\bibnamefont{Sethna}},
  \bibinfo{author}{\bibfnamefont{Y.}~\bibnamefont{Elhanati}},
  \bibinfo{author}{\bibfnamefont{C.~G.} \bibnamefont{Callan}},
  \bibinfo{author}{\bibfnamefont{A.~M.} \bibnamefont{Walczak}},
  \bibnamefont{and} \bibinfo{author}{\bibfnamefont{T.}~\bibnamefont{Mora}},
  \bibinfo{journal}{Bioinformatics} \textbf{\bibinfo{volume}{35}},
  \bibinfo{pages}{2974} (\bibinfo{year}{2019}).

\bibitem[{\citenamefont{Chu et~al.}(2019)\citenamefont{Chu, Bi, Emerson,
  Sherwood, Birnbaum, Robins, and Alm}}]{Chu2019}
\bibinfo{author}{\bibfnamefont{N.~D.} \bibnamefont{Chu}},
  \bibinfo{author}{\bibfnamefont{H.~S.} \bibnamefont{Bi}},
  \bibinfo{author}{\bibfnamefont{R.~O.} \bibnamefont{Emerson}},
  \bibinfo{author}{\bibfnamefont{A.~M.} \bibnamefont{Sherwood}},
  \bibinfo{author}{\bibfnamefont{M.~E.} \bibnamefont{Birnbaum}},
  \bibinfo{author}{\bibfnamefont{H.~S.} \bibnamefont{Robins}},
  \bibnamefont{and} \bibinfo{author}{\bibfnamefont{E.~J.} \bibnamefont{Alm}},
  \bibinfo{journal}{BMC Immunology} \textbf{\bibinfo{volume}{20}},
  \bibinfo{pages}{1} (\bibinfo{year}{2019}).

\bibitem[{\citenamefont{Britanova et~al.}(2014)\citenamefont{Britanova,
  Bolotin, Bogdanova, Turchaninova, Lebedev, Lukyanov, Mamedov, Merzlyak,
  Putintseva, Staroverov et~al.}}]{Britanova2014}
\bibinfo{author}{\bibfnamefont{O.~V.} \bibnamefont{Britanova}},
  \bibinfo{author}{\bibfnamefont{D.~A.} \bibnamefont{Bolotin}},
  \bibinfo{author}{\bibfnamefont{E.~A.} \bibnamefont{Bogdanova}},
  \bibinfo{author}{\bibfnamefont{M.~A.} \bibnamefont{Turchaninova}},
  \bibinfo{author}{\bibfnamefont{Y.~B.} \bibnamefont{Lebedev}},
  \bibinfo{author}{\bibfnamefont{S.}~\bibnamefont{Lukyanov}},
  \bibinfo{author}{\bibfnamefont{I.~Z.} \bibnamefont{Mamedov}},
  \bibinfo{author}{\bibfnamefont{E.~M.} \bibnamefont{Merzlyak}},
  \bibinfo{author}{\bibfnamefont{E.~V.} \bibnamefont{Putintseva}},
  \bibinfo{author}{\bibfnamefont{D.~B.} \bibnamefont{Staroverov}},
  \bibnamefont{et~al.}, \bibinfo{journal}{The Journal of Immunology}
  \textbf{\bibinfo{volume}{192}}, \bibinfo{pages}{2689} (\bibinfo{year}{2014}).

\bibitem[{\citenamefont{Shearer et~al.}(2003)\citenamefont{Shearer, Rosenblatt,
  Gelman, Oymopito, Plaeger, Stiehm, Wara, Douglas, Luzuriaga, McFarland
  et~al.}}]{Shearer2003}
\bibinfo{author}{\bibfnamefont{W.~T.} \bibnamefont{Shearer}},
  \bibinfo{author}{\bibfnamefont{H.~M.} \bibnamefont{Rosenblatt}},
  \bibinfo{author}{\bibfnamefont{R.~S.} \bibnamefont{Gelman}},
  \bibinfo{author}{\bibfnamefont{R.}~\bibnamefont{Oymopito}},
  \bibinfo{author}{\bibfnamefont{S.}~\bibnamefont{Plaeger}},
  \bibinfo{author}{\bibfnamefont{E.~R.} \bibnamefont{Stiehm}},
  \bibinfo{author}{\bibfnamefont{D.~W.} \bibnamefont{Wara}},
  \bibinfo{author}{\bibfnamefont{S.~D.} \bibnamefont{Douglas}},
  \bibinfo{author}{\bibfnamefont{K.}~\bibnamefont{Luzuriaga}},
  \bibinfo{author}{\bibfnamefont{E.~J.} \bibnamefont{McFarland}},
  \bibnamefont{et~al.}, \bibinfo{journal}{Journal of Allergy and Clinical
  Immunology} \textbf{\bibinfo{volume}{112}}, \bibinfo{pages}{973}
  (\bibinfo{year}{2003}).

\bibitem[{\citenamefont{Clauset et~al.}(2009)\citenamefont{Clauset, Shalizi,
  and Newman}}]{Clauset2009}
\bibinfo{author}{\bibfnamefont{A.}~\bibnamefont{Clauset}},
  \bibinfo{author}{\bibfnamefont{C.~R.} \bibnamefont{Shalizi}},
  \bibnamefont{and} \bibinfo{author}{\bibfnamefont{M.~E.~J.}
  \bibnamefont{Newman}}, \bibinfo{journal}{SIAM review}
  \textbf{\bibinfo{volume}{51}}, \bibinfo{pages}{661} (\bibinfo{year}{2009}).

\bibitem[{\citenamefont{Efron and Hastie}(2016)}]{Efron2016}
\bibinfo{author}{\bibfnamefont{B.}~\bibnamefont{Efron}} \bibnamefont{and}
  \bibinfo{author}{\bibfnamefont{T.}~\bibnamefont{Hastie}},
  \emph{\bibinfo{title}{{Computer Age Statistical Inference}}}
  (\bibinfo{year}{2016}).

\bibitem[{\citenamefont{Press et~al.}(2007)\citenamefont{Press, Teukolsky,
  Vetterling, and Flannery}}]{Press2007}
\bibinfo{author}{\bibfnamefont{W.~H.} \bibnamefont{Press}},
  \bibinfo{author}{\bibfnamefont{S.~a.} \bibnamefont{Teukolsky}},
  \bibinfo{author}{\bibfnamefont{W.~T.} \bibnamefont{Vetterling}},
  \bibnamefont{and} \bibinfo{author}{\bibfnamefont{B.~P.}
  \bibnamefont{Flannery}}, \emph{\bibinfo{title}{{Numerical Recipes 3rd
  Edition: The Art of Scientific Computing}}}, vol.~\bibinfo{volume}{1}
  (\bibinfo{publisher}{Cambridge University Press}, \bibinfo{year}{2007}).

\bibitem[{\citenamefont{Lewis and Shedler}(1979)}]{Lewis1979}
\bibinfo{author}{\bibfnamefont{P.~A.} \bibnamefont{Lewis}} \bibnamefont{and}
  \bibinfo{author}{\bibfnamefont{G.~S.} \bibnamefont{Shedler}},
  \bibinfo{journal}{Naval research logistics quarterly}
  \textbf{\bibinfo{volume}{26}}, \bibinfo{pages}{403} (\bibinfo{year}{1979}).

\bibitem[{\citenamefont{{De Boer} and Perelson}(2013)}]{DeBoer2013a}
\bibinfo{author}{\bibfnamefont{R.~J.} \bibnamefont{{De Boer}}}
  \bibnamefont{and} \bibinfo{author}{\bibfnamefont{A.~S.}
  \bibnamefont{Perelson}}, \bibinfo{journal}{Journal of Theoretical Biology}
  \textbf{\bibinfo{volume}{327}}, \bibinfo{pages}{45} (\bibinfo{year}{2013}).

\bibitem[{\citenamefont{Borghans et~al.}(2018)\citenamefont{Borghans,
  Tesselaar, and de~Boer}}]{Borghans2018}
\bibinfo{author}{\bibfnamefont{J.~A.} \bibnamefont{Borghans}},
  \bibinfo{author}{\bibfnamefont{K.}~\bibnamefont{Tesselaar}},
  \bibnamefont{and} \bibinfo{author}{\bibfnamefont{R.~J.}
  \bibnamefont{de~Boer}}, \bibinfo{journal}{Immunological Reviews}
  \textbf{\bibinfo{volume}{285}}, \bibinfo{pages}{233} (\bibinfo{year}{2018}).

\bibitem[{\citenamefont{den Braber et~al.}(2012)\citenamefont{den Braber,
  Mugwagwa, Vrisekoop, Westera, M{\"{o}}gling, {Bregje de Boer}, Willems,
  Schrijver, Spierenburg, Gaiser et~al.}}]{DenBraber2012}
\bibinfo{author}{\bibfnamefont{I.}~\bibnamefont{den Braber}},
  \bibinfo{author}{\bibfnamefont{T.}~\bibnamefont{Mugwagwa}},
  \bibinfo{author}{\bibfnamefont{N.}~\bibnamefont{Vrisekoop}},
  \bibinfo{author}{\bibfnamefont{L.}~\bibnamefont{Westera}},
  \bibinfo{author}{\bibfnamefont{R.}~\bibnamefont{M{\"{o}}gling}},
  \bibinfo{author}{\bibfnamefont{A.}~\bibnamefont{{Bregje de Boer}}},
  \bibinfo{author}{\bibfnamefont{N.}~\bibnamefont{Willems}},
  \bibinfo{author}{\bibfnamefont{E.~H.} \bibnamefont{Schrijver}},
  \bibinfo{author}{\bibfnamefont{G.}~\bibnamefont{Spierenburg}},
  \bibinfo{author}{\bibfnamefont{K.}~\bibnamefont{Gaiser}},
  \bibnamefont{et~al.}, \bibinfo{journal}{Immunity}
  \textbf{\bibinfo{volume}{36}}, \bibinfo{pages}{288} (\bibinfo{year}{2012}).

\bibitem[{\citenamefont{Macallan et~al.}(2017)\citenamefont{Macallan, Borghans,
  and Asquith}}]{Macallan2017}
\bibinfo{author}{\bibfnamefont{D.~C.} \bibnamefont{Macallan}},
  \bibinfo{author}{\bibfnamefont{J.~A.} \bibnamefont{Borghans}},
  \bibnamefont{and} \bibinfo{author}{\bibfnamefont{B.}~\bibnamefont{Asquith}},
  \bibinfo{journal}{Vaccines} \textbf{\bibinfo{volume}{5}}
  (\bibinfo{year}{2017}).

\bibitem[{\citenamefont{Hammarlund et~al.}(2003)\citenamefont{Hammarlund,
  Lewis, Hansen, Strelow, Nelson, Sexton, Hanifin, and
  Slifka}}]{Hammarlund2003}
\bibinfo{author}{\bibfnamefont{E.}~\bibnamefont{Hammarlund}},
  \bibinfo{author}{\bibfnamefont{M.~W.} \bibnamefont{Lewis}},
  \bibinfo{author}{\bibfnamefont{S.~G.} \bibnamefont{Hansen}},
  \bibinfo{author}{\bibfnamefont{L.~I.} \bibnamefont{Strelow}},
  \bibinfo{author}{\bibfnamefont{J.~A.} \bibnamefont{Nelson}},
  \bibinfo{author}{\bibfnamefont{G.~J.} \bibnamefont{Sexton}},
  \bibinfo{author}{\bibfnamefont{J.~M.} \bibnamefont{Hanifin}},
  \bibnamefont{and} \bibinfo{author}{\bibfnamefont{M.~K.}
  \bibnamefont{Slifka}}, \bibinfo{journal}{Nature Medicine}
  \textbf{\bibinfo{volume}{9}}, \bibinfo{pages}{1131} (\bibinfo{year}{2003}).

\bibitem[{\citenamefont{Akondy et~al.}(2017)\citenamefont{Akondy, Fitch,
  Edupuganti, Yang, Kissick, Kelvin, Alexe, Nagar, Mccausland, Abdelsamed
  et~al.}}]{Akondy2017}
\bibinfo{author}{\bibfnamefont{R.~S.} \bibnamefont{Akondy}},
  \bibinfo{author}{\bibfnamefont{M.}~\bibnamefont{Fitch}},
  \bibinfo{author}{\bibfnamefont{S.}~\bibnamefont{Edupuganti}},
  \bibinfo{author}{\bibfnamefont{S.}~\bibnamefont{Yang}},
  \bibinfo{author}{\bibfnamefont{H.~T.} \bibnamefont{Kissick}},
  \bibinfo{author}{\bibfnamefont{W.}~\bibnamefont{Kelvin}},
  \bibinfo{author}{\bibfnamefont{G.}~\bibnamefont{Alexe}},
  \bibinfo{author}{\bibfnamefont{S.}~\bibnamefont{Nagar}},
  \bibinfo{author}{\bibfnamefont{M.~M.} \bibnamefont{Mccausland}},
  \bibinfo{author}{\bibfnamefont{H.~A.} \bibnamefont{Abdelsamed}},
  \bibnamefont{et~al.}, \bibinfo{journal}{Nature}
  \textbf{\bibinfo{volume}{552}}, \bibinfo{pages}{pages362}
  (\bibinfo{year}{2017}).

\bibitem[{\citenamefont{Pogorelyy et~al.}(2017)\citenamefont{Pogorelyy,
  Elhanati, Marcou, Sycheva, Komech, Nazarov, Britanova, Chudakov, Mamedov,
  Lebedev et~al.}}]{Pogorelyy2017}
\bibinfo{author}{\bibfnamefont{M.~V.} \bibnamefont{Pogorelyy}},
  \bibinfo{author}{\bibfnamefont{Y.}~\bibnamefont{Elhanati}},
  \bibinfo{author}{\bibfnamefont{Q.}~\bibnamefont{Marcou}},
  \bibinfo{author}{\bibfnamefont{A.~L.} \bibnamefont{Sycheva}},
  \bibinfo{author}{\bibfnamefont{E.~A.} \bibnamefont{Komech}},
  \bibinfo{author}{\bibfnamefont{V.~I.} \bibnamefont{Nazarov}},
  \bibinfo{author}{\bibfnamefont{O.~V.} \bibnamefont{Britanova}},
  \bibinfo{author}{\bibfnamefont{D.~M.} \bibnamefont{Chudakov}},
  \bibinfo{author}{\bibfnamefont{I.~Z.} \bibnamefont{Mamedov}},
  \bibinfo{author}{\bibfnamefont{Y.~B.} \bibnamefont{Lebedev}},
  \bibnamefont{et~al.}, \bibinfo{journal}{PLoS Computational Biology}
  \textbf{\bibinfo{volume}{13}}, \bibinfo{pages}{e1005572}
  (\bibinfo{year}{2017}).

\bibitem[{\citenamefont{Tanno et~al.}(2020)\citenamefont{Tanno, Gould,
  Mcdaniel, Cao, Tanno, and Durrett}}]{Tanno2020}
\bibinfo{author}{\bibfnamefont{H.}~\bibnamefont{Tanno}},
  \bibinfo{author}{\bibfnamefont{T.~M.} \bibnamefont{Gould}},
  \bibinfo{author}{\bibfnamefont{J.~R.} \bibnamefont{Mcdaniel}},
  \bibinfo{author}{\bibfnamefont{W.}~\bibnamefont{Cao}},
  \bibinfo{author}{\bibfnamefont{Y.}~\bibnamefont{Tanno}}, \bibnamefont{and}
  \bibinfo{author}{\bibfnamefont{R.~E.} \bibnamefont{Durrett}},
  \bibinfo{journal}{Proceedings of the National Academy of Sciences}
  \textbf{\bibinfo{volume}{117}} (\bibinfo{year}{2020}).

\bibitem[{\citenamefont{Robins et~al.}(2009)\citenamefont{Robins, Campregher,
  Srivastava, Wacher, Turtle, Kahsai, Riddell, Warren, and
  Carlson}}]{Robins2009}
\bibinfo{author}{\bibfnamefont{H.~S.} \bibnamefont{Robins}},
  \bibinfo{author}{\bibfnamefont{P.~V.} \bibnamefont{Campregher}},
  \bibinfo{author}{\bibfnamefont{S.~K.} \bibnamefont{Srivastava}},
  \bibinfo{author}{\bibfnamefont{A.}~\bibnamefont{Wacher}},
  \bibinfo{author}{\bibfnamefont{C.~J.} \bibnamefont{Turtle}},
  \bibinfo{author}{\bibfnamefont{O.}~\bibnamefont{Kahsai}},
  \bibinfo{author}{\bibfnamefont{S.~R.} \bibnamefont{Riddell}},
  \bibinfo{author}{\bibfnamefont{E.~H.} \bibnamefont{Warren}},
  \bibnamefont{and} \bibinfo{author}{\bibfnamefont{C.~S.}
  \bibnamefont{Carlson}}, \bibinfo{journal}{Blood}
  \textbf{\bibinfo{volume}{114}}, \bibinfo{pages}{4099} (\bibinfo{year}{2009}).

\bibitem[{\citenamefont{Qi et~al.}(2014)\citenamefont{Qi, Liu, Cheng,
  Glanville, Zhang, Lee, Olshen, Weyand, Boyd, and Goronzy}}]{Qi2014}
\bibinfo{author}{\bibfnamefont{Q.}~\bibnamefont{Qi}},
  \bibinfo{author}{\bibfnamefont{Y.}~\bibnamefont{Liu}},
  \bibinfo{author}{\bibfnamefont{Y.}~\bibnamefont{Cheng}},
  \bibinfo{author}{\bibfnamefont{J.}~\bibnamefont{Glanville}},
  \bibinfo{author}{\bibfnamefont{D.}~\bibnamefont{Zhang}},
  \bibinfo{author}{\bibfnamefont{J.-Y.} \bibnamefont{Lee}},
  \bibinfo{author}{\bibfnamefont{R.~a.} \bibnamefont{Olshen}},
  \bibinfo{author}{\bibfnamefont{C.~M.} \bibnamefont{Weyand}},
  \bibinfo{author}{\bibfnamefont{S.~D.} \bibnamefont{Boyd}}, \bibnamefont{and}
  \bibinfo{author}{\bibfnamefont{J.~J.} \bibnamefont{Goronzy}},
  \bibinfo{journal}{Proceedings of the National Academy of Sciences of the
  United States of America} \textbf{\bibinfo{volume}{111}},
  \bibinfo{pages}{13139} (\bibinfo{year}{2014}).

\bibitem[{\citenamefont{Stumpf et~al.}(2005)\citenamefont{Stumpf, Wiuf, and
  May}}]{Stumpf2005}
\bibinfo{author}{\bibfnamefont{M.~P.} \bibnamefont{Stumpf}},
  \bibinfo{author}{\bibfnamefont{C.}~\bibnamefont{Wiuf}}, \bibnamefont{and}
  \bibinfo{author}{\bibfnamefont{R.~M.} \bibnamefont{May}},
  \bibinfo{journal}{Proceedings of the National Academy of Sciences}
  \textbf{\bibinfo{volume}{102}}, \bibinfo{pages}{4221 }
  (\bibinfo{year}{2005}).

\bibitem[{\citenamefont{{Puelma Touzel} et~al.}(2019)\citenamefont{{Puelma
  Touzel}, Walczak, and Mora}}]{PuelmaTouzel2019}
\bibinfo{author}{\bibfnamefont{M.}~\bibnamefont{{Puelma Touzel}}},
  \bibinfo{author}{\bibfnamefont{A.~M.} \bibnamefont{Walczak}},
  \bibnamefont{and} \bibinfo{author}{\bibfnamefont{T.}~\bibnamefont{Mora}},
  \bibinfo{journal}{arXiv preprint arXiv:1912.08304}  (\bibinfo{year}{2019}).

\bibitem[{\citenamefont{Levina and Priesemann}(2017)}]{Levina2017}
\bibinfo{author}{\bibfnamefont{A.}~\bibnamefont{Levina}} \bibnamefont{and}
  \bibinfo{author}{\bibfnamefont{V.}~\bibnamefont{Priesemann}},
  \bibinfo{journal}{Nature Communications} \textbf{\bibinfo{volume}{8}}
  (\bibinfo{year}{2017}).

\bibitem[{\citenamefont{Farber et~al.}(2014)\citenamefont{Farber, Yudanin, and
  Restifo}}]{Farber2014}
\bibinfo{author}{\bibfnamefont{D.~L.} \bibnamefont{Farber}},
  \bibinfo{author}{\bibfnamefont{N.~A.} \bibnamefont{Yudanin}},
  \bibnamefont{and} \bibinfo{author}{\bibfnamefont{N.~P.}
  \bibnamefont{Restifo}}, \bibinfo{journal}{Nature Reviews Immunology}
  \textbf{\bibinfo{volume}{14}}, \bibinfo{pages}{24} (\bibinfo{year}{2014}).

\bibitem[{\citenamefont{Mayer et~al.}(2019)\citenamefont{Mayer, Zhang,
  Perelson, and Wingreen}}]{Mayer2019b}
\bibinfo{author}{\bibfnamefont{A.}~\bibnamefont{Mayer}},
  \bibinfo{author}{\bibfnamefont{Y.}~\bibnamefont{Zhang}},
  \bibinfo{author}{\bibfnamefont{A.~S.} \bibnamefont{Perelson}},
  \bibnamefont{and} \bibinfo{author}{\bibfnamefont{N.~S.}
  \bibnamefont{Wingreen}}, \bibinfo{journal}{Proceedings of the National
  Academy of Sciences} \textbf{\bibinfo{volume}{116}}, \bibinfo{pages}{5914}
  (\bibinfo{year}{2019}).

\bibitem[{\citenamefont{Oakes et~al.}(2017)\citenamefont{Oakes, Heather, Best,
  Byng-Maddick, Husovsky, Ismail, Joshi, Maxwell, Noursadeghi, Riddell
  et~al.}}]{Oakes2017}
\bibinfo{author}{\bibfnamefont{T.}~\bibnamefont{Oakes}},
  \bibinfo{author}{\bibfnamefont{J.~M.} \bibnamefont{Heather}},
  \bibinfo{author}{\bibfnamefont{K.}~\bibnamefont{Best}},
  \bibinfo{author}{\bibfnamefont{R.}~\bibnamefont{Byng-Maddick}},
  \bibinfo{author}{\bibfnamefont{C.}~\bibnamefont{Husovsky}},
  \bibinfo{author}{\bibfnamefont{M.}~\bibnamefont{Ismail}},
  \bibinfo{author}{\bibfnamefont{K.}~\bibnamefont{Joshi}},
  \bibinfo{author}{\bibfnamefont{G.}~\bibnamefont{Maxwell}},
  \bibinfo{author}{\bibfnamefont{M.}~\bibnamefont{Noursadeghi}},
  \bibinfo{author}{\bibfnamefont{N.}~\bibnamefont{Riddell}},
  \bibnamefont{et~al.}, \bibinfo{journal}{Frontiers in Immunology}
  \textbf{\bibinfo{volume}{8}}, \bibinfo{pages}{1} (\bibinfo{year}{2017}).

\bibitem[{\citenamefont{Volkov et~al.}(2003)\citenamefont{Volkov, Banavar,
  Hubbell, and Maritan}}]{Volkov2003}
\bibinfo{author}{\bibfnamefont{I.}~\bibnamefont{Volkov}},
  \bibinfo{author}{\bibfnamefont{J.~R.} \bibnamefont{Banavar}},
  \bibinfo{author}{\bibfnamefont{S.~P.} \bibnamefont{Hubbell}},
  \bibnamefont{and} \bibinfo{author}{\bibfnamefont{A.}~\bibnamefont{Maritan}},
  \bibinfo{journal}{Nature} pp. \bibinfo{pages}{1035--1037}
  (\bibinfo{year}{2003}).

\bibitem[{\citenamefont{Desponds et~al.}(2016)\citenamefont{Desponds, Mora, and
  Walczak}}]{Desponds2016}
\bibinfo{author}{\bibfnamefont{J.}~\bibnamefont{Desponds}},
  \bibinfo{author}{\bibfnamefont{T.}~\bibnamefont{Mora}}, \bibnamefont{and}
  \bibinfo{author}{\bibfnamefont{A.~M.} \bibnamefont{Walczak}},
  \bibinfo{journal}{Proceedings of the National Academy of Sciences}
  \textbf{\bibinfo{volume}{113}}, \bibinfo{pages}{274} (\bibinfo{year}{2016}).

\bibitem[{\citenamefont{Desponds et~al.}(2017)\citenamefont{Desponds, Mayer,
  Mora, and Walczak}}]{Desponds2017}
\bibinfo{author}{\bibfnamefont{J.}~\bibnamefont{Desponds}},
  \bibinfo{author}{\bibfnamefont{A.}~\bibnamefont{Mayer}},
  \bibinfo{author}{\bibfnamefont{T.}~\bibnamefont{Mora}}, \bibnamefont{and}
  \bibinfo{author}{\bibfnamefont{A.~M.} \bibnamefont{Walczak}},
  \bibinfo{journal}{arXiv preprint arXiv:1703.00226}  (\bibinfo{year}{2017}).

\bibitem[{\citenamefont{Greef et~al.}(2020)\citenamefont{Greef, Oakes,
  Gerritsen, Ismail, Heather, Hermsen, Chain, Boer, James, Hermsen
  et~al.}}]{Greef2020}
\bibinfo{author}{\bibfnamefont{P.~C.~D.} \bibnamefont{Greef}},
  \bibinfo{author}{\bibfnamefont{T.}~\bibnamefont{Oakes}},
  \bibinfo{author}{\bibfnamefont{B.}~\bibnamefont{Gerritsen}},
  \bibinfo{author}{\bibfnamefont{M.}~\bibnamefont{Ismail}},
  \bibinfo{author}{\bibfnamefont{J.~M.} \bibnamefont{Heather}},
  \bibinfo{author}{\bibfnamefont{R.}~\bibnamefont{Hermsen}},
  \bibinfo{author}{\bibfnamefont{B.}~\bibnamefont{Chain}},
  \bibinfo{author}{\bibfnamefont{R.~J.~D.} \bibnamefont{Boer}},
  \bibinfo{author}{\bibfnamefont{M.}~\bibnamefont{James}},
  \bibinfo{author}{\bibfnamefont{R.}~\bibnamefont{Hermsen}},
  \bibnamefont{et~al.}, \bibinfo{journal}{eLife} \textbf{\bibinfo{volume}{9}},
  \bibinfo{pages}{e49900} (\bibinfo{year}{2020}).

\bibitem[{\citenamefont{Dodds et~al.}(2017)\citenamefont{Dodds, Dewhurst,
  Hazlehurst, Oort, Mitchell, Reagan, Williams, and Danforth}}]{Dodds2017}
\bibinfo{author}{\bibfnamefont{P.~S.} \bibnamefont{Dodds}},
  \bibinfo{author}{\bibfnamefont{D.~R.} \bibnamefont{Dewhurst}},
  \bibinfo{author}{\bibfnamefont{F.~F.} \bibnamefont{Hazlehurst}},
  \bibinfo{author}{\bibfnamefont{C.~M.~V.} \bibnamefont{Oort}},
  \bibinfo{author}{\bibfnamefont{L.}~\bibnamefont{Mitchell}},
  \bibinfo{author}{\bibfnamefont{A.~J.} \bibnamefont{Reagan}},
  \bibinfo{author}{\bibfnamefont{J.~R.} \bibnamefont{Williams}},
  \bibnamefont{and} \bibinfo{author}{\bibfnamefont{C.~M.}
  \bibnamefont{Danforth}}, \bibinfo{journal}{Physical Review E}
  \textbf{\bibinfo{volume}{052301}}, \bibinfo{pages}{1} (\bibinfo{year}{2017}).

\bibitem[{\citenamefont{{De Boer} and Perelson}(1995)}]{DeBoer1995}
\bibinfo{author}{\bibfnamefont{R.~J.} \bibnamefont{{De Boer}}}
  \bibnamefont{and} \bibinfo{author}{\bibfnamefont{A.~S.}
  \bibnamefont{Perelson}}, \bibinfo{journal}{Journal of Theoretical Biology}
  \textbf{\bibinfo{volume}{175}}, \bibinfo{pages}{567} (\bibinfo{year}{1995}).

\bibitem[{\citenamefont{{De Boer} and Perelson}(1994)}]{DeBoer1994}
\bibinfo{author}{\bibfnamefont{R.~J.} \bibnamefont{{De Boer}}}
  \bibnamefont{and} \bibinfo{author}{\bibfnamefont{A.~S.}
  \bibnamefont{Perelson}}, \bibinfo{journal}{Journal of Theoretical Biology}
  \textbf{\bibinfo{volume}{169}}, \bibinfo{pages}{375} (\bibinfo{year}{1994}).

\bibitem[{\citenamefont{{De Boer} et~al.}(2001)\citenamefont{{De Boer},
  Freitas, and Perelson}}]{DeBoer2001}
\bibinfo{author}{\bibfnamefont{R.~J.} \bibnamefont{{De Boer}}},
  \bibinfo{author}{\bibfnamefont{A.~A.} \bibnamefont{Freitas}},
  \bibnamefont{and} \bibinfo{author}{\bibfnamefont{A.~S.}
  \bibnamefont{Perelson}}, \bibinfo{journal}{Journal of Theoretical Biology}
  \textbf{\bibinfo{volume}{212}}, \bibinfo{pages}{333} (\bibinfo{year}{2001}).

\bibitem[{\citenamefont{Mayer et~al.}(2015)\citenamefont{Mayer,
  Balasubramanian, Mora, and Walczak}}]{Mayer2015}
\bibinfo{author}{\bibfnamefont{A.}~\bibnamefont{Mayer}},
  \bibinfo{author}{\bibfnamefont{V.}~\bibnamefont{Balasubramanian}},
  \bibinfo{author}{\bibfnamefont{T.}~\bibnamefont{Mora}}, \bibnamefont{and}
  \bibinfo{author}{\bibfnamefont{A.~M.~A.} \bibnamefont{Walczak}},
  \bibinfo{journal}{Proceedings of the National Academy of Sciences}
  \textbf{\bibinfo{volume}{112}}, \bibinfo{pages}{5950} (\bibinfo{year}{2015}).

\bibitem[{\citenamefont{Yule}(1924)}]{Yule1924}
\bibinfo{author}{\bibfnamefont{G.~U.} \bibnamefont{Yule}},
  \bibinfo{journal}{Phil. Trans. B} \textbf{\bibinfo{volume}{213}},
  \bibinfo{pages}{21} (\bibinfo{year}{1924}).

\bibitem[{\citenamefont{Luria and Delbr{\"{u}}ck}(1943)}]{Luria1943}
\bibinfo{author}{\bibfnamefont{S.~E.} \bibnamefont{Luria}} \bibnamefont{and}
  \bibinfo{author}{\bibfnamefont{M.}~\bibnamefont{Delbr{\"{u}}ck}},
  \bibinfo{journal}{Genetics} \textbf{\bibinfo{volume}{28}},
  \bibinfo{pages}{491} (\bibinfo{year}{1943}).

\bibitem[{\citenamefont{Barab{\'{a}}si and Albert}(1999)}]{Barabasi1999}
\bibinfo{author}{\bibfnamefont{A.~L.} \bibnamefont{Barab{\'{a}}si}}
  \bibnamefont{and} \bibinfo{author}{\bibfnamefont{R.}~\bibnamefont{Albert}},
  \bibinfo{journal}{Science} \textbf{\bibinfo{volume}{286}},
  \bibinfo{pages}{509} (\bibinfo{year}{1999}).

\bibitem[{\citenamefont{Sornette and Cont}(1997)}]{Sornette1997}
\bibinfo{author}{\bibfnamefont{D.}~\bibnamefont{Sornette}} \bibnamefont{and}
  \bibinfo{author}{\bibfnamefont{R.}~\bibnamefont{Cont}},
  \bibinfo{journal}{Journal de Physique} \textbf{\bibinfo{volume}{1}},
  \bibinfo{pages}{431 } (\bibinfo{year}{1997}).

\bibitem[{\citenamefont{Gabaix}(1999)}]{Gabaix1999}
\bibinfo{author}{\bibfnamefont{X.}~\bibnamefont{Gabaix}}, \bibinfo{journal}{The
  Quarterly Journal of Economics} \textbf{\bibinfo{volume}{114}},
  \bibinfo{pages}{739} (\bibinfo{year}{1999}).

\bibitem[{\citenamefont{Newman}(2005)}]{Newman2005}
\bibinfo{author}{\bibfnamefont{M.~E.} \bibnamefont{Newman}},
  \bibinfo{journal}{Contemporary Physics} \textbf{\bibinfo{volume}{46}},
  \bibinfo{pages}{323} (\bibinfo{year}{2005}).

\bibitem[{\citenamefont{Ferri}(2018)}]{Ferri2018}
\bibinfo{author}{\bibfnamefont{S.}~\bibnamefont{Ferri}},
  \emph{\bibinfo{title}{{Master thesis: Stochastic processes for natural
  evolutionary dynamics of T-cell repertoires}}} (\bibinfo{year}{2018}).

\end{thebibliography}

\end{document}


\title{Supporting information for Early life imprints the hierarchy of T cell clone sizes}\author{Mario U. Gaimann}
\affiliation{Lewis-Sigler Institute for Integrative Genomics, Princeton University}
\affiliation{Arnold Sommerfeld Center for Theoretical Physics and Center for NanoScience, Department of Physics, Ludwig-Maximilians-Universität München}
\author{Maximilian Nguyen}
\affiliation{Lewis-Sigler Institute for Integrative Genomics, Princeton University}
\author{Jonathan Desponds}
\affiliation{NSF-Simons Center for Quantitative Biology, Northwestern University}
\author{Andreas Mayer}
\affiliation{Lewis-Sigler Institute for Integrative Genomics, Princeton University}

\maketitle

\tableofcontents

\clearpage

\setcounter{figure}{0}
\renewcommand{\thefigure}{S\arabic{figure}}%
\renewcommand{\thesection}{\Alph{section}}
\renewcommand{\thesubsection}{\arabic{subsection}}

\section{Supporting Figures}

\begin{figure}[h!]
 \begin{center}
     \includegraphics{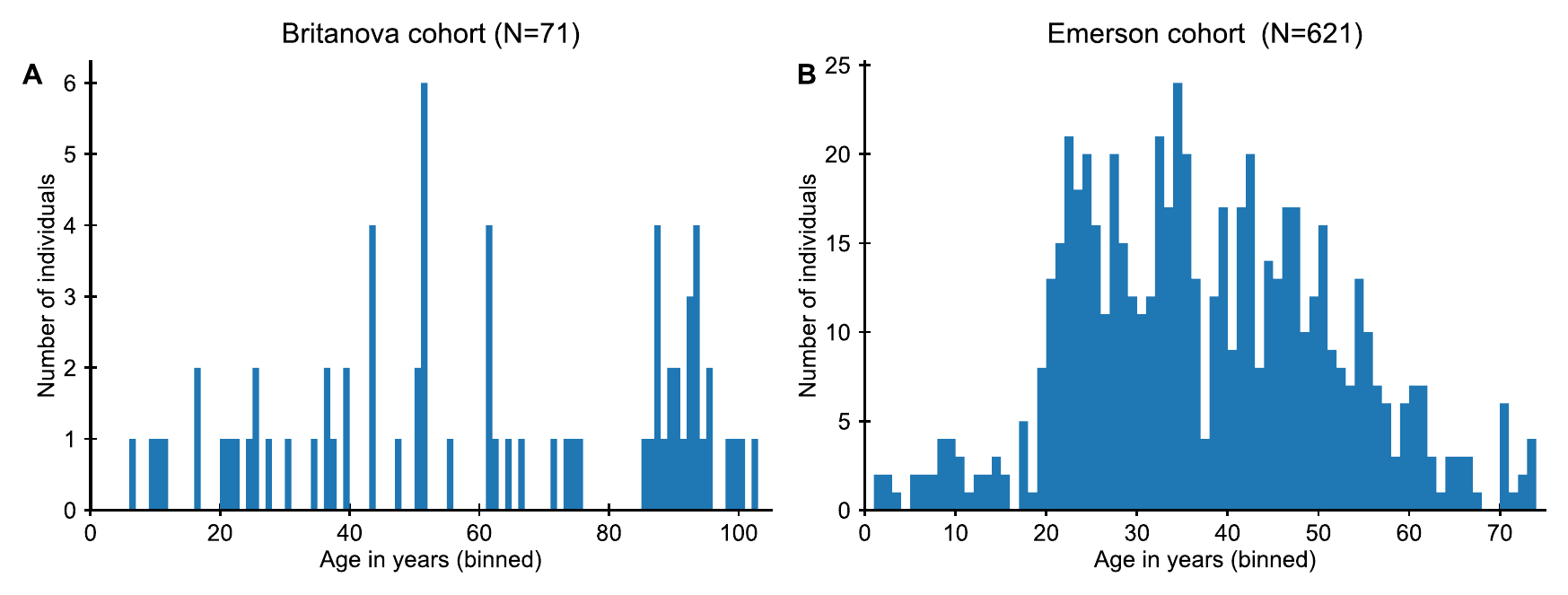}
 \end{center}
    \caption{{\bf Distribution of ages in the two cohort studies.}
  } \label{fig_cohortages}
\end{figure}

\begin{figure}
 \begin{center}
     \includegraphics[width=\textwidth]{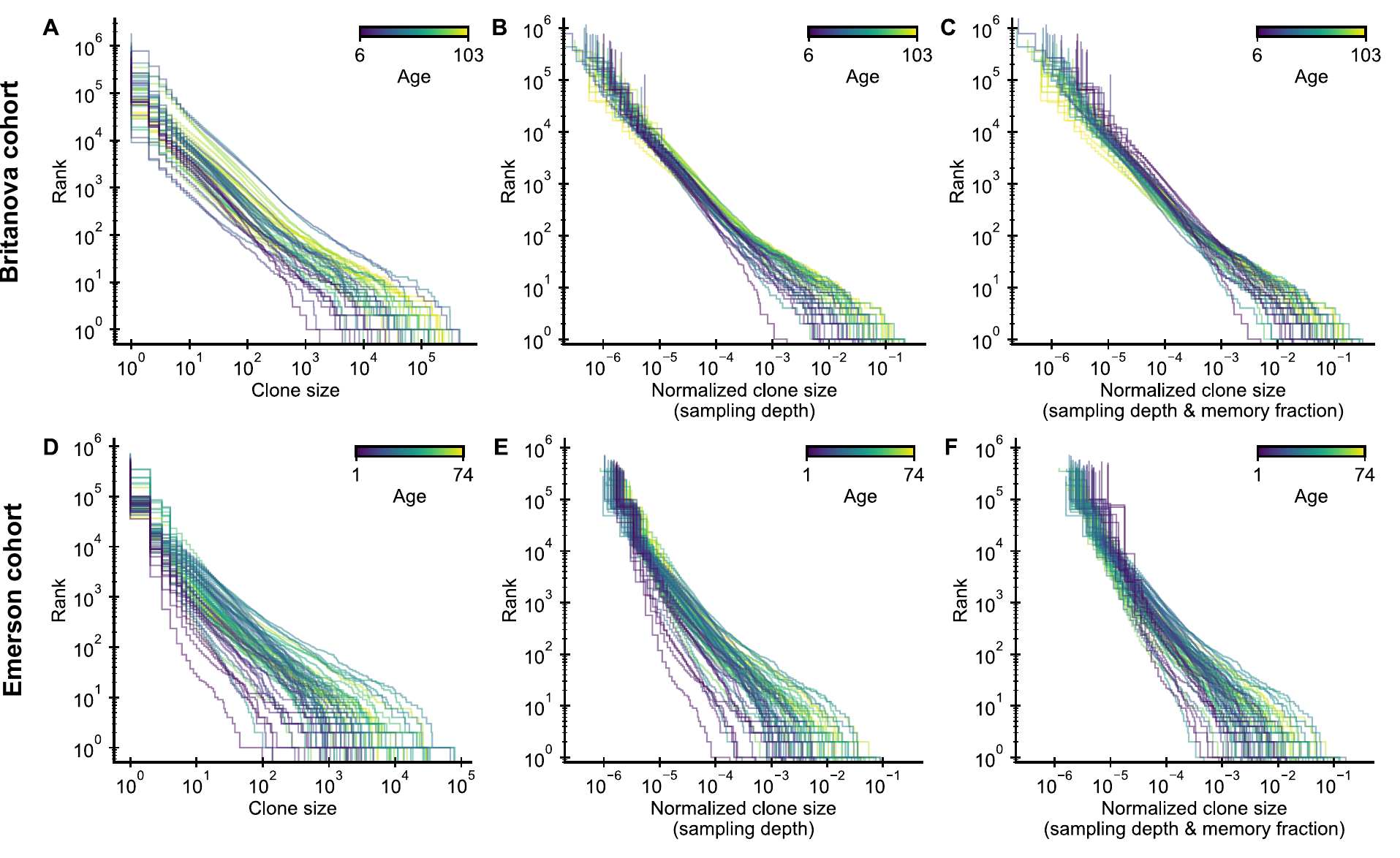}
 \end{center}
    \caption{{\bf Influence of normalization choice on clone size distributions (see Extended Methods~\ref{sec_dataanalysis}).} (A,D) Raw clone size distributions show large variability due to different sample sizes. (B,E) A normalization by sampling depth removes much of this variation. (C,F) A normalization by the fraction of memory cells at different ages further collapses the tails of the clone size distributions.
    Data sources: A-C \cite{Britanova2016}, D-F \cite{Emerson2017}.
  } \label{fig_clonesizes_stepbystep}
\end{figure}

\begin{figure}
 \begin{center}
     \includegraphics{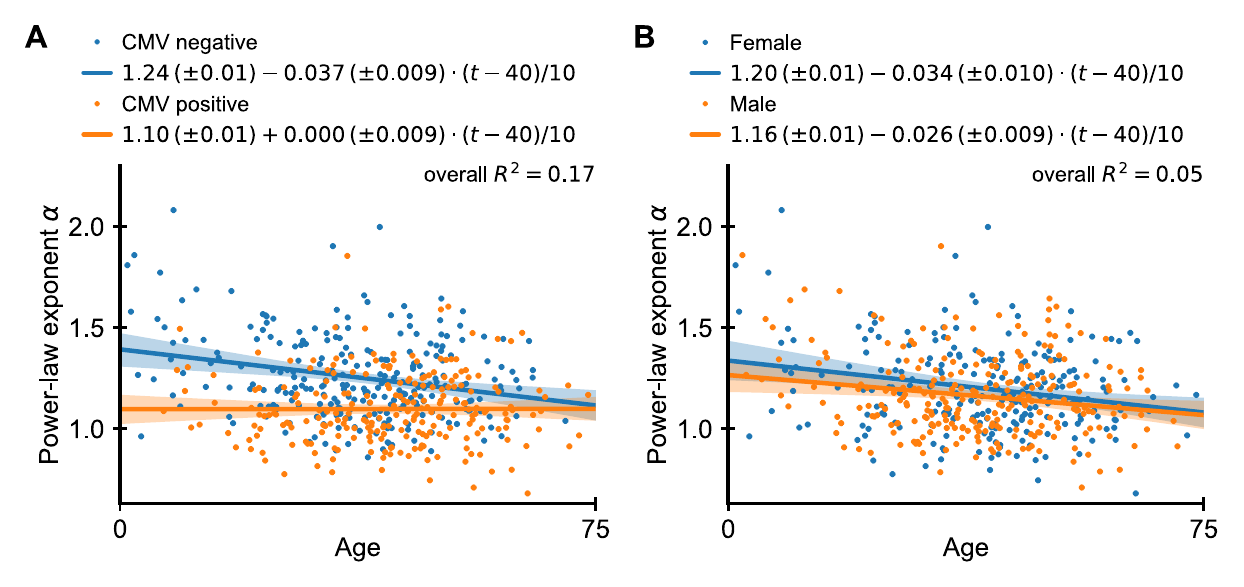}
 \end{center}
    \caption{{\bf Dependence of power-law exponent on age by cytomegalovirus (CMV) infection status and sex.}
       (A) Chronic infection with CMV drives large clonal expansions \cite{Sylwester2005,Lindau2019}. We thus repeated the analysis of Fig.~\ref{main-fig_statistics}E separating individuals based on their CMV infection status (fitted lines shown in legend, regression results displayed as offset + slope $\cdot$ (age in years - 40)/10). Overall, CMV positive individuals have a smaller $\alpha$ than uninfected individuals, which is independent of age. The average exponent in CMV negative individuals decreases slowly with age, and in old age coincides those of CMV positive individuals. Combining CMV infection status and age explained a significantly larger proportion of the variance in scaling exponents (17\%) than age alone. 
       (B) Many immune determinants differ markedly between the sexes \cite{Klein2016}. We thus analyzed whether $\alpha$ depends on sex. We find that the dependence on age is similar among the sexes, but men have on average a slightly smaller exponent than women indicating a more skewed repertoire organization. 
     Data source: Emerson \etal \cite{Emerson2017}.
    }
    \label{fig_exponent_cmv}
\end{figure}

\begin{figure}
 \begin{center}
     \includegraphics{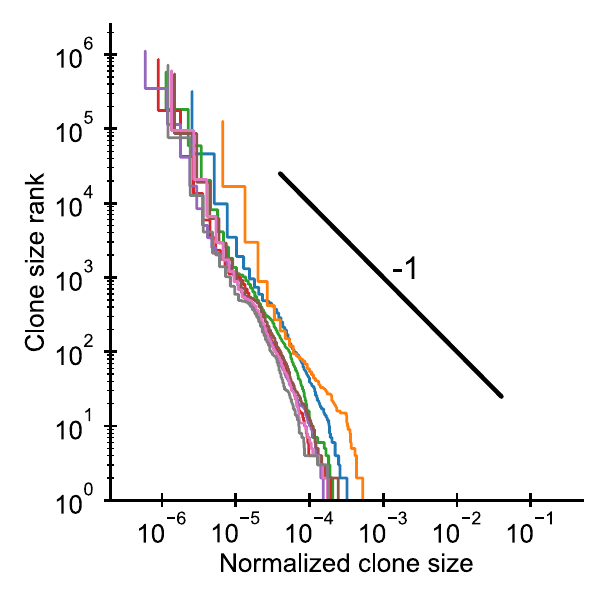}
 \end{center}
    \caption{{\bf Clone size distributions of human T cell receptor repertoires in cordblood.} Each line shows the distribution in one individual. The black line shows a power law with a slope of -1 for visual comparison. The fitted power-law exponents $\alpha=2.1\pm0.1$ (mean $\pm$ SE) are larger than in adult repertoires, but clone sizes are already remarkably broad. 
    Data source:  Britanova \etal \cite{Britanova2016}.
  } \label{fig_clonesizes_britanova_cordblood}
\end{figure}

\begin{figure}
 \begin{center}
     \includegraphics{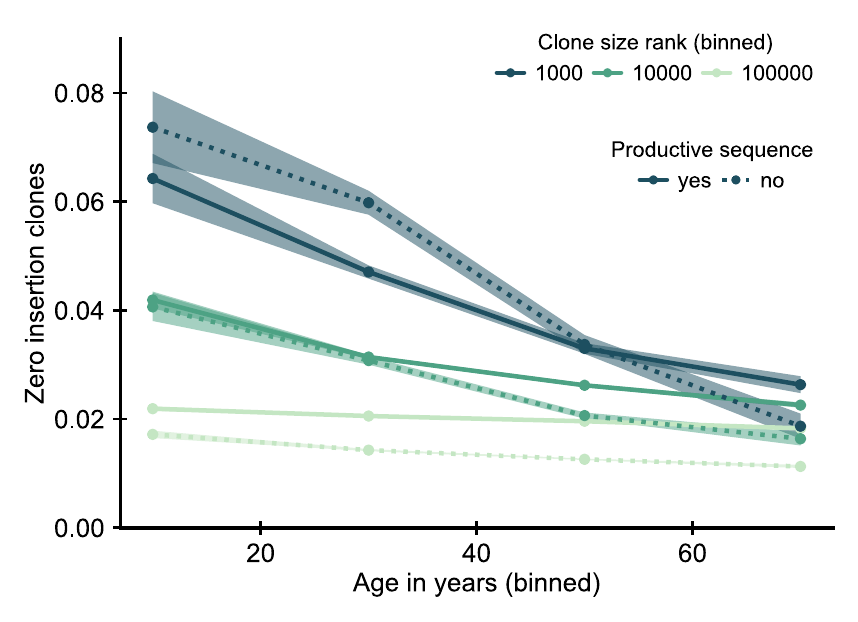}
 \end{center}
    \caption{{\bf Comparison of the relative fraction of zero insertion clones within productive and unproductive sequences.} Sequences with zero insertions code for a particular subset of all possible TCRs, and some of their enrichment might represent a peripheral selective advantage of this subset of receptors. We thus asked how the enrichment depends on whether the sequence used to define the clone represents a productive or unproductive rearrangement. An unproductive rearrangement, in which the recombination process introduces a frameshift or stop codon, can be rescued by a second productive rearrangement, but is not expressed and thus not selected upon. Under the adult recombination statistics an unproductive zero insertion sequence is likely to be paired with a productive sequence with many insertions, and thus we would not expect to see a similar enrichment for unproductive sequences if a general peripheral selective advantage was causing the enrichment. 
 Data source: Emerson \etal \cite{Emerson2017}.
    }
    \label{fig_zeroinsertion_out}
\end{figure}

\begin{figure}
 \begin{center}
     \includegraphics{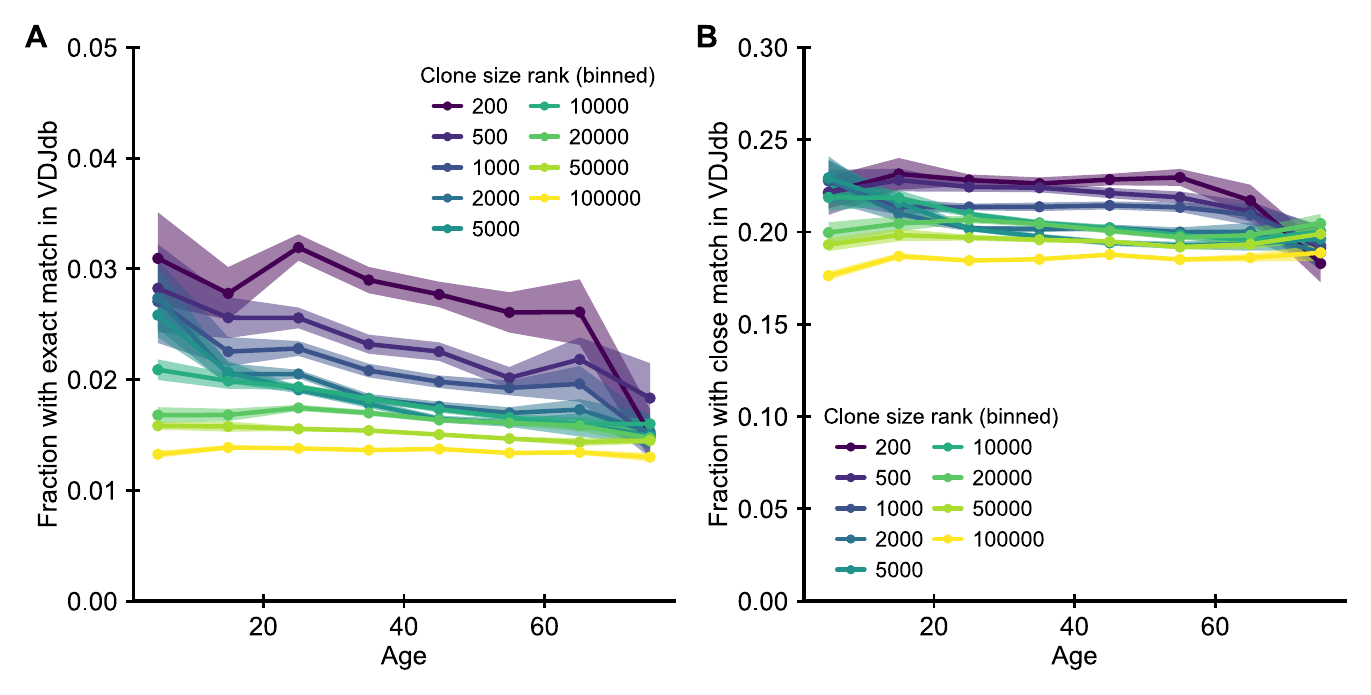}
 \end{center}
    \caption{{\bf Large clones are enriched in clones with known specificity.}
    (A) Fraction of clones with TCRs that have exact matches in the VDJdb \cite{Shugay2017} of known antigen specificities. (B) Fraction of clones with close matches (defined as nearest neighbor sequences in a Levenshtein distance sense, i.e. sequences with a single amino acid substitution, insertion or deletion).
    T cells known to be specific to particular antigens are enriched among the most abundant clones. However, there is little change in this enrichment as a function of age.
     Data source: Emerson \etal \cite{Emerson2017}.
    }
    \label{fig_invdjdb}
\end{figure}

\begin{figure}
 \begin{center}
     \includegraphics{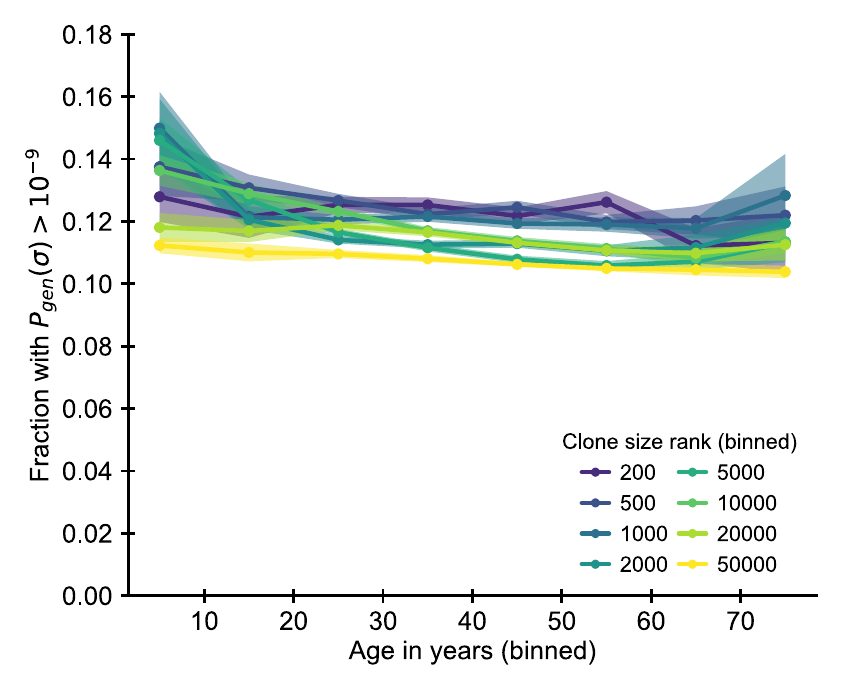}
 \end{center}
    \caption{{\bf Large clones are enriched in clones that are likely to be convergently recombined.}
    Fraction of clones with TCR sequences $\sigma$ with a probability of generation $P_{gen}(\sigma)$ higher than $10^{-9}$. The probability of generation was calculated based on the nucleotide sequence using a probabilistic model of recombination with default parameters for human TCR sequences \cite{Sethna2019a}. To remove confounding by the early expansionary dynamics we excluded zero insertion clones as most of these clones also have high probability of generation. We find that clones with high $P_{gen}$ are moderately more likely to be large. In comparison to the zero insertion clones, there is little change in their enrichment as a function of age.
 Data source: Emerson \etal \cite{Emerson2017}.
    }
    \label{fig_pgen}
\end{figure}

\begin{figure}
 \begin{center}
     \includegraphics{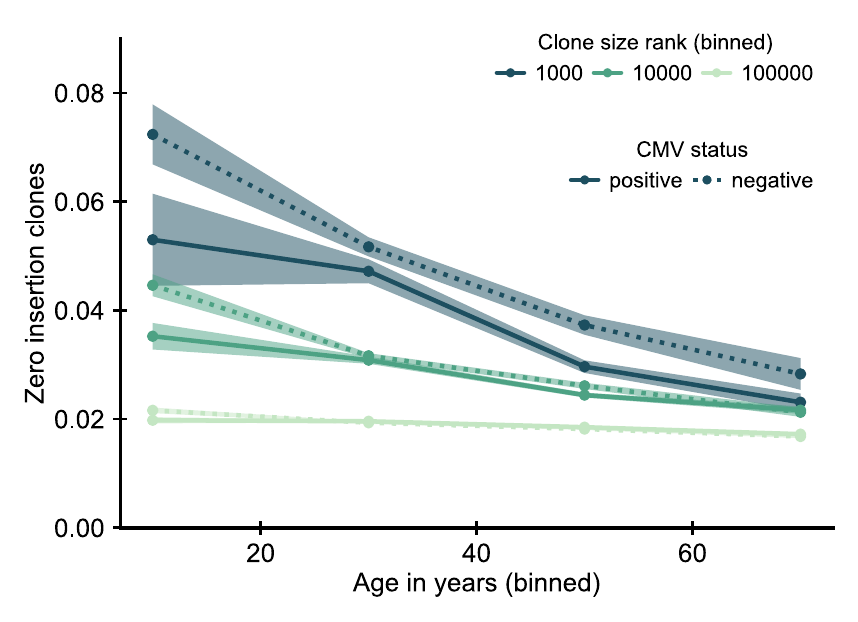}
 \end{center}
    \caption{{\bf Influence of CMV infection status on enrichment of zero insertion clones.}
    Data source: Emerson \etal \cite{Emerson2017}.
    }
    \label{fig_zeroinsertion_cmv}
\end{figure}

\begin{figure}
 \begin{center}
     \includegraphics{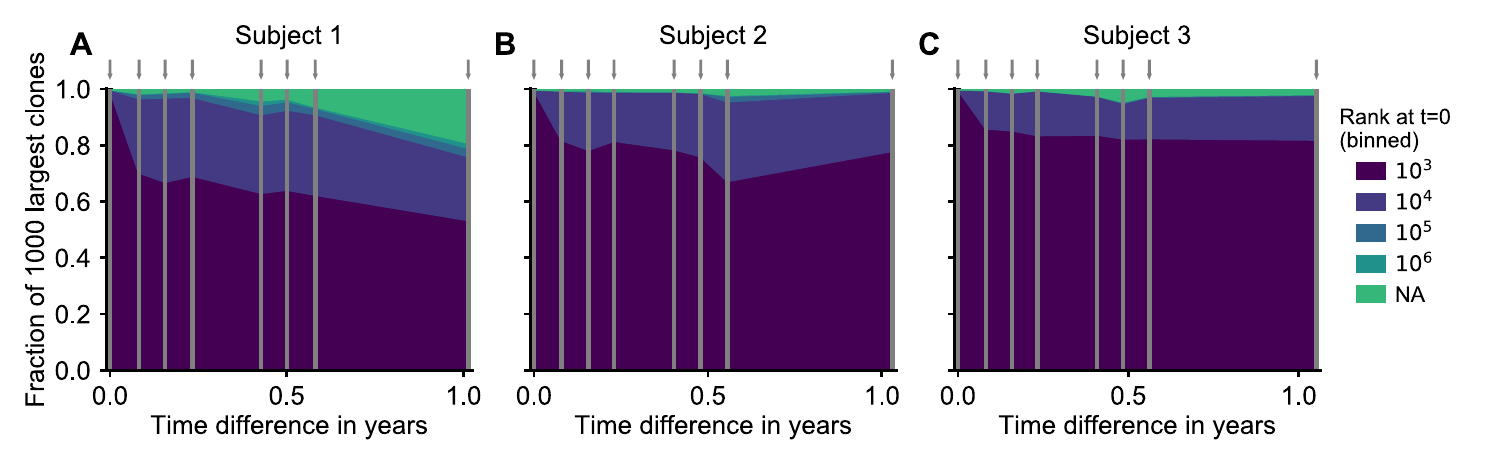}
 \end{center}
    \caption{{\bf Provenance of large T cell clones in a longitudinal study of T cell repertoire dynamics.} Longitudinal analysis of the origin of the 1000 largest clones at each time point (indicated by arrows) in three healthy adults over a one year time frame. For each clone we determined whether it was also sampled at the earliest time point, and if so at what clone size. The plot displays the fraction of clones that fall within a specific clone size rank bin at the first time point. At all times a majority of clones was already large initially. A small fraction was not detected at all at the first time point (ND) likely representing recently expanded clones. (Supplement to Fig.~\ref{main-figzeroinsertion}D which corresponds to panel C.)
    Data source: Chu \etal \cite{Chu2019}.
    }
    \label{fig_longitudinal_provenance}
\end{figure}

\begin{figure}
 \begin{center}
     \includegraphics{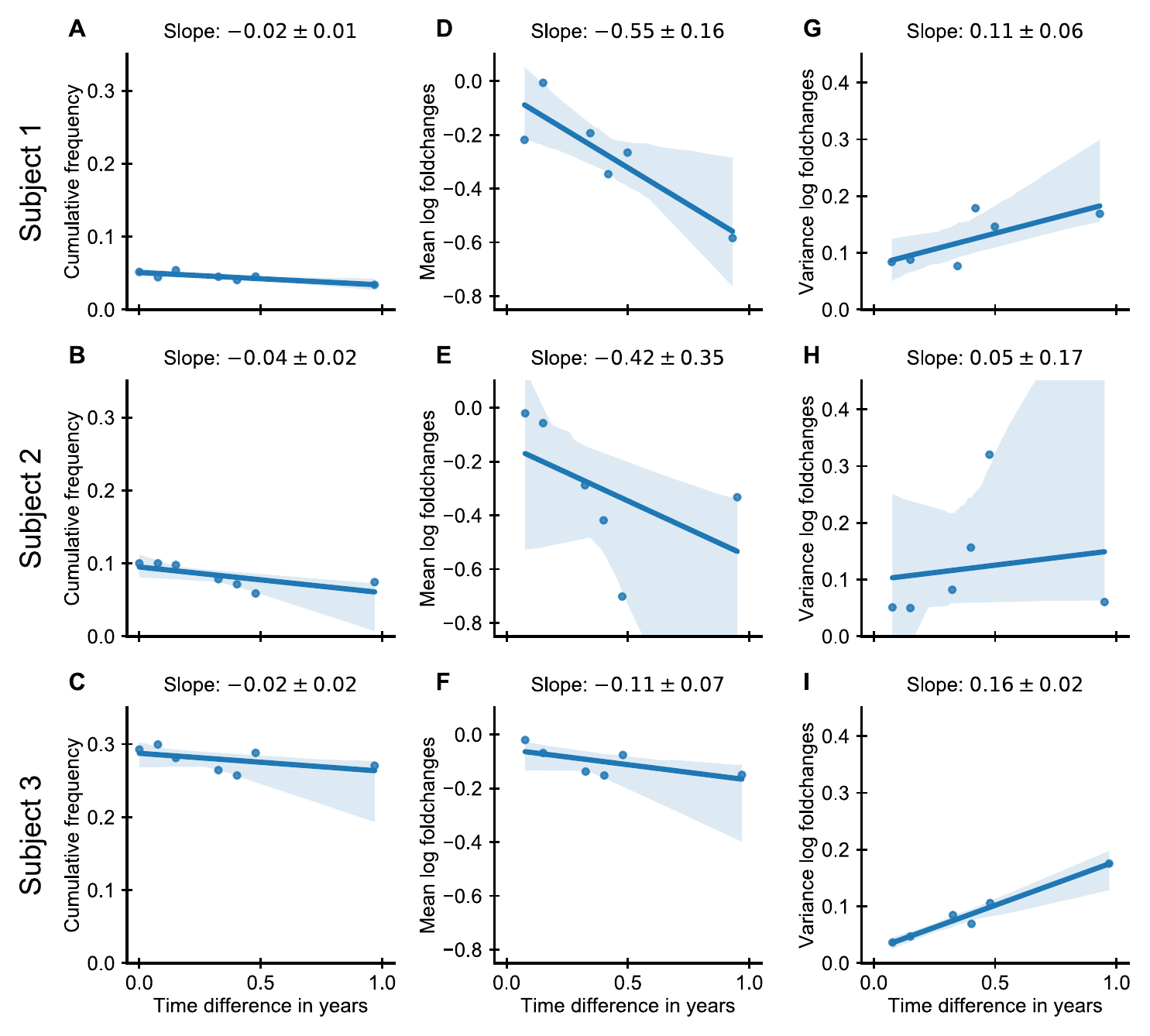}
 \end{center}
    \caption{{\bf Dynamics of large persistent T cell clones in a longitudinal study of T cell repertoire dynamics.} Dynamics of the 250 largest clones from second time point onwards excluding those not sampled at the first time point. (A-C) Fraction of the repertoire represented by these clones (sum of their normalized clone sizes); (D-F) mean and (G-I) variance of the log-foldchanges of their normalized clone sizes relative to time point 2. (Supplement to Fig.~\ref{main-figzeroinsertion}E which corresponds to panel I.)
Data source: Chu \etal \cite{Chu2019}.
    }
    \label{fig_longitudinal_msd}
\end{figure}

\begin{figure}
 \begin{center}
     \includegraphics[scale=1.2]{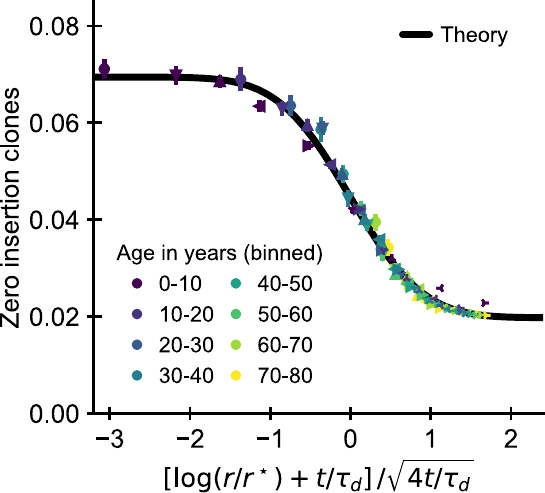}
 \end{center}
    \caption{{\bf Data collapse by parameter rescaling for the simulated cohort.} 
    Same data as in Fig.~\ref{main-figzeroinsertion}F displayed with a rescaled x-axis using fitted parameters $\tau_d = 10.2\pm0.4\, \mathrm{years}, r^\star = 1.19\pm0.08 \cdot 10^4$. The data collapses onto a sigmoidal function predicted by theory (SI Text~Eq.~\ref{eqerfc}) with fitted $p_{0, -} = 0.0695\pm0.0012$, $p_{0,+} = 0.0198\pm0.0003$ (black line).
    }
    \label{fig_mastercurve_model_collapse}
\end{figure}

\clearpage

\section{Extended Methods}

\subsection{Data sources}
\label{sec_datasources}

For all studies we used data, which was preprocessed as described in the original study. This data is publicly available from \url{https://doi.org/10.5281/zenodo.826447} (Britanova cohort), \url{https://doi.org/10.21417/B7001Z} (Emerson cohort), \url{https://doi.org/10.21417/PL2018JI} (data from \cite{Lindau2019}), and \url{https://doi.org/10.21417/B7J01X} (longitudinal study).

The Britanova cohort comprises $71$ individuals spanning ages $6-103$ years, as well as $8$ cord blood samples. The Emerson cohort spans ages $1-74$ years and consists of a training and validation set of $666$ and $120$ individuals, respectively. From the training set we excluded $111$ samples with missing age information and $62$ samples with a conflicting data format. We used only samples from the training set to analyze how the scaling law of repertoire organization changes with age (Fig.~\ref{main-fig_statistics}C,E). For the zero insertion enrichment analyses (Fig.~\ref{main-figzeroinsertion}B,C) we combined both the training and validation set together with separately published repertoire sequencing data from $8$ elderly individuals \cite{Lindau2019} generated using the same experimental pipeline (immunoSEQ, Adaptive Biotechnologies, Seattle) to achieve the broadest possible coverage of all age groups.

The longitudinal study by Chu \etal \cite{Chu2019} performed repertoire sequencing of peripheral blood from three healthy female volunteers (using the immunoSEQ pipeline) over 8 time points spanning a $\unsim 1$ year time frame. One individual in the study was in mid-adulthood (24-45 years, Subject 3 in the original study), while two were in early adulthood (18-24 years, Subject 1 and 2 in the original study). In the main text Fig.~\ref{main-figzeroinsertion}D we display data from the older individual as we expect dynamics of large clones to be masked less by measurement noise as the large clones increase in relative abundance with age. 

All studies from which we analyzed data sequenced the locus coding for the TCR CDR3 $\beta$-chain only, and we thus define clones as collections of cells sharing the same CDR3 $\beta$-chain. Clone sizes are defined as the number of distinct unique molecular identifiers (UMIs) sequenced (Britanova cohort), or based on sequencing reads (Emerson cohort). The definition of a clone solely based on the CDR3 $\beta$-chain neglects convergent recombination of the most easily produced receptors with different CDR3 $\alpha$-chains, but we expect convergent recombination to be sufficiently rare overall for this distinction not to qualitatively affect clone size distributions.

We also used flow cytometry data on the fraction of naive cells from Britanova \etal~\cite{Britanova2014} (available at \url{https://doi.org/10.1371/journal.pcbi.1005572.s016}) and from Shearer \etal~\cite{Shearer2003}.

\subsection{Data analysis}
\label{sec_dataanalysis}

\customparagraph{Fitting power-law exponents.}
We estimate the power-law exponent from sampled clone sizes $\{ C_i \}$, $i= 1, \dots, M$, which exceed a minimal size $C_{min}$ by numerically maximizing the log-likelihood of the data \cite{Clauset2009},
\begin{equation}
    \mathcal{L} = -M \ln \zeta(1+\alpha, C_{min}) - (1+\alpha) \sum_{i=1}^M \ln C_i,
\end{equation}
where $\zeta(x, k)$ is the incomplete Riemann zeta function. We use $C_{min} = 16$ for both cohorts, which provides a balance between minimizing bias of the estimated exponents induced by subsampling while not overly increasing the variance of the estimator by excluding most of the data (see Fig.~\ref{fig_powerlaw_trimming}).

\customparagraph{Fitting the zero insertion profiles.}
To fit the zero insertion fractions to the theory prediction (Eq.~\ref{eqerfc}) we determine the values for $r^\star$ and $\tau_d$ by a weighted least squares fit. We set $r$ and $t$ to the mid-value of each bin for the data. We weight each value by its empirical standard error with an additional model specification error that we set to a fixed value of $2\cdot 10^{-3}$. To demonstrate the feasibility of the parameter inference we reinferred the parameters from the simulated data and recovered those used as parameter values for the simulation.
We also fitted the values of $p_{0,-}$ and $p_{0,+}$, but we note that they are not used in the rescaling and are only needed to display the theoretical curve (Eq.~\ref{eqerfc}).

\customparagraph{Normalization of clone sizes. }
Variations in sampling depth can confound comparisons of clone sizes (SI Text~\ref{subsampling}).
Intuitively, if we sample more cells overall we also expect to sample proportionally more cells belonging to each given clone. This suggests to use the frequency with which cells are sampled from a given clone as a more robust measure, which can be empirically estimated by normalizing each clone size by the total sample size.
We further normalize clone sizes by the fraction of memory T cells found in people of different ages to account for the increase in memory cell fraction in peripheral blood with age (SI Text~\ref{phenotypes}).
Together these two normalization steps lead to a large degree of data collapse as compared to unnormalized clone sizes.

\customparagraph{Regression analyses.}
We determine 95\% confidence intervals on regression lines by bootstrapping using case resampling \cite{Efron2016}.

\subsection{Simulation procedures}
\label{secsimulations}

\customparagraph{Repertoire formation.}
To simulate the model efficiently at large scales we use a mean-field competition approximation (SI Text~\ref{si_meanfield}).
We verified the validity of the mean-field assumption by comparing them to full stochastic simulations of the coupled birth-death-immigration equations, which we simulated using the Gillespie algorithm~\cite{Press2007} (Fig.~\ref{fig_meanfield}).
In the mean field approximation the proliferation rate is time-dependent, which requires a specific procedure for sampling event times. The time interval until the next event depends on the total rate for all possible processes 
$
\lambda(t) = \theta + b(t) + d\,.
$
To sample an interval of time $\Delta t$ between two events from an inhomogeneous Poisson process of rate $\lambda(t)$ one can sample from a Poisson process with a rate function $\lambda^\star(t)$ fulfilling the majoration condition $\lambda^\star \geq \lambda(t)\, \forall t$ and then reject a proposed time interval $\Delta t^\star$ with a probability of
$
1-\lambda(t+\Delta t^\star)/\lambda^\star(t+\Delta t^\star)\,
$
 \cite{Lewis1979}.
The thinned set of event times follows the statistics of the Poisson process with rate $\lambda(t)$. Here, because competition is increasing with time, $\lambda(t)$ decreases monotonically. Therefore, the homogeneous Poisson process with a constant rate function $\lambda^\star(t) = \lambda(t_0)$, satisfies the majoration condition. Using this thinning technique we are able to efficiently sample the next event time while accounting for the time-dependence of the proliferation rate.

\customparagraph{Simulated cohort. }
As empirical evidence shows that the tail of the clone size distribution is almost exclusively driven by cells with memory phenotype (SI Text~\ref{phenotypes}), we focused on the clone size dynamics within the memory compartment. We assumed that the recruitment size for memory cells is independent of the prior naive cell dynamics, and we thus did not explicitly model the clone size dynamics within the naive compartment. Within the memory compartment we modeled clone size dynamics under the combined effect of early deterministic expansions during repertoire formation and fluctuating clonal growth rates according to Eq.~\ref{main-eqbirthfluctuating}. Given the large sizes of memory clones we expect demographic stochasticity to be negligible relative to clone size variability introduced by fluctuating selection. For tractability we thus ignored demographic fluctuations, which allowed us to combine the continuum solution to the deterministic clonal growth (Eq.~\ref{eqgrowthlaw}) with the stochastic propagator for the fluctuating dynamics (Eq.~\ref{eqpropagator}) to efficiently simulate the dynamics.
To study the enrichment of zero insertion clones in silico we assigned newly recruited memory clones as having zero insertions with a probability equal to the fraction $p_0(t)$ of zero insertion clones within the naive compartment. We assumed $p_0(t) = p_{0, -}$ before TdT expression turn-on at time $t^\dagger$ and $p_0(t) =  p_{0,-} t/t^\dagger + p_{0, +} (1-t/t^\dagger)$ for $t>t^\dagger$, where $t/t^\dagger$ is the fraction of naive clones produced since the switch to the adult recombination statistics.
Taken together, these simplifications lead to the following direct sampling scheme:
\begin{itemize}
    \item Sample the age $T$ of an individual uniformly from the range $[0, 80]$ years.
    \item Set the number of clones equal to $\theta T$ (rounded to the nearest integer), where $\theta$ is the rate of recruitment of new clones to the memory compartment.
    \item For each clone determine its recruitment time $t_i$ by drawing uniformly from the range $[0, T]$.
    \item Assign each clone as having zero insertions with a probability $$p_0(t) = \begin{cases} p_{0, -} & t<t^\dagger\\ p_{0,-} t/t^\dagger + p_{0, +} (1-t/t^\dagger) & \mathrm{otherwise} \end{cases}$$
        \item Sample the size $C_i(T)$ of each clone as follows (Eqs.~\ref{eqgrowthlaw} and \ref{eqpropagator}),
        \begin{equation}
            C_i = \exp(x_i), \quad x_i \sim \N\left(-d(T-t_i) + \frac{1}{1+\gamma} \log\left(\frac{e^{dT} -1}{e^{d t_i} -1}\right) - \sigma^2(T-t_i), 2\sigma^2 (T-t_i) \right),
        \end{equation}
        where $d, \gamma, \sigma^2$ are model parameters and $y \sim N(\mu, \sigma^2)$ indicates x being drawn from a normal distribution of mean $\mu$ and variance $\sigma^2$.

    \item Finally to mimick the experimental sampling depth of $N_{sample}$ reads we determine sampled clone sizes $\tilde C_i$ by Poisson sampling,
        \begin{equation}
            \tilde C_i \sim \Pois(N_{sample} \cdot C_i/N), \quad \text{with} \, N = \sum_i C_i,
        \end{equation}
        where $x \sim \Pois(\lambda)$ indicates x being drawn from a Poisson distribution of parameter $\lambda$.
\end{itemize}

\subsection{Parameter choices}
\label{si_parameters}

In the following we provide a summary of parameter choices we used to simulate repertoire dynamics along with additional motivation. 

Lifetimes of several years and several months have been measured by deuterium labelling for naive and memory T cells, respectively \cite{DeBoer2013a,Borghans2018}. 
Clonal turnover can be substantially slower than cellular turnover when proliferation balances most death (SI Text~\ref{si_neutral_timescale}). This has been shown to be the case for the maintenance of naive cells in human \cite{DenBraber2012}, where the aging-associated decline of the fraction of
T cells with T cell receptor excision circles (TRECs) suggests $\gamma \sim 0.1$. Similarly, memory T cell numbers decline much more slowly overall than suggested by the deuterium labelling literature, which is thought to be driven by homeostatic proliferation in the absence of reinfection \cite{Macallan2017}. For example, T cell memory has been observed to decline with half-lifes of $8-15$ years by following titers after small pox vaccination \cite{Hammarlund2003}. Additionally, the relatively short average lifetime of memory T cells likely masks substantial heterogeneity with a subset of more long-lived cells also contributing to the slower long-term decline of memory cells \cite{Akondy2017}.  
Another line of direct evidence for long clonal persistence has come from two studies of identical twins \cite{Pogorelyy2017,Tanno2020}, which have shown an excess sharing of identical clones decades after {\it in utero} blood exchange in monochorionic twins.

To simulate repertoire formation (Fig.~\ref{main-fig_model}B) we used the following set of parameters:
\begin{center}
\begin{tabular}{ c c c }
    parameter & explanation & value \\ \hline
    $d$ & death rate & 0.2/year  \\
    $\gamma$ & recruitment-to-proliferation ratio & 0.1  \\
    $\theta$ & recruitment rate & $10^6$/year \\
    $C_0$ & recruitment size & 1
\end{tabular}
\end{center}
We note that under mean-field competition the rate of recruitment $\theta$ only determines the overall number of clones, but does not influence the dynamics of an individual clone and thus the normalized clonal ranks. We thus used a rate smaller than suggested by estimates of thymic output, but importantly large enough to sufficiently sample from the tail of the clone size distribution. The dynamics can furthermore be non-dimensionalized by choosing units where the death rate is one. Therefore the qualitative nature of the results presented in Fig.~\ref{main-fig_model}B only depends on $\gamma$, in a way that is shown in Fig.~\ref{main-fig_model}D.

To study the enrichment of zero insertion clones in a simulated cohort (Fig.~\ref{main-figzeroinsertion}E) we used the same recruitment-to-proliferation ratio and death rate as in the previous simulation of repertoire formation. To determine the absolute number of large clones that have zero insertions in these simulations the choice of the recruitment rate $\theta$ is important. Based on order-of-magnitude estimates of the clonal diversity of the memory compartment \cite{Robins2009,Qi2014} we chose a value of $\theta = 10^5$/year. Additionally, we chose a fraction of zero insertion clones within the early naive compartment of $p_{0, -}=0.07$ (roughly equal to their overall fraction in cord blood \cite{Pogorelyy2017}) and in the late naive compartment equal to $p_{0, +} = 0.02$ (roughly equal to their overall fraction in adult blood). Finally, we used $t^\dagger = 0.05$ years for the time of the recombination switch, which together with the choice of $\theta$ produces $\unsim 10^4$ excess zero insertion clones recruited during repertoire formation in line with the enrichment data in the $<10$ years age group (Fig.~\ref{main-figzeroinsertion}B).
All parameters are summarized in the following table:
\begin{center}
\begin{tabular}{ c c c }
    parameter & explanation & value \\ \hline
    $\sigma^2$ & magnitude of clone size fluctuations & $0.08$/year \\ 
    $d$ & death rate & 0.2/year  \\
    $\gamma$ & recruitment-to-proliferation ratio & 0.1  \\
    $\theta$ & recruitment rate & $10^5$/year \\
    $p_{0,-}$ & Zero insertion fraction early in life & 0.07\\
    $p_{0,+}$ & Adult zero insertion fraction & 0.02 \\ 
    $t^\dagger$ & Time of recombination statistics switch & $0.05$ years \\
    $N_{sample}$ & simulated sample size & $5 \cdot 10^5$
\end{tabular}
\end{center}

\section{Subsampling scaling}
\label{subsampling}

Only a small fraction of the $\unsim 10^{12}$ T cells in the human body are sampled by repertoire sequencing. What effect does subsampling have on the clone size distribution?
In the following we discuss how subsampling affects the distribution of sampled clone sizes and we discuss analysis techniques for robust inferences and data visualization despite variations in sampling depth.

\subsection{Inference of scaling exponent}
Given a clone of size $C$ in the repertoire, the number of reads from that clone $\tilde C$ follows a distribution $P(\tilde C | C)$. The form of $P(\tilde C | C)$ depends on the sampling process. To build intuition let us consider the simplest case, in which every cell is sampled independently with a probability $\eta$, the subsampling fraction. Then the sampling distribution is binomial
\begin{equation}
    P(\tilde C| C) = {C \choose \tilde C} \eta^{\tilde C} (1-\eta)^{C-\tilde C}.
\end{equation}
The mean of this distribution is
\begin{equation} \label{eqsubsamplemean}
    \langle \tilde C \rangle = \eta C,
\end{equation}
which implies that sampled clone sizes are on average smaller by a factor $\eta$ than the actual clone size.
In the practically relevant limit where the sampling fraction is small, $\eta \ll 1$, we can further simplify and assume that the counts from the large clones follow a Poisson distribution.
In the Poisson limit the sampled clone size varies around its mean value with a coefficient of variation that scales as an inverse of the square root of the mean sampled count, 
\begin{equation}
    c_v = \frac{\sqrt{\langle \left(\tilde C - \langle \tilde C \rangle \right)^2 \rangle}}{\langle \tilde C \rangle}  = \frac{1}{\sqrt{\eta C}}.
\end{equation}
Importantly, the stochastic sampling introduces a subsampling scale, $\tilde C = \eta C \sim 1$, at the clone size $C = 1/\eta$, from which on average we expect a single sampled cell. Due to the existence of this scale subsampling breaks scale-invariance: even if $P(C)$ follows a perfect power law, the distribution of sampled counts
\begin{equation}
    P(\tilde C) = \sum_C P(C) P(\tilde C | C)
\end{equation}
deviates from power-law scaling close to $\tilde C = 1$. This intuition can be made rigorous using a generating function formalism \cite{Stumpf2005}: for example for $P(C) = C^{-2}/\zeta(2)$ one obtains for $\tilde C>1$ 
\begin{equation} \label{eqsubsampledscaling}
    P(\tilde C) \sim \frac{1}{\tilde C (\tilde C - 1)}.
\end{equation}
As expected the scaling with an exponent $-2$ is recovered asymptotically, but subsampling leads to a deviation from scaling when $\tilde C$ is close to 1.

\begin{figure}
 \begin{center}
     \includegraphics{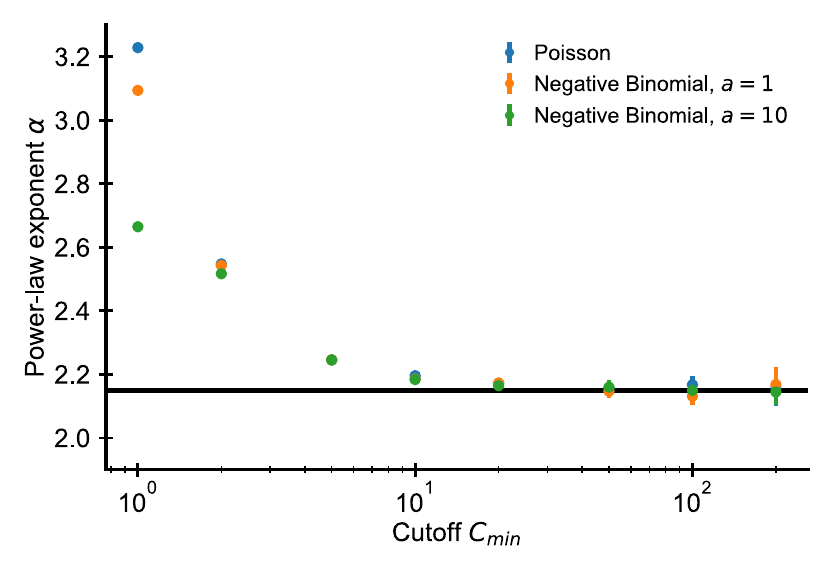}
 \end{center}
    \caption{{\bf Estimated power-law exponents converge to correct value using trimming method.} Fitted exponent as a function of the cutoff choice in simulated data (errorbars $\pm 2 \, \cdot \, \mathrm{SE}$ over 50 independent draws). The fitted exponent changes drastically for small $C_{min}$ before levelling off indicating deviations from true power-law scaling at the smallest clone sizes. Such a deviation is expected due to subsampling despite the true power-law scaling in the underlying distribution (see text). Simulations: $10^7$ clones were drawn from a discrete power-law distribution with $\alpha=2.15$. A sample of size $5\cdot10^5$ cells was then drawn from the underlying power law based on a Poisson (blue dots) or negative binomial sampling (orange and green dots show two choices of the overdispersion coefficient $a$).
  } \label{fig_powerlaw_trimming_simulations}
\end{figure}

\begin{figure}
 \begin{center}
     \includegraphics{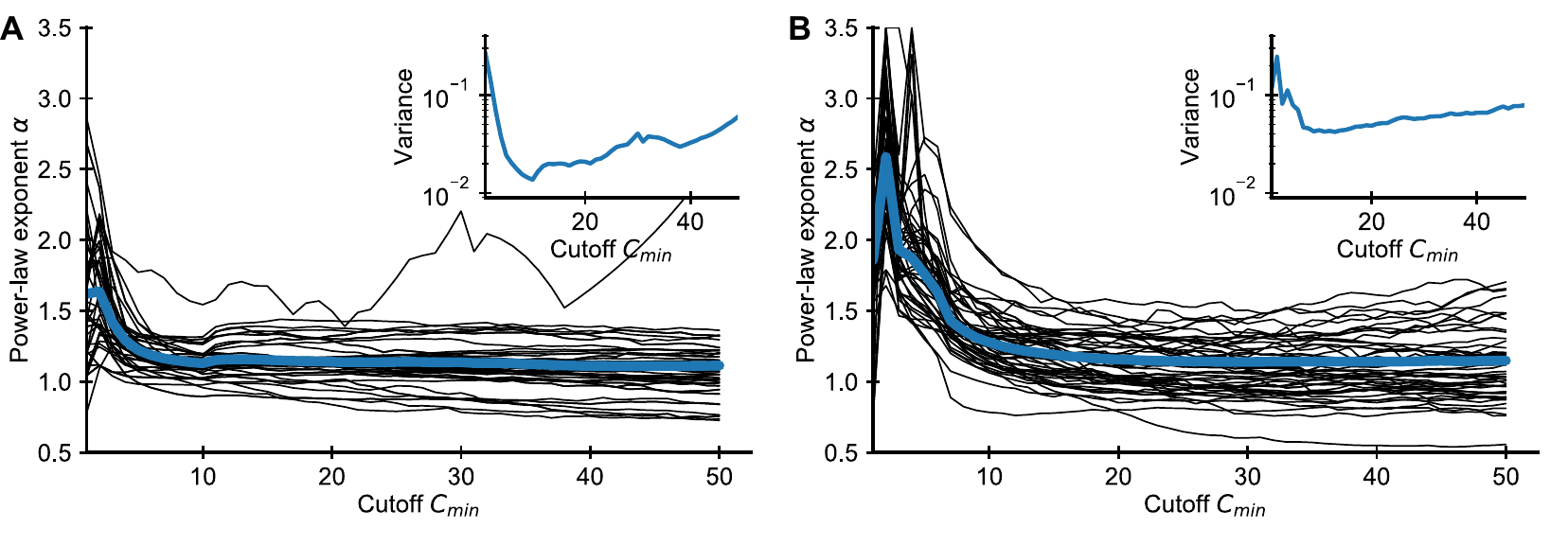}
 \end{center}
    \caption{{\bf Influence of choice of $C_{min}$ on fitted power-law exponent for empirical data.} Fitted exponent as a function of the cutoff choice (black lines: 50 random repertoires, blue line: mean) in the (A) Britanova \etal and (B) Emerson \etal datasets. The fitted exponent changes drastically for small $C_{min}$ before levelling off indicating deviations from true power-law scaling at the smallest clone sizes, similarly to those seen in simulated data (Fig.~\ref{fig_powerlaw_trimming_simulations}). To alleviate the bias induced by finite sampling we choose a cutoff value $C_{min}$, for which the power-law exponent estimates have levelled off. For large $C_{min}$ the variance of fitted exponent increases as more and more data is excluded from the fit (A, B inset), which sets a practical upper bound for choosing $C_{min}$.
  } \label{fig_powerlaw_trimming}
\end{figure}

The deviation from scaling due to subsampling leads to biases in naive estimates of the scaling exponent. How can we determine a power-law exponent in a way that is robust to subsampling? When the sampling distribution is known or can be inferred from replicate sequencing the exponent can be inferred using maximum likelihood estimation of a model with an underlying power law distribution of clone sizes convolved with the sampling probability \cite{PuelmaTouzel2019}. Here, we propose a simpler approach that does not require precise knowledge of the sampling process. We exploit the fact that the deviations from scaling vanish asymptotically for large $\tilde C$ (Eq.~\ref{eqsubsampledscaling}), by excluding small clones below some minimal size $C_{min}$ from the fitting. The power-law exponent is expected to converge as we increase $C_{min}$, which we confirm using simulated data (Fig.~\ref{fig_powerlaw_trimming_simulations}, blue line).
We can also consider more realistic models for the sampling process that account for overdispersion, i.e. their coefficient of variation exceeds the minimal value of one set by Poisson sampling. Mechanistically, such overdispersion arises for a number of reasons, most importantly because in practice we are not actually directly counting cells: in the DNA-based sequencing pipeline every cell can give rise to multiple sequencing reads due to the polymerase chain reaction amplification step, and in the mRNA-based sequencing pipeline despite the addition of unique molecular identifiers several of them can originate from different mRNA molecules from the same cell. As long as the number of reads from each cell is independently and identically distributed the law of large numbers ensures that the relative frequencies of large clones converge. We thus expect that the trimming method of fitting only to counts greater than $C_{min}$ also works for overdispersed sampling. We test the trimming method on simulated data, in which the sampling follows a negative binomial distribution with mean $\mu$ and variance $\mu + a \mu^2$ (which reduces to Poisson sampling for $a=0$). We find that trimming allows a correct estimate of $\alpha$ (Fig.~\ref{fig_powerlaw_trimming_simulations}, orange and green line).
Applying the same method to the empirical data we find that the fitted exponents also depend on $C_{min}$ (Fig.~\ref{fig_powerlaw_trimming}). In practice, we chose $C_{min} = 16$ to balance a trade-off between minimizing bias and variance, which increases as more of the data is excluded from the fit (Fig.~\ref{fig_powerlaw_trimming} insets).

\subsection{Graphical display of subsampled distributions}

\begin{figure}
 \begin{center}
     \includegraphics{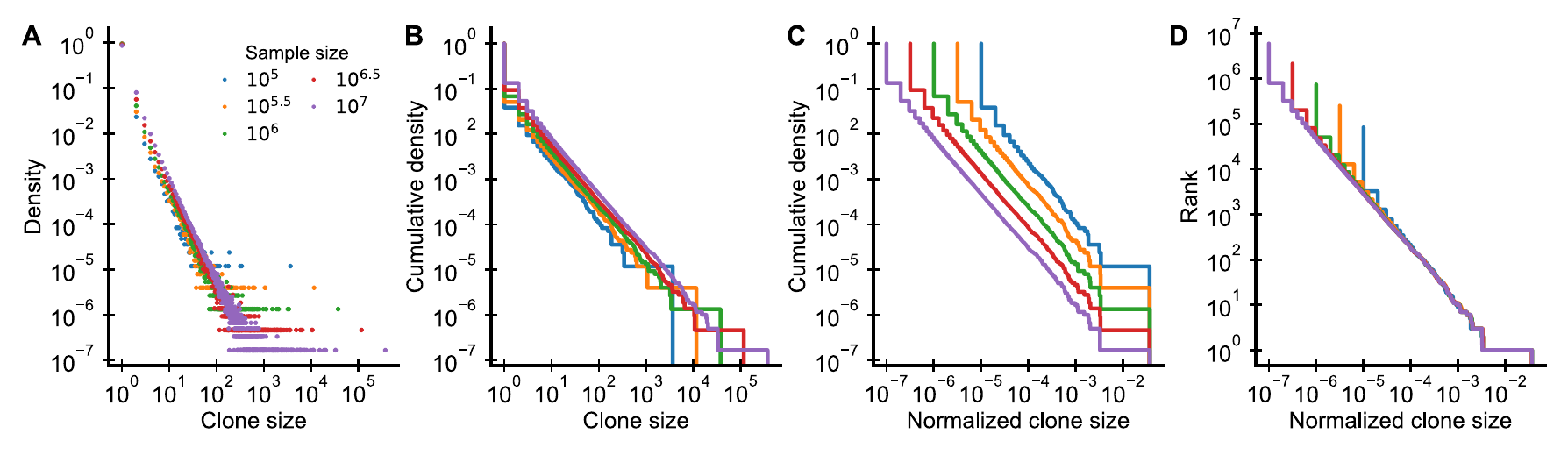}
 \end{center}
    \caption{{\bf Graphical display of subsampled power-law distributions.} (A-D) show various ways of displaying clone size distributions obtained by subsampling an underlying clone size distribution consisting of $10^8$ clones drawn according to $P(C) \sim C^{-2.2}$ to various sampling depths. (A) The empirical probability density function of clone sizes, (B) its cumulative density, as well as (C) the cumulative density of normalized clone sizes are not invariant under changes of the sampling depth. Only the tail behavior of relative frequencies of finding cells from large clones is reproducibly captured, which makes rank-frequency plots (displays of unnormalized cumulative distributions of normalized clone sizes) the method of choice for collapsing clone size distributions at various sampling depths.
  } \label{fig_subsampling}
\end{figure}

The intuition we have built about how subsampling affects clone size distributions can help us choose an appropriate method for displaying subsampled data (Fig.~\ref{fig_subsampling}). Which graphical representation of the clone size distribution minimizes the influence of variations in sampling depth?

The shift of the mean clone size (Eq.~\ref{eqsubsamplemean}) suggests that we should normalize sampled clone sizes by the sampling fraction $\eta$, as has been noted elsewhere \cite{Levina2017}. While experimentally we do not know the sampling fraction, we can instead simply divide the clone sizes by the total sample size (Fig.~\ref{fig_subsampling}C,D). This normalization is particularly intuitive as it corresponds to using the relative frequencies of cells in different clones. While the absolute number of cells in a large clone increases with more sampling, the fraction of all sampled cells that are part of a particular clone remains constant on average.

Plots of the cumulative distribution of clone sizes make it easier to visually assess the tail behavior of the distribution (Fig.~\ref{fig_subsampling}B) than plots of the probability density (Fig.~\ref{fig_subsampling}A). However, even after normalizing clone sizes by the sample size there remains a very visible shift between the cumulative distributions at different sampling depths (Fig.~\ref{fig_subsampling}C). This shift arises because the implicit normalization by the total number of unique clones makes the sampled cumulative distribution depend heavily on sampling depth. As sampling increases so does the total number of unique clones that will be sequenced. This suggests that we might do better by simply omitting the normalization. Ranking clones by their normalized size yields precisely such an unnormalized cumulative distribution. Taken together, by both scaling clone sizes by the sample size and resisting the temptation to normalize the ranks, we can collapse distributions sampled at different depths (Fig.~\ref{fig_subsampling}D). 

\begin{figure}
 \begin{center}
     \includegraphics{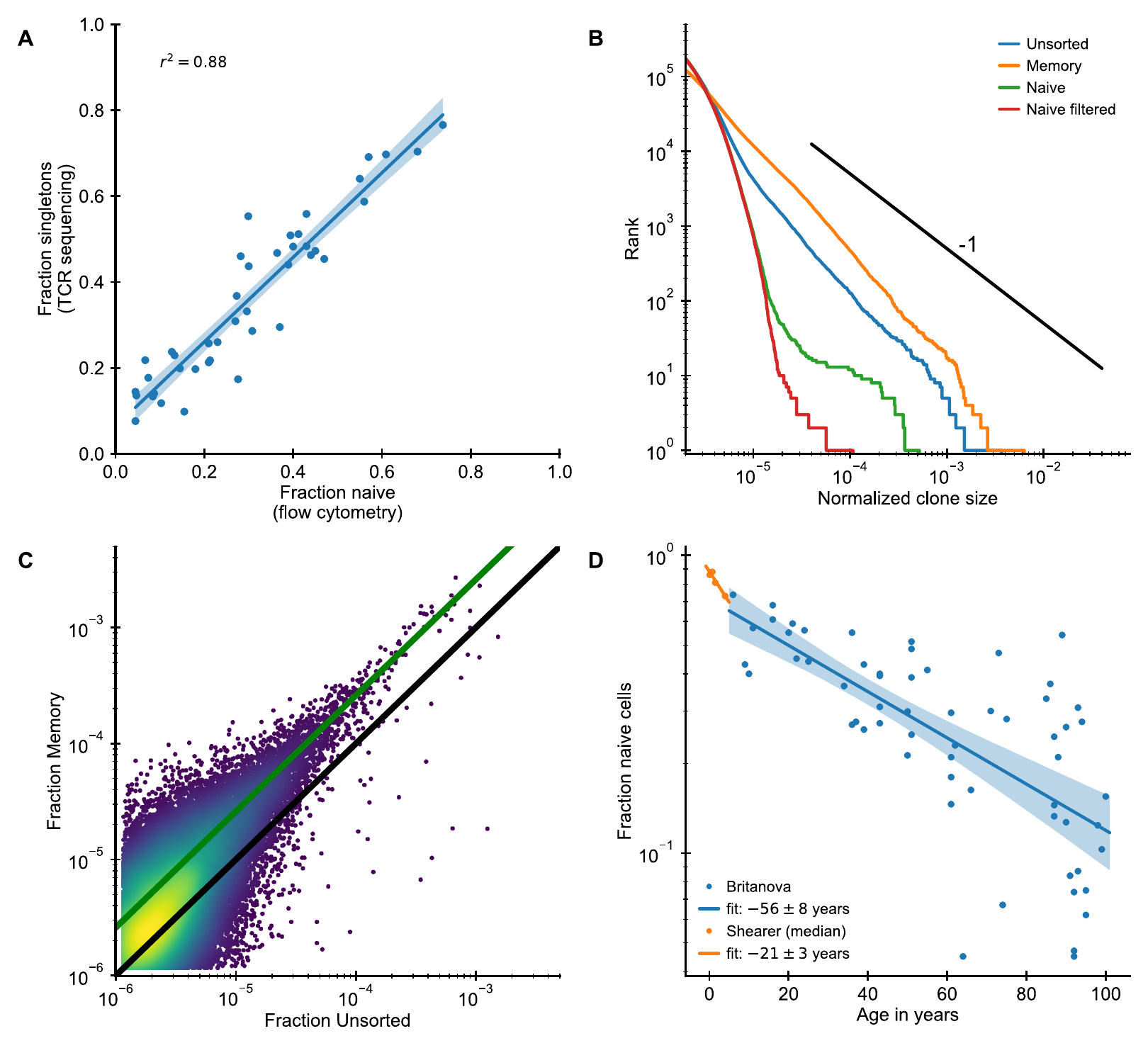}
 \end{center}
    \caption{{\bf The large clones in unsorted peripheral blood are predominantly of memory phenotype.}
    (A) The naive cell fraction as determined by flow cytometry and the fraction of singletons are closely correlated in the Britanova cohort. To diminish the influence of sampling depth variations we computationally subsampled all repertoires to an equal sample size of $5\cdot10^5$ counts.
    (B,C) Analysis of unsorted (TCR sequencing from all peripheral blood mononuclear cells), memory (CD3$^+$, CD45RO$^+$), and naive (CD3$^+$, CD45RA$^+$) blood samples from the same individual (Data source: \cite{Chu2019}). (A) Clone size distributions in the different T cell compartments. Filtering naive clones that are also found in the memory compartment removes most large naive clones. (B) Frequency of large clones in the memory sample is shifted upwards relative to their frequency within the unsorted sample. Color represents logarithm of local kernel density estimate in regions with overplotting. The solid lines are guides to the eye (black line represents equal frequency, green line 2.6-fold higher frequency in the memory compartment).
    (D) Fraction of naive cells decreases with age (Data source: \cite{Britanova2016}) starting in early infancy (Data source: \cite{Shearer2003}). Legend shows fitted time constant of exponential decay ($\pm$ SE).
    }
    \label{fig_naive}
\end{figure}

\section{Relation between clone size and cellular phenotypes}
\label{phenotypes}

In both cohorts all T cells from peripheral blood were sequenced irrespective of their phenotypes.
Antigenic challenges drive large clonal expansions and we thus expect clones with effector or memory cells to be larger than naive clones all else being equal \cite{Farber2014,Mayer2019b}. This has generally been confirmed by TCR repertoire sequencing studies \cite{Oakes2017}, but there have also been some reports \cite{Qi2014,Pogorelyy2017} of expanded naive clones with similar sizes to the largest memory clones. Given this unclear picture from the literature we analyzed the relative contribution of naive and memory cells to clones of different sizes.

Overall, we might expect that naive clones dominate the clone size distribution at the smallest sizes. To test this idea we compared sequencing and flow cytometry data from the Britanova cohort and founnd that the fraction of naive cells in different individuals explains a remarkably high $88\%$ of variability in the number of clones sequenced only once after subsampling all repertoires to the same size (Fig.~\ref{fig_naive}A). To further determine how cells from clones of different sizes partition phenotypically we analyzed data from a study in which T cells were sequenced both in unsorted blood as well as after sorting into naive and memory cells \cite{Chu2019}. We find that the sizes of large clones follow the same scaling in unsorted blood and in the memory compartment (Fig.~\ref{fig_naive}B). Within the naive compartment most clones are small, in particular when excluding clones from which cells are also found in the memory compartment (Fig.~\ref{fig_naive}B, red line). We note from the plot that all of the largest 200 clones in unsorted blood have memory phenotype cells, and less than one percent of the top 1000 clones are not found within the memory compartment. This rules out that the enrichment of zero insertion clones among the most abundant clones found in Fig.~\ref{main-figzeroinsertion} is driven by naive clones as has been suggested in a previous study \cite{Pogorelyy2017}. The relative frequency of a clone within the memory compartment is larger by a constant fold-factor (Fig.~\ref{fig_naive}C), likely reflecting an increased relative frequency of the large clones when excluding naive cells from the denominator.

To correct for the decrease of naive cells with age (Fig.~\ref{fig_naive}D) \cite{Shearer2003,Britanova2016} we normalize clonal frequencies in unsorted peripheral blood by the mean fraction of memory cells expected at different ages fit to the flow cytometry data. We find that this normalization collapses the tails of empirical clone size distributions (Fig.~\ref{fig_clonesizes_stepbystep}C,F).

\section{Modeling neutral repertoire dynamics}
\label{si_neutral}

In the following we review results on the neutral dynamics of clone sizes in which the continuous recruitment of new clones is balanced by a net negative growth of already established clones $b<d$. These models have a long history in ecology \cite{Volkov2003}, and have also been proposed as null models in the context of T cell dynamics previously \cite{Desponds2016,Desponds2017,Greef2020}. While the results can be found in the literature their inclusion serves to introduce a parametrization highlighting the recruitment-to-proliferation ratio $\gamma$ as a key quantity governing clonal dynamics.

\subsection{Steady state clone size distribution}
At steady state the probability distribution $P(C)$ of clone sizes $C$ needs to fulfill the balance condition 
\begin{equation}
    b C P(C) = d (C+1) P(C+1),
\end{equation}
for all $C>C_0$ which yields
\begin{equation} \label{eqneutral_db}
    P(C) \propto \frac{1}{C} \left(\frac{b}{d}\right)^C = \frac{1}{C} \exp \left(-C \log (d/b)  \right).
\end{equation}
This distribution is characterized by power-law scaling with an exponent of $1$ for small clone sizes, and, importantly, has an exponential cutoff at $C^\star = 1/\log(d/b)$. 
In contradiction with this model experiments point towards power-law scaling with an exponent $\unsim 2$ (Note that $P(C) \sim C^{-\alpha-1}$ when $\mathrm{rank} \sim C^{-\alpha}$). Additionally the size of large clones seen experimentally is incompatible with the predicted exponential cutoff as we discuss below. 

The total repertoire size follows the following continuum equation
\begin{equation}
    \frac{\ud N}{\ud t} =   (b - d) N + \theta C_0,
\end{equation}
such that at steady-state, $\frac{\ud N}{\ud t} = 0$, the repertoire has a total size 
\begin{equation}
    N_\infty  = \frac{\theta C_0}{d-b}.
\end{equation}
For a more interpretable alternative parametrization we introduce the recruitment-to-proliferation ratio for the maintenance of cells at steady state
\begin{equation} \label{eqgammageneral}
    \gamma = \frac{\theta C_0}{b N_\infty} = \frac{d}{b} - 1.
\end{equation}
Using this relation to rewrite Eq.~\ref{eqneutral_db} we obtain
\begin{equation} \label{eqneutral}
    P(C) \propto \frac{1}{C} \exp \left(-C \log (1+\gamma)  \right),
\end{equation}
implying a cutoff clone size of $C^\star = 1/\log(1+\gamma)$.
The largest clones represent on the order of one percent of the repertoire, which assuming independent sampling from the underlying repertoire would correspond to $\unsim 10^{10}$ cells in the complete repertoire. For small $\gamma$ we can expand $C^\star \approx 1/\gamma$, so in order to have a cutoff clone size $C^\star$ of this order of magnitude one would need to have an unreasonably small $\gamma \sim 10^{-10}$.

\subsection{Relaxation time scale}
\label{si_neutral_timescale}
Over what timescale do transiently expanded clones disappear? The time scale $\tau_c = \frac{1}{d-b}$ for deterministic clonal decay can be much larger than the lifespan $1/d$ of a single cell when birth and death are closely balanced. Rewriting the birth rate in terms of $\gamma$ and $d$ we obtain 
\begin{equation} \label{eqtauc}
    \tau_c = \frac{1+\gamma}{d\, \gamma},
\end{equation}
demonstrating that for $\gamma \ll 1$ clonal dynamics is a factor of $1/\gamma$ slower than cellular dynamics.

\section{Modeling repertoire formation}

\subsection{Mechanistic motivation for the competition function}
\label{si_1oNmechanism}

We consider a population of $N$ T cells that proliferate at a rate proportional to the concentration $S$ of a set of stimuli (stimulatory cytokines), $b \propto S$. We assume that the cytokines are produced by other cells at some fixed rate $p$ and degraded at a basal rate $q$. We further assume that competition between T cells is mediated by their consumption of cytokines. The dynamics of $S$ is then described by
\begin{equation}
    \frac{\ud S}{\ud t} = p - q S - k S N,
\end{equation}
where $-k S N$ is a mass action term describing how T cells lower cytokine levels.
Assuming a separation of timescales in which cytokine concentrations change quickly we obtain the quasi steady state approximation
\begin{equation}
    S = \frac{p}{q+k N}.
\end{equation}
When the consumption term dominates relative to basal decay, $kN \gg q$, we obtain $b \propto S \propto 1/N$.

\begin{figure}
 \begin{center}
     \includegraphics{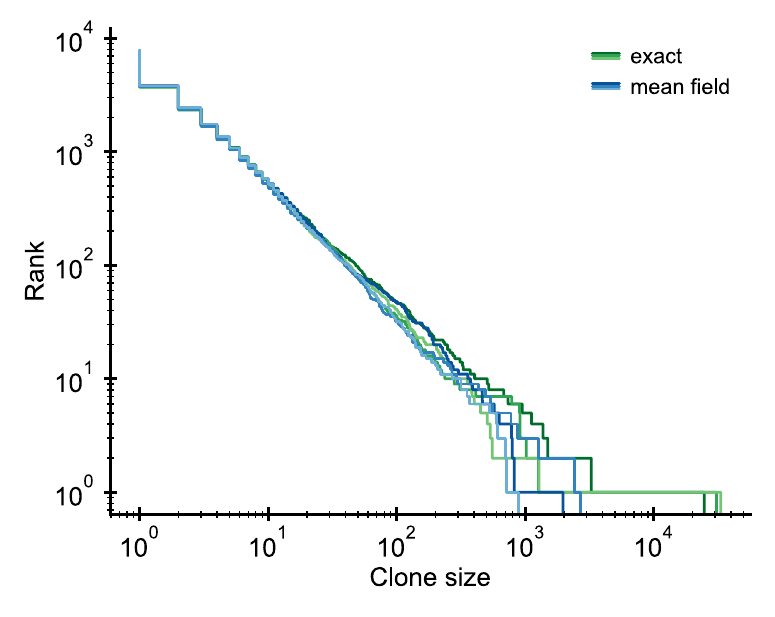}
 \end{center}
    \caption{{\bf Validation of the mean-field approximation.} Comparison of full stochastic simulations and simulations using mean-field competition.
    Parameter: $b_0 = 2 \cdot 10^4$/year, $d=0.2$/year, $\theta = 2 \cdot 10^3$/year (implying $\gamma$ = 0.1), simulation length 5 years.
    }
\label{fig_meanfield}
\end{figure}

\subsection{Mean-field competition approximation}
\label{si_meanfield}
We simplify the full stochastic model (Eqs.~\ref{main-eqbirth}-\ref{main-eqimmigration}) using a mean-field approximation for the competition, which decouples the dynamics of individual clones while retaining the full stochasticity on the clonal level. This approximation replaces the dependence of the proliferation rate on $N$ by a dependence on its continuum theory average given by Eq.~\ref{eqtotalpop}.
We exactly simulated a system of reduced size to validate the mean-field approximation (see Sec.~\ref{secsimulations}). The distributions of the exact and mean-field simulations agree to within stochasticity (Fig.~\ref{fig_meanfield}), with the exception of the largest clone, which is larger in the exact simulations as has been discussed elsewhere \cite{Dodds2017}.

\subsection{Continuum theory of clonal growth}
\label{si_continuum_theory}

To obtain insight into why the model produces power-law scaling we present a simple continuum theory of early clonal dynamics. We approximate the clone size dynamics of the $i$-th clone $C_i$ as
\begin{equation} \label{eqearlydynamics}
    \frac{\ud C_i}{\ud t} = \left(\frac{b_0}{N(t)} -d\right) C_i,
\end{equation}
with $C_i(t_i) = C_0$ at the time of recruitment $t_i$.
The total repertoire size $N = \sum_i C_i$ evolves according to
\begin{equation}
\frac{\ud N}{\ud t} =   b_0 - d  N  +\theta C_0,
\end{equation}
whose solution is given by
\begin{equation}
\label{eqtotalpop}
    N(t) = (b_0+\theta C_0)\left(1-e^{-d \, t}\right)/d.
\end{equation}
For times large compared to $1/d$ the total repertoire size given in Eq.~\ref{eqtotalpop} reaches a steady-state,
\begin{equation}
N_\infty = (b_0 + \theta C_0)/d,
\end{equation}
because competition for proliferation signals acts as a homeostatic regulator. 
By combining Eq.~\ref{eqtotalpop} and Eq.~\ref{eqearlydynamics} we derive the clonal growth law
\begin{equation} \label{eqgrowthlaw}
    C_i(t) = C_0 \left(\frac{e^{d t}-1}{e^{d t_i} -1}\right)^{1/(1+\gamma)} e^{-d(t-t_i)},
\end{equation}
where $\gamma$ as in SI Text~\ref{si_neutral} is the recruitment-to-proliferation ratio which in this model is given by $\gamma = \theta C_0/b_0$.
To simplify we expand the growth law at leading order for small times, $t_i < t \ll 1/d$, to obtain
\begin{equation} \label{eqctappendix}
    C_i(t) = C_0 \left(\frac{t}{t_i}\right)^{1/(1+\gamma)}.
\end{equation}
This expression can also be derived directly by noting that early repertoire growth is linear $N(t) \approx (b_0 + \theta C_0) t$, and that the early dynamics is dominated by proliferation and not death such that 
\begin{equation}  \label{eqdyncloneearly}
    \frac{\ud C_i}{\ud t} = \frac{1}{(1+\gamma)t} C_i,
\end{equation}
which is solved by Eq.~\ref{eqctappendix}.
Given the constant recruitment of new clones the distribution of the $t_i$'s is uniform, which with Eq.~\ref{eqctappendix} implies a clone size distribution
\begin{equation} \label{eqmfpowerlaw}
    P(C) = P(t_i(C)) \left| \frac{\ud t_i}{\ud C} \right| \propto C^{-2 - \gamma}
\end{equation}
that follows power-law scaling with an adjustable exponent that depends on $\gamma$.
Note that the exponent for $P(C)$ differs by one from the exponent for the rank \cite{Clauset2009}, which is a complementary cumulative distribution, and thus $\alpha = 1+\gamma$.

\subsection{Steady-state distribution}
\label{si_steadystate_repertoireformation}
To derive the power-law scaling we have expanded the total repertoire size for small times (or death rates). How does the clone size distribution change later in life? At large times the division rate $b_0/N(t)$ falls below the constant death rate $d$ as the steady-state repertoire size $N_\infty$ is approached following Eq.~\ref{eqtotalpop}. In this model this happens at a time $t^\star \simeq \log (1+1/\gamma)/d$, after which the large clones experience a deterministic force towards extinction. 
For times $t \gg t^\star$ the model effectively reduces to the neutral birth-death dynamics considered in SI Text~\ref{si_neutral}. (The growth rate fluctuations produced by variations of the total population size around steady state asymptotically vanish for large $N_\infty$.) We thus expect the steady-state clone size distribution to be equivalent to that of the neutral model (Eq.~\ref{eqneutral}).
Indeed this distribution accurately describes the distribution of small clones in old age (Fig.~\ref{main-fig_model}B). The neutral distribution is not compatible with data as discussed before. However, the timescale over which large early founded clones vanish is long (SI Text.~\ref{si_neutral_timescale}) such that a tail of large clones resulting from the early growth dynamics can be maintained much beyond $t^\star$ until $t \gg \tau_c$.

\subsection{Relaxations of model assumptions}
\label{secmodelrelaxations}

For tractability and interpretability we have kept the model presented in the main text deliberately simple. Here, we explore how a saturation of the proliferation rate, competition for specific resources, or variations in the recruitment size modify clone size distributions.

\begin{figure}
 \begin{center}
     \includegraphics{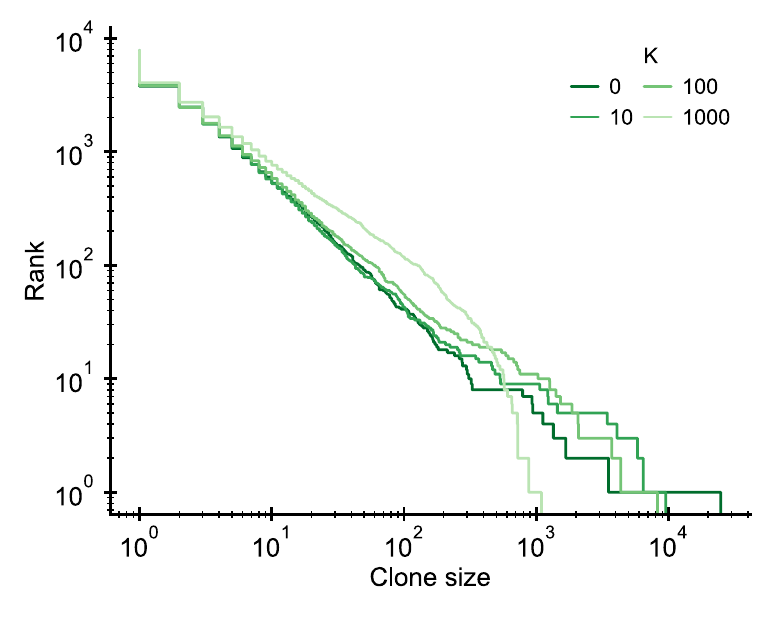}
 \end{center}
    \caption{{\bf Saturation of proliferation rate.} Influence of a saturation of the proliferation rate, $b = b_0/(K+N)$, on the clone size distribution. The saturation induces a change of the scaling behavior at the largest clone sizes.
    Parameter: $b_0 = 2 \cdot 10^4$/year, $d=0.2$/year, $\theta = 2 \cdot 10^3$/year (implying $\gamma$ = 0.1), simulation length 5 years.
    }
\label{fig_relaxations}
\end{figure}

\customparagraph{Saturation of proliferation rate.}
Cellular growth is not arbitrarily fast, which is not accounted for in the simple model in which cells proliferate very rapidly early in life. To understand how such a saturation effect influences clone size distributions we introduce an upper limit on birth rate that limits proliferation in the absence of competition. Following \cite{DeBoer1995} we set the clonal birth rate to $b(t) = b_0 / (K + N)$ for some constant $K$, which sets the repertoire size below which competition is negligible. Given this choice the birth rate remains limited to a value $b_0/K$ even in the absence of any competitors.
Increasing $K$ leads to deviation in the scaling of the largest clones (Fig.~\ref{fig_relaxations}), but the same scaling remains at intermediate clone sizes.
In the model early clonal growth is exponential until the total repertoire has reached size $N(t) \sim K$, which explains the different distribution of the largest clones. However, the number of clones that are recruited during this phase grows only logarithmically with $K$ due to the exponential increase in total repertoire size.

\begin{figure}
 \begin{center}
     \includegraphics{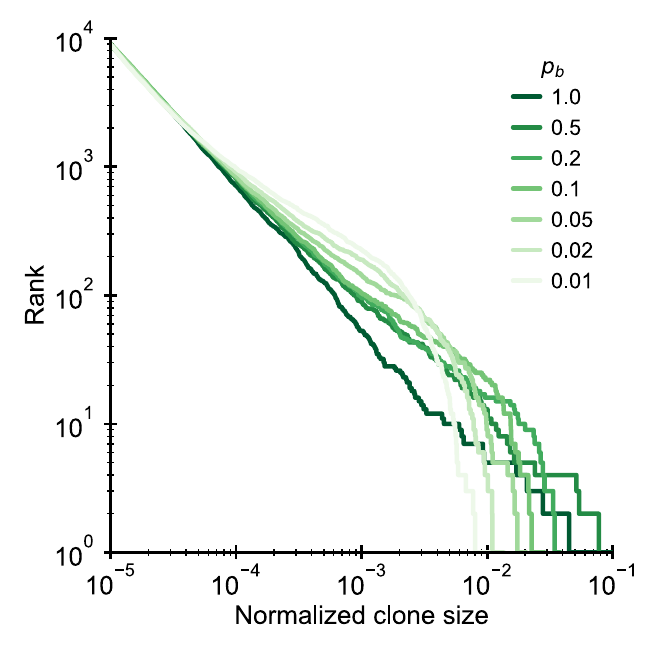}
 \end{center}
    \caption{{\bf Competition for specific resources.} Clone size distributions in a simulated model where clones compete for specific antigens to which they bind with a probability $p_b$.
    Parameter: $b_0 = 10^4$/year, $\theta = 10^3$/year (implying $\gamma$ = 0.1), $N_a=1000$, $d=0$, simulation length $10$ years.
  } \label{figspecific}
\end{figure}

\customparagraph{Competition for specific resources.}
T cells respond to stimuli from peptide-MHC complexes, which could also act as limiting resources. T cells then compete only with those cells specific to the same antigens in contrast to the global competition considered previously. To assess how assumptions about the mechanisms of competition influence our results we simulated the repertoire formation process using a classical description of competition for antigens \cite{DeBoer1994,DeBoer2001,Mayer2015}.
We consider a fixed number of antigens $N_a$ and encode the specificity of the $M$ clones in a matrix $K$ of size $M \times N_a$, where $K_{ij} = 1$ if clone $i$ recognizes the antigen $j$ and $K_{ij} = 0$ otherwise.
We draw the entries of $K$ independently with a fixed binding probability $p_b$.
We assume that the proliferation rate of a cell of the $i$-th clone is proportional to the amount of antigenic stimulation:
\begin{equation} \label{eqbi}
    b_i = \frac{b_0}{N_a} \sum_{j=1}^{N_a} K_{ij} F_j
\end{equation}
where the availability of antigen $j$ is given by
\begin{equation}
    F_j= \frac{1}{1+\sum_i K_{ij} C_i}.
\end{equation}
The normalization of Eq.~\ref{eqbi} ensures that total proliferation is comparable to a global resource model with the same parameters independent of $N_a$.
For computational tractability we simulated the clone size dynamics without taking into account demographic stochasticity in proliferation and death of cells. While more specific competition (smaller $p_b$) leads to a deviation in the distribution of the largest clones, we find that clone size distributions are heavy tailed independently of the choice of $p_b$ and all display the same scaling at intermediate clone sizes (Fig.~\ref{figspecific}). 

\begin{figure}
 \begin{center}
     \includegraphics{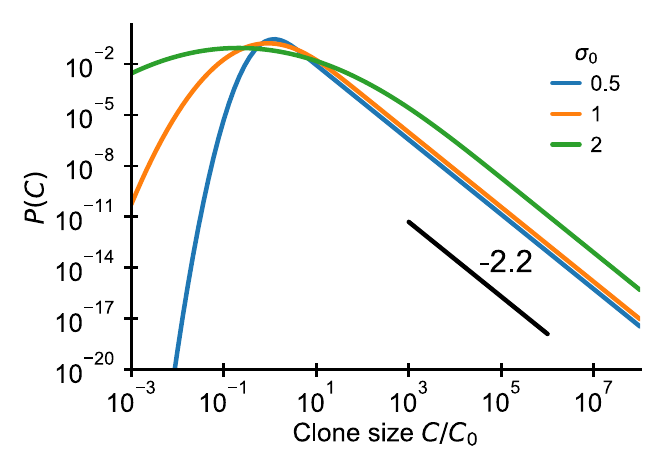}
 \end{center}
    \caption{{\bf Variation of recruitment size.} Clone size distributions resulting from a variable recruitment size and repertoire growth (Eq.~\ref{eqrecruitmentsizevariation}). The black line shows a power law with a slope of -2.2 for visual comparison.
    Parameter: $\gamma = 0.2$
  } \label{fig_introsizevar}
\end{figure}

\customparagraph{Variations of the recruitment size.}
The numbers of cells $C_0$ that are recruited might also be variable. In particular, this will be the case when we think about the memory compartment, in which $C_0$ represents the number of cells from a clone recruited into memory following infection. To understand how such variations modify the dynamics of repertoire formation we derive an analytical prediction in the case where the distribution of recruitment sizes, $P(C_0)$, is lognormal. Given a lognormal distribution with parameters $\mu_0$ and $\sigma_0$ the mean introduction size is given by $\overline{C_0}= e^{\mu_0+\sigma_0^2/2}$. To keep the mean introduction size constant while changing the variability of clone sizes, we use a parametrization in terms of $\overline{C_0}$ and $\sigma_0$ and set $\mu_0 = \log(\overline{C_0}) -\sigma_0^2/2$. To determine the clone size distribution resulting from early repertoire growth we integrate the continuum theory prediction, $P(C/C_0)\propto (C/C_0)^{-2-\gamma}$ over the distribution of $C_0$:
\begin{align} \label{eqrecruitmentsizevariation}
    P(C) & \propto \int_0^C dC_0 \left( C/ \overline{C_0}\right)^{-2-\gamma} \frac{1}{\sigma_0 C_0/\overline{C_0} \sqrt{2 \pi}} e^{-(\log \left(C_0/\overline{C_0}\right) + \sigma_0^2/2)^2/(2\sigma_0^2)} \nonumber \\
    & = \left(C/\overline{C_0}\right)^{-2-\gamma} \cdot e^{\frac{1}{2} (\gamma+2) (\gamma+1) \sigma_0^2} \frac{1}{2} \erfc\left(\frac{-2\log (C/\overline{C_0})+(3+2\gamma) \sigma_0^2}{2\sqrt{2} \sigma_0}\right).
\end{align}
The complementary error function $\erfc(x)$ saturates for $x \ll -1$ and thus the distribution follows the same power-law $P(C) \sim C^{-2-\gamma}$ for large clones, $\log C/\overline{C_0} \gg \sigma_0 \left(\sqrt{2} + \frac{3+2\gamma}{2}\sigma_0\right)$, while it deviates for smaller clones within the range of recruitment sizes (Fig.~\ref{fig_introsizevar}).

\subsection{Relation to mechanisms generating power laws in other growth processes}
\label{si_preferentialattachment}

The origin of power law scaling during repertoire formation is reminiscent of a class of stochastic processes widely studied in the literature as a mechanism underlying power-law distributions found in diverse contexts \cite{Yule1924,Luria1943,Barabasi1999}, which has been rediscovered multiple times since the pioneering work of Yule on speciation \cite{Yule1924}.
Common to these processes is that the distribution of types at a given point is the result of a balance between the growth of existing types and the addition of new types.
The different models depending on their context differ in (i) the growth rate $r(t)$ of the number of units of each already existing type and (ii) the rate function $\theta(t)$ at which new types are introduced. They all share the same basic mathematical mechanism that produces a power law distribution of types as we review below. The three maybe most well-known instances of this class of processes are the following:
\begin{itemize}
    \item the Yule model of speciation~\cite{Yule1924}, in which (i) species within a genus speciate at some constant rate, and (ii) new genus is created at a rate proportional to the number of already existing genera.
\item the Luria-Delbrück model of bacterial population genetics during exponential growth~\cite{Luria1943}
       in which
        (i) each cell divides at a constant rate, and (ii) new alleles arise through random mutation at a constant rate per cell division.
\item the Barabási-Albert (BA) model of network growth~\cite{Barabasi1999}, 
    in which at every time step (ii) a new node is added, and is (i) linked to $m$ already existing nodes with a probability proportional to the number links that a chosen node already has.
\end{itemize}

Despite differing in their assumptions about the functional form of the growth and innovation rates we show in the following that these different models all share a common mathematical basis. 
To provide a common terminology we will use the language of urn models and refer to different types as urns and to the different number of units of each types as balls in each urn. In an attempt to unify the different models we develop a continuum theory for these growth-innovation processes. To do so we rescale time to
\begin{equation}
    \tau = \int_{0}^{t} \theta(t') \ud t',
\end{equation}
such that new urns are added at unit rate, $\theta(\tau) = 1$.
The number of balls in each urn then grows according to
\begin{equation}
    \frac{\ud C_i}{\ud \tau} = \frac{\ud C_i}{\ud t} \frac{\ud t}{\ud \tau}
    = \frac{r(t(\tau))}{\theta(t(\tau))} C_i =: \zeta(\tau)C_i.
\end{equation}
The key to the power-law scaling in all these models is the existence of a regime in which
\begin{equation} \label{eqzeta}
    \zeta(\tau) = \frac{1}{\alpha \tau},
\end{equation}
i.e. the growth rate scales inversely with rescaled time with a proportionality factor $1/\alpha$.
Eq.~\ref{eqzeta} has the same form as Eq.~\ref{eqdyncloneearly} that we derived for our model of repertoire formation. Thus following the derivation of Eq.~\ref{eqmfpowerlaw} within SI Text~\ref{si_continuum_theory} we obtain a subexponential growth of balls in already existing urns, which leads to a power law scaling of the distribution of balls per urn,
\begin{equation} 
    P(C) \propto C^{-\alpha-1},
\end{equation}
with an adjustable exponent that depends on $\alpha$.

Before deriving how Eq.~\ref{eqzeta} arises in specific contexts let us first remark on a general consequence of this form of effective growth law: The total number of balls added to all existing urns per rescaled time unit is constant.
To derive this let us assume that each new urn is populated by $C_0$ balls, then the total number of balls $N(t) = \sum_i C_i$ grows according to
\begin{equation}
    \frac{\ud N}{\ud \tau} = \frac{1}{\alpha\tau} N + C_0,
\end{equation}
which is solved by
\begin{equation}
    N(\tau) = \frac{C_0 \alpha}{\alpha-1} \tau.
\end{equation}
Multiplying by the growth rate Eq.~\ref{eqzeta} yields a constant,
\begin{equation}
    N(\tau) \zeta(\tau) = \frac{C_0}{\alpha-1},
\end{equation}
thus showing the equivalency between the assumed growth rate depency on rescaled time and the constancy of how many balls are added per rescaled time unit.

In the Yule process we have $r(t) = r$ and recruitment is proportional to the number of genera, $\theta(t) = s G(t)$, which grow at a (generally different) rate $s$, $G(t) = G_0 e^{st}$. By integration we obtain $\tau(t) = G_0 \left(e^{st} -1 \right)$, which leads to $\zeta(\tau) = \frac{r}{s\tau + s G_0}$. Thus $\zeta(\tau) \approx \frac{r}{s \tau}$ when the number of newly created genera exceeds the initial number $\tau \gg G_0$. The exponent of the power-law is determined by the ratio of the growth of genera and species, $\alpha = s/r$. 

In the Luria-Delbrück model we have $r(t) = r$; recruitment is proportional to the total population size $\theta(t) = \mu r N(t)$, where $\mu$ is the mutation probability per replication and where $N(t) = N_0 e^{r t}$. By integration we obtain $\tau(t) = \mu N_0 \left(e^{rt} - 1\right)$, which leads to $\zeta(\tau) = \frac{1}{\tau + \mu N_0}$. Thus $\zeta(\tau) = \frac{1}{\tau}$ when $\tau \gg \mu N_0$. In contrast to Yule's model the power-law exponent is fixed at $\alpha = 1$, because the same growth process governs the increase in $\theta(t)$ and in cell numbers. 

In the Barabasi-Albert model the introduction rate $\theta(t) = 1$ is constant, but $r(t)$ decreases with time. The $m$ newly added links attach preferentially to those nodes that already have a large degree. The growth rate $r(t) = m/N(t)$ of a node thus decreases proportionally to the total degree $N = 2mt$ of all present nodes. We have $r(t) = 1/(2t)$, which implies $\zeta(\tau) = \frac{1}{2 \tau}$ and $\alpha=2$.

\section{Modeling long-term repertoire dynamics with fluctuating clonal growth rates}
\label{secffmodeldynamics}

\begin{figure}
 \begin{center}
     \includegraphics{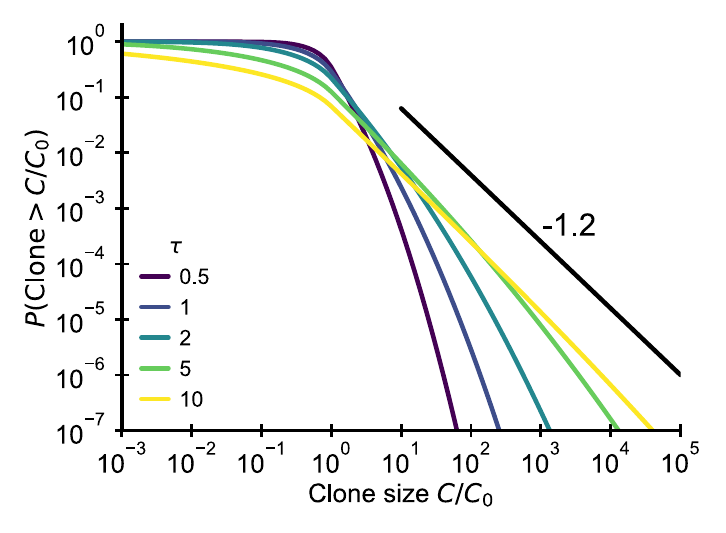}
 \end{center}
    \caption{{\bf Fluctuating fitness model out-of-steady state.} Analytical predictions for the clone size distributions in a geometric Brownian motion fluctuating fitness model (Integral of Eq.~\ref{eqrescaled}) as a function of effective age $\tau = T\sigma^2$. The black line shows the asymptotic prediction for the steady-state scaling. Parameter: $\alpha = 1.2$
  } \label{figfluct}
\end{figure}

\subsection{Slow convergence to steady-state scaling}
\label{si_ffconvergence}
Multiplicative stochastic processes are a classical generative mechanisms for heavy-tailed distributions \cite{Sornette1997,Gabaix1999,Newman2005}. In the context of lymphocyte dynamics this mechanism has first been proposed by Desponds \etal \cite{Desponds2016}, who argued that fluctuations in antigen availability can lead to multiplicative stochastic dynamics producing power-law scaling at steady state. Here, we expand on this ealier work by analyzing a simple fluctuating fitness model out-of-steady-state. Our analytical results show that the emergence of scaling can be slow when the fluctuation amplitude is small.

We opted to treat proliferation rate fluctuations as temporally uncorrelated for computational tractability (Eq.~\ref{main-eqbirthfluctuating}). Correlations in proliferation rate fluctuations are clearly an important feature of short term dynamics -- e.g. to describe the quick expansion and contraction during and following acute infection over a timescales of days and weeks, respectively \cite{Mayer2019b}. However, given finite correlation times we expect to be able to capture dynamics over the long timescales which we are interested in here, with uncorrelated noise with an effective net fluctuation strength that averages over the short-term dynamics.

In this limit clone sizes follow a geometric Brownian motion, i.e. $x = \log C/C_0$ follows the Langevin equation
\begin{equation} \label{eqlogdiffusion}
    \frac{\ud x_i}{\ud t} = f_0 + \sqrt{2} \sigma \eta_i,
\end{equation}
with initial condition $x(t_i) = 0$, where $\sigma$ sets the fluctuation strength and where $\langle \eta_i(t) \eta_j(t') \rangle = \delta_{ij} \delta(t-t')$.
A negative mean fitness $f_0<0$ balances the recruitment of new clones and the net expansion induced by the fluctuating term. 
In general, we might want to include also demographic noise and the extinction of clones as an absorbing boundary condition \cite{Desponds2016}, but here for simplicity we will neglect those effects.
Eq.~\ref{eqlogdiffusion} is a diffusion equation for the logarithmic clone size $x$ and has the well-known Green's function
\begin{equation} \label{eqpropagator}
    G(x, y, t) = \frac{1}{\sqrt{4 \pi \sigma^2 t}} e^{-\frac{(x-y-f_0 t)^2}{4\sigma^2 t}},
\end{equation}
which describes how the distribution spreads out from an initial $\delta$-distribution centered at size $y$.
The clone size distribution at time $T$ is given by
\begin{equation}
    P(x, T) = \int_{0}^T \ud t P(t) G(x, 0, t),
\end{equation}
where $t$ is the clonal age. For a constant immigration rate $t$ is uniformly distributed and we obtain by integration
\begin{equation}
    P(x, T) = \frac{e^{\frac{f_0 x (1-\theta(x))}{\sigma^2}} \erfc\left( \frac{|x|-f_0 T}{\sqrt{4 T \sigma^2}}\right) - e^{\frac{f_0 x \theta(x)}{\sigma^2}} \erfc\left( \frac{|x|+f_0 T}{\sqrt{4 T \sigma^2}}\right)} {2 f_0 T},
\end{equation}
where $\theta(x)$ is the Heaviside step function, $\theta(x) = 0$ for $x<0$ and $\theta(x) = 1$ otherwise.
For large $T$ and $x>0$ this reduces to
\begin{equation}
    P(x) \to e^{\frac{f_0 x}{\sigma^2}} / (-f_0 T),
\end{equation}
which implies
\begin{equation}
    P(C) \sim C^{-(1+\alpha)} \quad \text{with}\, \alpha = - f_0 / \sigma^2,
\end{equation}
recovering the steady-state result from \cite{Desponds2016}. 

Setting $f_0 = - \alpha \sigma^2$ and rescaling age as $\tau = T \sigma^2$, we can rewrite the finite time solution as 
\begin{equation} \label{eqrescaled}
    P(x, \tau) = \frac{e^{-\alpha x \theta(x)} \erfc\left( \frac{|x|+\tau\alpha}{\sqrt{4 \tau}}\right) - e^{-\alpha x (1-\theta(x))} \erfc\left( \frac{|x|-\alpha \tau}{\sqrt{4 \tau}}\right)} {2 \alpha \tau}.
\end{equation}
Plotting the cumulative distribution of clone sizes at different effective ages (Fig.~\ref{figfluct}) we observe that the convergence of clone size distributions is slow when $\sigma^2$ is small. Based on estimates for the fluctuation strength from longitudinal data (Fig.~\ref{main-figzeroinsertion}D) we would expect significant deviations from the steady state power-law scaling that persist into adulthood. Thus this mechanism alone is unable to account for the observed power-law scaling in data.

\subsection{A note on the scaling exponent}
A minimal requirement for the existence of a steady state is $f_0 < 0$ ensuring that clones eventually die to balance the recruitment of new clones. This condition still allows such multiplicative processes to produce power-laws with arbitrary exponents as noted before \cite{Desponds2016}. Here, we propose that the parameters should fulfill a stronger condition. In particular, it seems reasonable to require that the large clones do not deterministically take up a larger fraction of the overall repertoire, or equivalently that their expected change in clone size should not exceed one. The mean of the lognormal distribution of clone size change is given by $e^{f_0 + \sigma^2}$, and thus we find the stronger condition
\begin{equation}
    -f_0 < \sigma^2.
\end{equation}
Importantly, it follows that exponents in the vicinity of $\alpha=-1$ arise without fine-tuning as long as the timescale of expected net clonal decay is large compared to the diffusion timescale.

Another perspective on the parameterization is provided by noting that the Langevin equation for $C$ (not $x=\log C$) in the Stratonovic convention includes an extra drift term $-\sigma^2$, to keep $\langle \Delta C \rangle$ independent of the choice of $\sigma$. Alternatively, in the Ito convention the extra drift term arises by Ito's lemma when transforming the equation from $C$ to $x$

\subsection{Predictions for longitudinal fluctuations in clone sizes}
\label{si_longitudinal_modeling}
To quantify longitudinal fluctuations we calculate the mean and variance of log-clonesize changes with respect to a reference time $t_0$. From the model we according to Eq.~\ref{eqpropagator} expect
\begin{align}
    \langle x(t) - x(t_0) \rangle &= f_0 t \\
    \langle (x(t) - x(t_0)  - \langle x(t) - x(t_0) \rangle)^2 \rangle &= 2 \sigma^2 t.
\end{align}
The variance of log-clonesize changes in empirical data involves an additional term $\sigma_S^2$ accounting for sample-to-sample variability. This term is expected not to depend on the time difference, and we can thus determine $\sigma^2$ by linear regression with an intercept that captures the sampling variability $\sigma_S^2$ (Fig.~\ref{main-figlongitudinalmain}B).

We note that a similar approach has been independently proposed in unpublished work by Ferri \cite{Ferri2018}.

\subsection{Relaxation of the zero insertion distribution}
\label{seclongtermdynamics}

Here, we solve for the relaxation dynamics of the zero insertion distribution in a simplified setting. Throughout we use log clone sizes $x = \log C$ for notational convenience. We posit that at time $0$ the power-law distribution $P(x, 0) = \alpha e^{-\alpha x}$ is already established and we further assume that the $r^\star$ largest clones have zero insertion probability $p_{0,-}$ and all smaller or later added clones have probability $p_{0,+}$.
Then the probability that a clone of a given size $x$ has zero insertions is given by
\begin{equation} \label{eqp0}
    P_0(x, t) = \Delta p_0 f_{early}(x, t) + p_{0,+}
\end{equation}
where $\Delta p_0 = p_{0,-}-p_{0,+}$ and $f_{early}(x, t)$ is the fraction of clones of size $x$ and time $t$ that derive from the $r^\star$ largest clones at time $0$.

In the following we determine an analytical formula for $f_{early}(x, t)$ under the assumption that the dynamics leaves the distribution unchanged $P(x, t) = P(x, 0)$. We then have
\begin{equation}
    f_{early}(x, t) = \frac{\int_{x_{min}}^\infty \ud y \, e^{-\alpha y} G(x, y, t)}{e^{-\alpha x}}, 
\end{equation}
where $G(x, y, t)$ as before is the Green's function of the fluctuating proliferation rate dynamics and $x_{min}$ is defined such that the total number of clones times $P(x>x_{min})$ equals $r^\star$. By integration one obtains
\begin{equation}
    f_{early}(x, t) =  \frac{1}{2} e^{\alpha t (f_0+\alpha \sigma^2)} \erfc\left(\frac{x_{min} - x + t (f_0+2\alpha \sigma^2) }{\sqrt{4 \sigma^2 t}}\right),
\end{equation}
which after setting $f_0 = - \alpha \sigma^2$ reduces to
\begin{equation} \label{eqfearly}
    f_{early}(x, t) =  \frac{1}{2} \erfc\left(\frac{x_{min} - x + \alpha \sigma^2 t }{\sqrt{4 \sigma^2 t}}\right).
\end{equation}
To convert clone size into ranks, we note that $\mathrm{rank} \sim e^{-\alpha x}$ and thus $x_{min} - x \sim \frac{1}{\alpha} \log\left(\frac{r}{r^\star}\right)$.
In combination with Eqs.~\ref{eqfearly} and \ref{eqp0} we thus obtain
\begin{equation}
    P_0(r, t) = \frac{\Delta p_0}{2} \erfc \left( \frac{\frac{1}{\alpha}\log\left(r/r^\star\right) + \alpha \sigma^2 t}{\sqrt{4 \sigma^2 t}} \right) + p_{0,+}.
\end{equation}
Defining a characteristic timescale for the diffusive dynamics as $\tau_d = 1/(\alpha \sigma)^2$ we can simplify this expression to
\begin{equation} \label{eqerfc}
    P_0(r, t) = \frac{\Delta p_0}{2} \erfc \left( \frac{\log\left(r/r^\star\right) + t/\tau_d}{2 \sqrt{t/\tau_d}} \right) + p_{0,+}.
\end{equation}

\bibliography{library}